\documentclass[12pt]{article}
\usepackage{epsfig,amsfonts,amssymb}
\usepackage{hyperref}
\usepackage{cite}
\input epsf.sty
\usepackage{cite}

\newcommand{\hab}{}

\newcommand{\pii}{\pi}

\newcommand{\vq}{\xi}

\newcommand{\tree}{}

\def\ZZZ{{\hbox{ Z\kern-1.6mm Z}}}
\def\RRR{{\hbox{ R\kern-2.4mm R}}}
\def\CCC{{\hbox{ C\kern-2.0mm C}}}
\def\zzz{{\hbox{z\kern-1mm z}}}

\newcommand{\ten}{{(10)}}
\newcommand{\bet}{{( b )}}

\newcommand{\qq}{k}
\newcommand{\pp}{l}
\newcommand{\nn}{\nonumber \\}

\newcommand{\vt}{\vartheta}

\newcommand{\vtau} {\vec \tau}
\newcommand{\vj} {\vec J}
\newcommand{\vxi} {\vec \xi}
\newcommand{\vu} {\vec u}
\newcommand{\htau} {\vec \eta}
\newcommand{\vc}{\vec\chi}
\newcommand{\vpsi} {\vec \psi}

\newcommand{\qeq}{{\hbox{=\kern-2.3mm ? \kern.5mm }}}
\renewcommand{\qeq}{=}

\newcommand{\rrho}{r}
\newcommand{\bA}{{\bf A}}
\newcommand{\tx}{\wt x}
\newcommand{\bG}{{\bf G}}
\newcommand{\bF}{{\bar F}}
\newcommand{\bbb}{{\bar b}}
\newcommand{\gam}{\tau}
\newcommand{\eps}{\epsilon}
\newcommand{\vareps}{\varepsilon}
\newcommand{\ra}{\rangle}
\newcommand{\la}{\langle}
\newcommand{\T}{\chi_{T}(k)}
\newcommand{\Tm}{\chi_{T}(k')}
\newcommand{\Cn}{{\cal C}_n}
\newcommand{\vp}{\varphi}
\newcommand{\ve}{\varepsilon}
\newcommand{\tl}{\lambda}
\newcommand{\dt}{(\vec \nabla T)^2}
\newcommand{\hp}{{\wh\Phi}}
\newcommand{\hq}{{\wh Q_B}}
\newcommand{\he}{{\wh\eta_0}}
\newcommand{\ha}{{\wh{A}}}
\newcommand{\lllb}{\Bigl\langle\Bigl\langle}
\newcommand{\rrrb}{\Bigr\rangle\Bigr\rangle}
\newcommand{\tf}{\wt f}
\newcommand{\sss}{{\cal L}_{av}}
\newcommand{\bx}{\bar x}
\newcommand{\bw}{\bar w}
\newcommand{\ws}{{\wt\sigma}}
\newcommand{\wrh}{{\wt\rho}}
\newcommand{\wv}{{\wt v}}

\newcommand{\vv} {\bar v}
\newcommand{\uu} {\bar u}

\newcommand{\K}{{\rm K_1}}
\newcommand{\Kt}{{\rm \widetilde K_1}}

\newcommand{\B}{b'}
\newcommand{\C}{c\,'}
\newcommand{\bB}{\bar b'}
\newcommand{\Bu}{B_{\vec u}}
\newcommand{\VV}{{\cal V}}
\newcommand{\BB}{{\cal B}}
\newcommand{\DD}{{\cal D}}
\newcommand{\BBB}{{\cal B}}
\newcommand{\II}{{\cal I}}
\newcommand{\AAA}{{\cal A}}
\newcommand{\GG}{{\cal G}}
\newcommand{\KK}{{\cal K}}
\newcommand{\fff}{{\bf f}}
\newcommand{\ccc}{{\bf c}}
\newcommand{\FF}{{\cal F}}
\newcommand{\JJ}{{\cal J}}
\newcommand{\HH}{{\cal H}}
\newcommand{\MM}{{\cal M}}
\newcommand{\CC}{{\cal C}}
\newcommand{\bC}{{\bf C}}
\newcommand{\OO}{{\cal O}}
\newcommand{\QQ}{{\cal Q}}
\newcommand{\PP}{{\cal P}}
\newcommand{\EE}{{\cal E}}
\newcommand{\LL}{{\cal L}}

\newcommand{\XX}{{\cal X}}

 \newcommand{\rrr}{\rangle\rangle}
\newcommand{\half}{{1\over 2}}
\newcommand{\wt}{\widetilde}
\newcommand{\wh}{\widehat}
\newcommand{\wc}{\wt}
\newcommand{\wb}{\bar}
\newcommand{\RR}{{\cal R}}
\newcommand{\NN}{{\cal N}}
\newcommand{\TT}{{\cal T}}
\newcommand{\bg}{\bar g}
\newcommand{\ba}{\bar a}
\newcommand{\bc}{\bar c}
\newcommand{\bd}{\bar d}
\newcommand{\bb}{\bar b}
\newcommand{\bT}{\bar \Theta}
\newcommand{\SSS}{{\cal S}}
\newcommand{\tlx}{\left(\tilde \lambda ; X^0(0) \right)}
\newcommand{\al}{\alpha}

\newcommand{\tk}{\tilde \kappa}

\newcommand{\ppp}{\prime\prime}

\newcommand{\omk}{\omega_n(\vec k)}
\newcommand{\onk}{\omega^{(N)}_{\vec k_\perp}}
\newcommand{\tI}{\wt\II}
\newcommand{\hI}{\wh\II}
\newcommand{\nI}{\II}

\newcommand{\cp}{\check\Phi}
\newcommand{\cps}{\Psi}
\newcommand{\crh}{\check\rho}
\newcommand{\cs}{\check\sigma}
\newcommand{\cv}{\check v}
\newcommand{\com}{\check\Omega}

\newcommand{\be}{\begin{equation}}
\newcommand{\ee}{\end{equation}}
\newcommand{\ben}{\begin{eqnarray}\displaystyle}
\newcommand{\een}{\end{eqnarray}}

\newcommand{\refb}[1]{(\ref{#1})}
\newcommand{\p}{\partial}
\newcommand{\sectiono}[1]{\section{#1}\setcounter{equation}{0}}
\newcommand{\subsectiono}[1]{\subsection{#1}\setcounter{equation}{0}}

\newcommand{\zet}{\zeta}

\newcommand{\gsim}{\stackrel{>}{\sim}}
\newcommand{\lsim}{\stackrel{<}{\sim}}

\newcommand{\Lamb}{\Lambda}

\def\one{{\hbox{ 1\kern-.8mm l}}}
\def\zero{{\hbox{ 0\kern-1.5mm 0}}}

\def\wa{{\wh a}}
\def\wb{{\wh b}}
\def\wc{{\wh c}}
\def\wc{\check}
\def\wdd{{\wh d}}

\newcommand{\bi}{{\bf i}}

\renewcommand{\theequation}{\thesection.\arabic{equation}}

\newcommand{\bea}[1]{\begin{eqnarray}\label{#1} }
\newcommand{\eea}{\end{eqnarray}}

\newcommand{\wJ}{\wt J}
\newcommand{\bN}{{\bf N}}

\newcommand{\aaa}{b}

\newcommand{\eqref}{\refb}

\newcommand{\un}{{\rm u}}

\newcommand{\dotalpha}{{\dot{\alpha}}}

\newcommand{\dotbeta}{{\dot{\beta}}}

\newcommand{\dotgamma}{{\dot{\gamma}}}

\newcommand{\dalpha}{\beta}

\newcommand{\Vm}{V}

\newcommand{\gb}{G}

\newcommand{\q}{e}

\newcommand{\PPP}{{\cal P}}

\newcommand{\gold}{\VV_{\rm G}}

\newcommand{\goldc}{\VV^c_{\rm G}}

\usepackage{bm}
\usepackage[table]{xcolor}
\def\rpnote#1{{\color{magenta} #1}}
\def\arnote#1{{\color{blue} #1}}
\def\asnote#1{{\color{red} #1}}

\newcommand{\scalar}{\VV_{\rm S}} 

\newcommand{\wscalar}{\wt\VV_{\rm B}}

\newcommand{\fermion}{\VV_{\rm F}} 

\newcommand{\wfermion}{\wt\VV_{\rm F}}  

\newcommand{\wts}{\wt\Sigma}

\newcommand{\wtsp}{\wt\Sigma^c}

\newcommand{\four}{(4)}

\newcommand{\cL} {\{\hskip -4pt\{}
\newcommand{\cR} {\}\hskip -4pt\}}
\newcommand{\sL} {[\hskip -1.5pt[}
\newcommand{\sR} {]\hskip -1.5pt]}
\newcommand{\oR}{{\overline{\RR}}}

\def\figsubtle{

\def\JPicScale{0.5}
\ifx\JPicScale\undefined\def\JPicScale{1}\fi
\unitlength \JPicScale mm

}

\def\figsofttwotree{

\def\JPicScale{0.6}
\ifx\JPicScale\undefined\def\JPicScale{1}\fi
\unitlength \JPicScale mm
\begin{picture}(160,90)(0,0)

\linethickness{0.3mm}
\put(109.89,44.75){\line(0,1){0.5}}
\multiput(109.88,45.75)(0.02,-0.5){1}{\line(0,-1){0.5}}
\multiput(109.84,46.24)(0.03,-0.5){1}{\line(0,-1){0.5}}
\multiput(109.8,46.74)(0.05,-0.5){1}{\line(0,-1){0.5}}
\multiput(109.73,47.23)(0.07,-0.49){1}{\line(0,-1){0.49}}
\multiput(109.65,47.72)(0.08,-0.49){1}{\line(0,-1){0.49}}
\multiput(109.55,48.21)(0.1,-0.49){1}{\line(0,-1){0.49}}
\multiput(109.43,48.7)(0.12,-0.48){1}{\line(0,-1){0.48}}
\multiput(109.3,49.18)(0.13,-0.48){1}{\line(0,-1){0.48}}
\multiput(109.15,49.65)(0.15,-0.48){1}{\line(0,-1){0.48}}
\multiput(108.99,50.12)(0.16,-0.47){1}{\line(0,-1){0.47}}
\multiput(108.81,50.59)(0.18,-0.46){1}{\line(0,-1){0.46}}
\multiput(108.62,51.04)(0.1,-0.23){2}{\line(0,-1){0.23}}
\multiput(108.41,51.5)(0.1,-0.23){2}{\line(0,-1){0.23}}
\multiput(108.18,51.94)(0.11,-0.22){2}{\line(0,-1){0.22}}
\multiput(107.94,52.38)(0.12,-0.22){2}{\line(0,-1){0.22}}
\multiput(107.69,52.8)(0.13,-0.21){2}{\line(0,-1){0.21}}
\multiput(107.42,53.22)(0.13,-0.21){2}{\line(0,-1){0.21}}
\multiput(107.14,53.63)(0.14,-0.21){2}{\line(0,-1){0.21}}
\multiput(106.84,54.04)(0.15,-0.2){2}{\line(0,-1){0.2}}
\multiput(106.54,54.43)(0.1,-0.13){3}{\line(0,-1){0.13}}
\multiput(106.21,54.81)(0.11,-0.13){3}{\line(0,-1){0.13}}
\multiput(105.88,55.18)(0.11,-0.12){3}{\line(0,-1){0.12}}
\multiput(105.53,55.53)(0.12,-0.12){3}{\line(0,-1){0.12}}
\multiput(105.18,55.88)(0.12,-0.12){3}{\line(1,0){0.12}}
\multiput(104.81,56.21)(0.12,-0.11){3}{\line(1,0){0.12}}
\multiput(104.43,56.54)(0.13,-0.11){3}{\line(1,0){0.13}}
\multiput(104.04,56.84)(0.13,-0.1){3}{\line(1,0){0.13}}
\multiput(103.63,57.14)(0.2,-0.15){2}{\line(1,0){0.2}}
\multiput(103.22,57.42)(0.21,-0.14){2}{\line(1,0){0.21}}
\multiput(102.8,57.69)(0.21,-0.13){2}{\line(1,0){0.21}}
\multiput(102.38,57.94)(0.21,-0.13){2}{\line(1,0){0.21}}
\multiput(101.94,58.18)(0.22,-0.12){2}{\line(1,0){0.22}}
\multiput(101.5,58.41)(0.22,-0.11){2}{\line(1,0){0.22}}
\multiput(101.04,58.62)(0.23,-0.1){2}{\line(1,0){0.23}}
\multiput(100.59,58.81)(0.23,-0.1){2}{\line(1,0){0.23}}
\multiput(100.12,58.99)(0.46,-0.18){1}{\line(1,0){0.46}}
\multiput(99.65,59.15)(0.47,-0.16){1}{\line(1,0){0.47}}
\multiput(99.18,59.3)(0.48,-0.15){1}{\line(1,0){0.48}}
\multiput(98.7,59.43)(0.48,-0.13){1}{\line(1,0){0.48}}
\multiput(98.21,59.55)(0.48,-0.12){1}{\line(1,0){0.48}}
\multiput(97.72,59.65)(0.49,-0.1){1}{\line(1,0){0.49}}
\multiput(97.23,59.73)(0.49,-0.08){1}{\line(1,0){0.49}}
\multiput(96.74,59.8)(0.49,-0.07){1}{\line(1,0){0.49}}
\multiput(96.24,59.84)(0.5,-0.05){1}{\line(1,0){0.5}}
\multiput(95.75,59.88)(0.5,-0.03){1}{\line(1,0){0.5}}
\multiput(95.25,59.89)(0.5,-0.02){1}{\line(1,0){0.5}}
\put(94.75,59.89){\line(1,0){0.5}}
\multiput(94.25,59.88)(0.5,0.02){1}{\line(1,0){0.5}}
\multiput(93.76,59.84)(0.5,0.03){1}{\line(1,0){0.5}}
\multiput(93.26,59.8)(0.5,0.05){1}{\line(1,0){0.5}}
\multiput(92.77,59.73)(0.49,0.07){1}{\line(1,0){0.49}}
\multiput(92.28,59.65)(0.49,0.08){1}{\line(1,0){0.49}}
\multiput(91.79,59.55)(0.49,0.1){1}{\line(1,0){0.49}}
\multiput(91.3,59.43)(0.48,0.12){1}{\line(1,0){0.48}}
\multiput(90.82,59.3)(0.48,0.13){1}{\line(1,0){0.48}}
\multiput(90.35,59.15)(0.48,0.15){1}{\line(1,0){0.48}}
\multiput(89.88,58.99)(0.47,0.16){1}{\line(1,0){0.47}}
\multiput(89.41,58.81)(0.46,0.18){1}{\line(1,0){0.46}}
\multiput(88.96,58.62)(0.23,0.1){2}{\line(1,0){0.23}}
\multiput(88.5,58.41)(0.23,0.1){2}{\line(1,0){0.23}}
\multiput(88.06,58.18)(0.22,0.11){2}{\line(1,0){0.22}}
\multiput(87.62,57.94)(0.22,0.12){2}{\line(1,0){0.22}}
\multiput(87.2,57.69)(0.21,0.13){2}{\line(1,0){0.21}}
\multiput(86.78,57.42)(0.21,0.13){2}{\line(1,0){0.21}}
\multiput(86.37,57.14)(0.21,0.14){2}{\line(1,0){0.21}}
\multiput(85.96,56.84)(0.2,0.15){2}{\line(1,0){0.2}}
\multiput(85.57,56.54)(0.13,0.1){3}{\line(1,0){0.13}}
\multiput(85.19,56.21)(0.13,0.11){3}{\line(1,0){0.13}}
\multiput(84.82,55.88)(0.12,0.11){3}{\line(1,0){0.12}}
\multiput(84.47,55.53)(0.12,0.12){3}{\line(1,0){0.12}}
\multiput(84.12,55.18)(0.12,0.12){3}{\line(0,1){0.12}}
\multiput(83.79,54.81)(0.11,0.12){3}{\line(0,1){0.12}}
\multiput(83.46,54.43)(0.11,0.13){3}{\line(0,1){0.13}}
\multiput(83.16,54.04)(0.1,0.13){3}{\line(0,1){0.13}}
\multiput(82.86,53.63)(0.15,0.2){2}{\line(0,1){0.2}}
\multiput(82.58,53.22)(0.14,0.21){2}{\line(0,1){0.21}}
\multiput(82.31,52.8)(0.13,0.21){2}{\line(0,1){0.21}}
\multiput(82.06,52.38)(0.13,0.21){2}{\line(0,1){0.21}}
\multiput(81.82,51.94)(0.12,0.22){2}{\line(0,1){0.22}}
\multiput(81.59,51.5)(0.11,0.22){2}{\line(0,1){0.22}}
\multiput(81.38,51.04)(0.1,0.23){2}{\line(0,1){0.23}}
\multiput(81.19,50.59)(0.1,0.23){2}{\line(0,1){0.23}}
\multiput(81.01,50.12)(0.18,0.46){1}{\line(0,1){0.46}}
\multiput(80.85,49.65)(0.16,0.47){1}{\line(0,1){0.47}}
\multiput(80.7,49.18)(0.15,0.48){1}{\line(0,1){0.48}}
\multiput(80.57,48.7)(0.13,0.48){1}{\line(0,1){0.48}}
\multiput(80.45,48.21)(0.12,0.48){1}{\line(0,1){0.48}}
\multiput(80.35,47.72)(0.1,0.49){1}{\line(0,1){0.49}}
\multiput(80.27,47.23)(0.08,0.49){1}{\line(0,1){0.49}}
\multiput(80.2,46.74)(0.07,0.49){1}{\line(0,1){0.49}}
\multiput(80.16,46.24)(0.05,0.5){1}{\line(0,1){0.5}}
\multiput(80.12,45.75)(0.03,0.5){1}{\line(0,1){0.5}}
\multiput(80.11,45.25)(0.02,0.5){1}{\line(0,1){0.5}}
\put(80.11,44.75){\line(0,1){0.5}}
\multiput(80.11,44.75)(0.02,-0.5){1}{\line(0,-1){0.5}}
\multiput(80.12,44.25)(0.03,-0.5){1}{\line(0,-1){0.5}}
\multiput(80.16,43.76)(0.05,-0.5){1}{\line(0,-1){0.5}}
\multiput(80.2,43.26)(0.07,-0.49){1}{\line(0,-1){0.49}}
\multiput(80.27,42.77)(0.08,-0.49){1}{\line(0,-1){0.49}}
\multiput(80.35,42.28)(0.1,-0.49){1}{\line(0,-1){0.49}}
\multiput(80.45,41.79)(0.12,-0.48){1}{\line(0,-1){0.48}}
\multiput(80.57,41.3)(0.13,-0.48){1}{\line(0,-1){0.48}}
\multiput(80.7,40.82)(0.15,-0.48){1}{\line(0,-1){0.48}}
\multiput(80.85,40.35)(0.16,-0.47){1}{\line(0,-1){0.47}}
\multiput(81.01,39.88)(0.18,-0.46){1}{\line(0,-1){0.46}}
\multiput(81.19,39.41)(0.1,-0.23){2}{\line(0,-1){0.23}}
\multiput(81.38,38.96)(0.1,-0.23){2}{\line(0,-1){0.23}}
\multiput(81.59,38.5)(0.11,-0.22){2}{\line(0,-1){0.22}}
\multiput(81.82,38.06)(0.12,-0.22){2}{\line(0,-1){0.22}}
\multiput(82.06,37.62)(0.13,-0.21){2}{\line(0,-1){0.21}}
\multiput(82.31,37.2)(0.13,-0.21){2}{\line(0,-1){0.21}}
\multiput(82.58,36.78)(0.14,-0.21){2}{\line(0,-1){0.21}}
\multiput(82.86,36.37)(0.15,-0.2){2}{\line(0,-1){0.2}}
\multiput(83.16,35.96)(0.1,-0.13){3}{\line(0,-1){0.13}}
\multiput(83.46,35.57)(0.11,-0.13){3}{\line(0,-1){0.13}}
\multiput(83.79,35.19)(0.11,-0.12){3}{\line(0,-1){0.12}}
\multiput(84.12,34.82)(0.12,-0.12){3}{\line(0,-1){0.12}}
\multiput(84.47,34.47)(0.12,-0.12){3}{\line(1,0){0.12}}
\multiput(84.82,34.12)(0.12,-0.11){3}{\line(1,0){0.12}}
\multiput(85.19,33.79)(0.13,-0.11){3}{\line(1,0){0.13}}
\multiput(85.57,33.46)(0.13,-0.1){3}{\line(1,0){0.13}}
\multiput(85.96,33.16)(0.2,-0.15){2}{\line(1,0){0.2}}
\multiput(86.37,32.86)(0.21,-0.14){2}{\line(1,0){0.21}}
\multiput(86.78,32.58)(0.21,-0.13){2}{\line(1,0){0.21}}
\multiput(87.2,32.31)(0.21,-0.13){2}{\line(1,0){0.21}}
\multiput(87.62,32.06)(0.22,-0.12){2}{\line(1,0){0.22}}
\multiput(88.06,31.82)(0.22,-0.11){2}{\line(1,0){0.22}}
\multiput(88.5,31.59)(0.23,-0.1){2}{\line(1,0){0.23}}
\multiput(88.96,31.38)(0.23,-0.1){2}{\line(1,0){0.23}}
\multiput(89.41,31.19)(0.46,-0.18){1}{\line(1,0){0.46}}
\multiput(89.88,31.01)(0.47,-0.16){1}{\line(1,0){0.47}}
\multiput(90.35,30.85)(0.48,-0.15){1}{\line(1,0){0.48}}
\multiput(90.82,30.7)(0.48,-0.13){1}{\line(1,0){0.48}}
\multiput(91.3,30.57)(0.48,-0.12){1}{\line(1,0){0.48}}
\multiput(91.79,30.45)(0.49,-0.1){1}{\line(1,0){0.49}}
\multiput(92.28,30.35)(0.49,-0.08){1}{\line(1,0){0.49}}
\multiput(92.77,30.27)(0.49,-0.07){1}{\line(1,0){0.49}}
\multiput(93.26,30.2)(0.5,-0.05){1}{\line(1,0){0.5}}
\multiput(93.76,30.16)(0.5,-0.03){1}{\line(1,0){0.5}}
\multiput(94.25,30.12)(0.5,-0.02){1}{\line(1,0){0.5}}
\put(94.75,30.11){\line(1,0){0.5}}
\multiput(95.25,30.11)(0.5,0.02){1}{\line(1,0){0.5}}
\multiput(95.75,30.12)(0.5,0.03){1}{\line(1,0){0.5}}
\multiput(96.24,30.16)(0.5,0.05){1}{\line(1,0){0.5}}
\multiput(96.74,30.2)(0.49,0.07){1}{\line(1,0){0.49}}
\multiput(97.23,30.27)(0.49,0.08){1}{\line(1,0){0.49}}
\multiput(97.72,30.35)(0.49,0.1){1}{\line(1,0){0.49}}
\multiput(98.21,30.45)(0.48,0.12){1}{\line(1,0){0.48}}
\multiput(98.7,30.57)(0.48,0.13){1}{\line(1,0){0.48}}
\multiput(99.18,30.7)(0.48,0.15){1}{\line(1,0){0.48}}
\multiput(99.65,30.85)(0.47,0.16){1}{\line(1,0){0.47}}
\multiput(100.12,31.01)(0.46,0.18){1}{\line(1,0){0.46}}
\multiput(100.59,31.19)(0.23,0.1){2}{\line(1,0){0.23}}
\multiput(101.04,31.38)(0.23,0.1){2}{\line(1,0){0.23}}
\multiput(101.5,31.59)(0.22,0.11){2}{\line(1,0){0.22}}
\multiput(101.94,31.82)(0.22,0.12){2}{\line(1,0){0.22}}
\multiput(102.38,32.06)(0.21,0.13){2}{\line(1,0){0.21}}
\multiput(102.8,32.31)(0.21,0.13){2}{\line(1,0){0.21}}
\multiput(103.22,32.58)(0.21,0.14){2}{\line(1,0){0.21}}
\multiput(103.63,32.86)(0.2,0.15){2}{\line(1,0){0.2}}
\multiput(104.04,33.16)(0.13,0.1){3}{\line(1,0){0.13}}
\multiput(104.43,33.46)(0.13,0.11){3}{\line(1,0){0.13}}
\multiput(104.81,33.79)(0.12,0.11){3}{\line(1,0){0.12}}
\multiput(105.18,34.12)(0.12,0.12){3}{\line(1,0){0.12}}
\multiput(105.53,34.47)(0.12,0.12){3}{\line(0,1){0.12}}
\multiput(105.88,34.82)(0.11,0.12){3}{\line(0,1){0.12}}
\multiput(106.21,35.19)(0.11,0.13){3}{\line(0,1){0.13}}
\multiput(106.54,35.57)(0.1,0.13){3}{\line(0,1){0.13}}
\multiput(106.84,35.96)(0.15,0.2){2}{\line(0,1){0.2}}
\multiput(107.14,36.37)(0.14,0.21){2}{\line(0,1){0.21}}
\multiput(107.42,36.78)(0.13,0.21){2}{\line(0,1){0.21}}
\multiput(107.69,37.2)(0.13,0.21){2}{\line(0,1){0.21}}
\multiput(107.94,37.62)(0.12,0.22){2}{\line(0,1){0.22}}
\multiput(108.18,38.06)(0.11,0.22){2}{\line(0,1){0.22}}
\multiput(108.41,38.5)(0.1,0.23){2}{\line(0,1){0.23}}
\multiput(108.62,38.96)(0.1,0.23){2}{\line(0,1){0.23}}
\multiput(108.81,39.41)(0.18,0.46){1}{\line(0,1){0.46}}
\multiput(108.99,39.88)(0.16,0.47){1}{\line(0,1){0.47}}
\multiput(109.15,40.35)(0.15,0.48){1}{\line(0,1){0.48}}
\multiput(109.3,40.82)(0.13,0.48){1}{\line(0,1){0.48}}
\multiput(109.43,41.3)(0.12,0.48){1}{\line(0,1){0.48}}
\multiput(109.55,41.79)(0.1,0.49){1}{\line(0,1){0.49}}
\multiput(109.65,42.28)(0.08,0.49){1}{\line(0,1){0.49}}
\multiput(109.73,42.77)(0.07,0.49){1}{\line(0,1){0.49}}
\multiput(109.8,43.26)(0.05,0.5){1}{\line(0,1){0.5}}
\multiput(109.84,43.76)(0.03,0.5){1}{\line(0,1){0.5}}
\multiput(109.88,44.25)(0.02,0.5){1}{\line(0,1){0.5}}

\linethickness{0.8mm}
\put(10,45){\line(1,0){48}}
\linethickness{0.8mm}
\put(58,45){\line(1,0){22}}
\linethickness{0.8mm}
\multiput(109,50)(0.18,0.12){283}{\line(1,0){0.18}}
\linethickness{0.8mm}
\linethickness{0.3mm}
\put(45,10){\line(0,1){35}}
\linethickness{0.3mm}
\multiput(110,90)(0.18,-0.12){163}{\line(1,0){0.18}}
\linethickness{0.8mm}
\multiput(70,15)(0.12,0.16){125}{\line(0,1){0.16}}
\linethickness{0.8mm}
\put(95,10){\line(0,1){20}}
\linethickness{0.8mm}
\multiput(105,35)(0.12,-0.12){125}{\line(1,0){0.12}}
\put(10,50){\makebox(0,0)[cc]{$p_i$}}

\put(40,15){\makebox(0,0)[cc]{$k_1$}}

\put(65,50){\makebox(0,0)[cc]{$p_i+k_1$}}

\put(150,85){\makebox(0,0)[cc]{$p_j$}}

\put(105,85){\makebox(0,0)[cc]{$k_2$}}

\put(130,52){\makebox(0,0)[cc]{$p_j+k_2$}}

\put(95,45){\makebox(0,0)[cc]{$\Gamma$}}

\put(105,15){\makebox(0,0)[cc]{$\cdot$}}

\put(108,18){\makebox(0,0)[cc]{$\cdot$}}

\end{picture}

}

\def\figsoftthreetree{

\def\JPicScale{0.6}
\ifx\JPicScale\undefined\def\JPicScale{1}\fi
\unitlength \JPicScale mm
\begin{picture}(205,80)(0,0)

\linethickness{0.3mm}
\put(180.32,44.75){\line(0,1){0.5}}
\multiput(180.31,45.75)(0.02,-0.5){1}{\line(0,-1){0.5}}
\multiput(180.27,46.25)(0.03,-0.5){1}{\line(0,-1){0.5}}
\multiput(180.22,46.75)(0.05,-0.5){1}{\line(0,-1){0.5}}
\multiput(180.16,47.25)(0.07,-0.5){1}{\line(0,-1){0.5}}
\multiput(180.08,47.74)(0.08,-0.49){1}{\line(0,-1){0.49}}
\multiput(179.98,48.24)(0.1,-0.49){1}{\line(0,-1){0.49}}
\multiput(179.87,48.72)(0.11,-0.49){1}{\line(0,-1){0.49}}
\multiput(179.74,49.21)(0.13,-0.48){1}{\line(0,-1){0.48}}
\multiput(179.59,49.69)(0.15,-0.48){1}{\line(0,-1){0.48}}
\multiput(179.43,50.16)(0.16,-0.47){1}{\line(0,-1){0.47}}
\multiput(179.25,50.63)(0.18,-0.47){1}{\line(0,-1){0.47}}
\multiput(179.06,51.1)(0.1,-0.23){2}{\line(0,-1){0.23}}
\multiput(178.85,51.55)(0.1,-0.23){2}{\line(0,-1){0.23}}
\multiput(178.63,52)(0.11,-0.22){2}{\line(0,-1){0.22}}
\multiput(178.4,52.44)(0.12,-0.22){2}{\line(0,-1){0.22}}
\multiput(178.14,52.88)(0.13,-0.22){2}{\line(0,-1){0.22}}
\multiput(177.88,53.3)(0.13,-0.21){2}{\line(0,-1){0.21}}
\multiput(177.6,53.72)(0.14,-0.21){2}{\line(0,-1){0.21}}
\multiput(177.31,54.13)(0.15,-0.2){2}{\line(0,-1){0.2}}
\multiput(177,54.53)(0.1,-0.13){3}{\line(0,-1){0.13}}
\multiput(176.69,54.91)(0.11,-0.13){3}{\line(0,-1){0.13}}
\multiput(176.36,55.29)(0.11,-0.13){3}{\line(0,-1){0.13}}
\multiput(176.01,55.66)(0.11,-0.12){3}{\line(0,-1){0.12}}
\multiput(175.66,56.01)(0.12,-0.12){3}{\line(0,-1){0.12}}
\multiput(175.29,56.36)(0.12,-0.11){3}{\line(1,0){0.12}}
\multiput(174.91,56.69)(0.13,-0.11){3}{\line(1,0){0.13}}
\multiput(174.53,57)(0.13,-0.11){3}{\line(1,0){0.13}}
\multiput(174.13,57.31)(0.13,-0.1){3}{\line(1,0){0.13}}
\multiput(173.72,57.6)(0.2,-0.15){2}{\line(1,0){0.2}}
\multiput(173.3,57.88)(0.21,-0.14){2}{\line(1,0){0.21}}
\multiput(172.88,58.14)(0.21,-0.13){2}{\line(1,0){0.21}}
\multiput(172.44,58.4)(0.22,-0.13){2}{\line(1,0){0.22}}
\multiput(172,58.63)(0.22,-0.12){2}{\line(1,0){0.22}}
\multiput(171.55,58.85)(0.22,-0.11){2}{\line(1,0){0.22}}
\multiput(171.1,59.06)(0.23,-0.1){2}{\line(1,0){0.23}}
\multiput(170.63,59.25)(0.23,-0.1){2}{\line(1,0){0.23}}
\multiput(170.16,59.43)(0.47,-0.18){1}{\line(1,0){0.47}}
\multiput(169.69,59.59)(0.47,-0.16){1}{\line(1,0){0.47}}
\multiput(169.21,59.74)(0.48,-0.15){1}{\line(1,0){0.48}}
\multiput(168.72,59.87)(0.48,-0.13){1}{\line(1,0){0.48}}
\multiput(168.24,59.98)(0.49,-0.11){1}{\line(1,0){0.49}}
\multiput(167.74,60.08)(0.49,-0.1){1}{\line(1,0){0.49}}
\multiput(167.25,60.16)(0.49,-0.08){1}{\line(1,0){0.49}}
\multiput(166.75,60.22)(0.5,-0.07){1}{\line(1,0){0.5}}
\multiput(166.25,60.27)(0.5,-0.05){1}{\line(1,0){0.5}}
\multiput(165.75,60.31)(0.5,-0.03){1}{\line(1,0){0.5}}
\multiput(165.25,60.32)(0.5,-0.02){1}{\line(1,0){0.5}}
\put(164.75,60.32){\line(1,0){0.5}}
\multiput(164.25,60.31)(0.5,0.02){1}{\line(1,0){0.5}}
\multiput(163.75,60.27)(0.5,0.03){1}{\line(1,0){0.5}}
\multiput(163.25,60.22)(0.5,0.05){1}{\line(1,0){0.5}}
\multiput(162.75,60.16)(0.5,0.07){1}{\line(1,0){0.5}}
\multiput(162.26,60.08)(0.49,0.08){1}{\line(1,0){0.49}}
\multiput(161.76,59.98)(0.49,0.1){1}{\line(1,0){0.49}}
\multiput(161.28,59.87)(0.49,0.11){1}{\line(1,0){0.49}}
\multiput(160.79,59.74)(0.48,0.13){1}{\line(1,0){0.48}}
\multiput(160.31,59.59)(0.48,0.15){1}{\line(1,0){0.48}}
\multiput(159.84,59.43)(0.47,0.16){1}{\line(1,0){0.47}}
\multiput(159.37,59.25)(0.47,0.18){1}{\line(1,0){0.47}}
\multiput(158.9,59.06)(0.23,0.1){2}{\line(1,0){0.23}}
\multiput(158.45,58.85)(0.23,0.1){2}{\line(1,0){0.23}}
\multiput(158,58.63)(0.22,0.11){2}{\line(1,0){0.22}}
\multiput(157.56,58.4)(0.22,0.12){2}{\line(1,0){0.22}}
\multiput(157.12,58.14)(0.22,0.13){2}{\line(1,0){0.22}}
\multiput(156.7,57.88)(0.21,0.13){2}{\line(1,0){0.21}}
\multiput(156.28,57.6)(0.21,0.14){2}{\line(1,0){0.21}}
\multiput(155.87,57.31)(0.2,0.15){2}{\line(1,0){0.2}}
\multiput(155.47,57)(0.13,0.1){3}{\line(1,0){0.13}}
\multiput(155.09,56.69)(0.13,0.11){3}{\line(1,0){0.13}}
\multiput(154.71,56.36)(0.13,0.11){3}{\line(1,0){0.13}}
\multiput(154.34,56.01)(0.12,0.11){3}{\line(1,0){0.12}}
\multiput(153.99,55.66)(0.12,0.12){3}{\line(0,1){0.12}}
\multiput(153.64,55.29)(0.11,0.12){3}{\line(0,1){0.12}}
\multiput(153.31,54.91)(0.11,0.13){3}{\line(0,1){0.13}}
\multiput(153,54.53)(0.11,0.13){3}{\line(0,1){0.13}}
\multiput(152.69,54.13)(0.1,0.13){3}{\line(0,1){0.13}}
\multiput(152.4,53.72)(0.15,0.2){2}{\line(0,1){0.2}}
\multiput(152.12,53.3)(0.14,0.21){2}{\line(0,1){0.21}}
\multiput(151.86,52.88)(0.13,0.21){2}{\line(0,1){0.21}}
\multiput(151.6,52.44)(0.13,0.22){2}{\line(0,1){0.22}}
\multiput(151.37,52)(0.12,0.22){2}{\line(0,1){0.22}}
\multiput(151.15,51.55)(0.11,0.22){2}{\line(0,1){0.22}}
\multiput(150.94,51.1)(0.1,0.23){2}{\line(0,1){0.23}}
\multiput(150.75,50.63)(0.1,0.23){2}{\line(0,1){0.23}}
\multiput(150.57,50.16)(0.18,0.47){1}{\line(0,1){0.47}}
\multiput(150.41,49.69)(0.16,0.47){1}{\line(0,1){0.47}}
\multiput(150.26,49.21)(0.15,0.48){1}{\line(0,1){0.48}}
\multiput(150.13,48.72)(0.13,0.48){1}{\line(0,1){0.48}}
\multiput(150.02,48.24)(0.11,0.49){1}{\line(0,1){0.49}}
\multiput(149.92,47.74)(0.1,0.49){1}{\line(0,1){0.49}}
\multiput(149.84,47.25)(0.08,0.49){1}{\line(0,1){0.49}}
\multiput(149.78,46.75)(0.07,0.5){1}{\line(0,1){0.5}}
\multiput(149.73,46.25)(0.05,0.5){1}{\line(0,1){0.5}}
\multiput(149.69,45.75)(0.03,0.5){1}{\line(0,1){0.5}}
\multiput(149.68,45.25)(0.02,0.5){1}{\line(0,1){0.5}}
\put(149.68,44.75){\line(0,1){0.5}}
\multiput(149.68,44.75)(0.02,-0.5){1}{\line(0,-1){0.5}}
\multiput(149.69,44.25)(0.03,-0.5){1}{\line(0,-1){0.5}}
\multiput(149.73,43.75)(0.05,-0.5){1}{\line(0,-1){0.5}}
\multiput(149.78,43.25)(0.07,-0.5){1}{\line(0,-1){0.5}}
\multiput(149.84,42.75)(0.08,-0.49){1}{\line(0,-1){0.49}}
\multiput(149.92,42.26)(0.1,-0.49){1}{\line(0,-1){0.49}}
\multiput(150.02,41.76)(0.11,-0.49){1}{\line(0,-1){0.49}}
\multiput(150.13,41.28)(0.13,-0.48){1}{\line(0,-1){0.48}}
\multiput(150.26,40.79)(0.15,-0.48){1}{\line(0,-1){0.48}}
\multiput(150.41,40.31)(0.16,-0.47){1}{\line(0,-1){0.47}}
\multiput(150.57,39.84)(0.18,-0.47){1}{\line(0,-1){0.47}}
\multiput(150.75,39.37)(0.1,-0.23){2}{\line(0,-1){0.23}}
\multiput(150.94,38.9)(0.1,-0.23){2}{\line(0,-1){0.23}}
\multiput(151.15,38.45)(0.11,-0.22){2}{\line(0,-1){0.22}}
\multiput(151.37,38)(0.12,-0.22){2}{\line(0,-1){0.22}}
\multiput(151.6,37.56)(0.13,-0.22){2}{\line(0,-1){0.22}}
\multiput(151.86,37.12)(0.13,-0.21){2}{\line(0,-1){0.21}}
\multiput(152.12,36.7)(0.14,-0.21){2}{\line(0,-1){0.21}}
\multiput(152.4,36.28)(0.15,-0.2){2}{\line(0,-1){0.2}}
\multiput(152.69,35.87)(0.1,-0.13){3}{\line(0,-1){0.13}}
\multiput(153,35.47)(0.11,-0.13){3}{\line(0,-1){0.13}}
\multiput(153.31,35.09)(0.11,-0.13){3}{\line(0,-1){0.13}}
\multiput(153.64,34.71)(0.11,-0.12){3}{\line(0,-1){0.12}}
\multiput(153.99,34.34)(0.12,-0.12){3}{\line(0,-1){0.12}}
\multiput(154.34,33.99)(0.12,-0.11){3}{\line(1,0){0.12}}
\multiput(154.71,33.64)(0.13,-0.11){3}{\line(1,0){0.13}}
\multiput(155.09,33.31)(0.13,-0.11){3}{\line(1,0){0.13}}
\multiput(155.47,33)(0.13,-0.1){3}{\line(1,0){0.13}}
\multiput(155.87,32.69)(0.2,-0.15){2}{\line(1,0){0.2}}
\multiput(156.28,32.4)(0.21,-0.14){2}{\line(1,0){0.21}}
\multiput(156.7,32.12)(0.21,-0.13){2}{\line(1,0){0.21}}
\multiput(157.12,31.86)(0.22,-0.13){2}{\line(1,0){0.22}}
\multiput(157.56,31.6)(0.22,-0.12){2}{\line(1,0){0.22}}
\multiput(158,31.37)(0.22,-0.11){2}{\line(1,0){0.22}}
\multiput(158.45,31.15)(0.23,-0.1){2}{\line(1,0){0.23}}
\multiput(158.9,30.94)(0.23,-0.1){2}{\line(1,0){0.23}}
\multiput(159.37,30.75)(0.47,-0.18){1}{\line(1,0){0.47}}
\multiput(159.84,30.57)(0.47,-0.16){1}{\line(1,0){0.47}}
\multiput(160.31,30.41)(0.48,-0.15){1}{\line(1,0){0.48}}
\multiput(160.79,30.26)(0.48,-0.13){1}{\line(1,0){0.48}}
\multiput(161.28,30.13)(0.49,-0.11){1}{\line(1,0){0.49}}
\multiput(161.76,30.02)(0.49,-0.1){1}{\line(1,0){0.49}}
\multiput(162.26,29.92)(0.49,-0.08){1}{\line(1,0){0.49}}
\multiput(162.75,29.84)(0.5,-0.07){1}{\line(1,0){0.5}}
\multiput(163.25,29.78)(0.5,-0.05){1}{\line(1,0){0.5}}
\multiput(163.75,29.73)(0.5,-0.03){1}{\line(1,0){0.5}}
\multiput(164.25,29.69)(0.5,-0.02){1}{\line(1,0){0.5}}
\put(164.75,29.68){\line(1,0){0.5}}
\multiput(165.25,29.68)(0.5,0.02){1}{\line(1,0){0.5}}
\multiput(165.75,29.69)(0.5,0.03){1}{\line(1,0){0.5}}
\multiput(166.25,29.73)(0.5,0.05){1}{\line(1,0){0.5}}
\multiput(166.75,29.78)(0.5,0.07){1}{\line(1,0){0.5}}
\multiput(167.25,29.84)(0.49,0.08){1}{\line(1,0){0.49}}
\multiput(167.74,29.92)(0.49,0.1){1}{\line(1,0){0.49}}
\multiput(168.24,30.02)(0.49,0.11){1}{\line(1,0){0.49}}
\multiput(168.72,30.13)(0.48,0.13){1}{\line(1,0){0.48}}
\multiput(169.21,30.26)(0.48,0.15){1}{\line(1,0){0.48}}
\multiput(169.69,30.41)(0.47,0.16){1}{\line(1,0){0.47}}
\multiput(170.16,30.57)(0.47,0.18){1}{\line(1,0){0.47}}
\multiput(170.63,30.75)(0.23,0.1){2}{\line(1,0){0.23}}
\multiput(171.1,30.94)(0.23,0.1){2}{\line(1,0){0.23}}
\multiput(171.55,31.15)(0.22,0.11){2}{\line(1,0){0.22}}
\multiput(172,31.37)(0.22,0.12){2}{\line(1,0){0.22}}
\multiput(172.44,31.6)(0.22,0.13){2}{\line(1,0){0.22}}
\multiput(172.88,31.86)(0.21,0.13){2}{\line(1,0){0.21}}
\multiput(173.3,32.12)(0.21,0.14){2}{\line(1,0){0.21}}
\multiput(173.72,32.4)(0.2,0.15){2}{\line(1,0){0.2}}
\multiput(174.13,32.69)(0.13,0.1){3}{\line(1,0){0.13}}
\multiput(174.53,33)(0.13,0.11){3}{\line(1,0){0.13}}
\multiput(174.91,33.31)(0.13,0.11){3}{\line(1,0){0.13}}
\multiput(175.29,33.64)(0.12,0.11){3}{\line(1,0){0.12}}
\multiput(175.66,33.99)(0.12,0.12){3}{\line(0,1){0.12}}
\multiput(176.01,34.34)(0.11,0.12){3}{\line(0,1){0.12}}
\multiput(176.36,34.71)(0.11,0.13){3}{\line(0,1){0.13}}
\multiput(176.69,35.09)(0.11,0.13){3}{\line(0,1){0.13}}
\multiput(177,35.47)(0.1,0.13){3}{\line(0,1){0.13}}
\multiput(177.31,35.87)(0.15,0.2){2}{\line(0,1){0.2}}
\multiput(177.6,36.28)(0.14,0.21){2}{\line(0,1){0.21}}
\multiput(177.88,36.7)(0.13,0.21){2}{\line(0,1){0.21}}
\multiput(178.14,37.12)(0.13,0.22){2}{\line(0,1){0.22}}
\multiput(178.4,37.56)(0.12,0.22){2}{\line(0,1){0.22}}
\multiput(178.63,38)(0.11,0.22){2}{\line(0,1){0.22}}
\multiput(178.85,38.45)(0.1,0.23){2}{\line(0,1){0.23}}
\multiput(179.06,38.9)(0.1,0.23){2}{\line(0,1){0.23}}
\multiput(179.25,39.37)(0.18,0.47){1}{\line(0,1){0.47}}
\multiput(179.43,39.84)(0.16,0.47){1}{\line(0,1){0.47}}
\multiput(179.59,40.31)(0.15,0.48){1}{\line(0,1){0.48}}
\multiput(179.74,40.79)(0.13,0.48){1}{\line(0,1){0.48}}
\multiput(179.87,41.28)(0.11,0.49){1}{\line(0,1){0.49}}
\multiput(179.98,41.76)(0.1,0.49){1}{\line(0,1){0.49}}
\multiput(180.08,42.26)(0.08,0.49){1}{\line(0,1){0.49}}
\multiput(180.16,42.75)(0.07,0.5){1}{\line(0,1){0.5}}
\multiput(180.22,43.25)(0.05,0.5){1}{\line(0,1){0.5}}
\multiput(180.27,43.75)(0.03,0.5){1}{\line(0,1){0.5}}
\multiput(180.31,44.25)(0.02,0.5){1}{\line(0,1){0.5}}

\linethickness{0.8mm}
\put(0,45){\line(1,0){40}}
\linethickness{0.8mm}
\put(40,45){\line(1,0){50}}
\linethickness{0.8mm}
\put(90,45){\line(1,0){60}}
\linethickness{0.3mm}
\put(25,10){\line(0,1){35}}
\linethickness{0.3mm}
\put(75,10){\line(0,1){35}}
\linethickness{0.8mm}
\multiput(170,60)(0.15,0.12){167}{\line(1,0){0.15}}
\linethickness{0.8mm}
\multiput(175,35)(0.18,-0.12){167}{\line(1,0){0.18}}
\linethickness{0.8mm}
\multiput(175,55)(0.3,0.12){83}{\line(1,0){0.3}}
\put(5,50){\makebox(0,0)[cc]{$p_i$}}

\put(30,15){\makebox(0,0)[cc]{$k_1$}}

\put(80,15){\makebox(0,0)[cc]{$k_2$}}

\put(50,50){\makebox(0,0)[cc]{$p_i+k_1$}}

\put(120,50){\makebox(0,0)[cc]{$p_i+k_1+k_2$}}

\put(190,45){\makebox(0,0)[cc]{$\cdot$}}

\put(185,35){\makebox(0,0)[cc]{$\cdot$}}

\put(165,45){\makebox(0,0)[cc]{$\Gamma$}}

\end{picture}

}

\def\figsoftonefieldtree{

\def\JPicScale{0.6}
\ifx\JPicScale\undefined\def\JPicScale{1}\fi
\unitlength \JPicScale mm
\begin{picture}(135,85)(0,0)

\linethickness{0.3mm}
\put(110.36,49.75){\line(0,1){0.5}}
\multiput(110.34,50.75)(0.02,-0.5){1}{\line(0,-1){0.5}}
\multiput(110.31,51.26)(0.03,-0.5){1}{\line(0,-1){0.5}}
\multiput(110.26,51.76)(0.05,-0.5){1}{\line(0,-1){0.5}}
\multiput(110.2,52.25)(0.07,-0.5){1}{\line(0,-1){0.5}}
\multiput(110.11,52.75)(0.08,-0.5){1}{\line(0,-1){0.5}}
\multiput(110.02,53.24)(0.1,-0.49){1}{\line(0,-1){0.49}}
\multiput(109.9,53.73)(0.11,-0.49){1}{\line(0,-1){0.49}}
\multiput(109.77,54.22)(0.13,-0.49){1}{\line(0,-1){0.49}}
\multiput(109.62,54.7)(0.15,-0.48){1}{\line(0,-1){0.48}}
\multiput(109.46,55.18)(0.16,-0.48){1}{\line(0,-1){0.48}}
\multiput(109.29,55.65)(0.18,-0.47){1}{\line(0,-1){0.47}}
\multiput(109.09,56.11)(0.1,-0.23){2}{\line(0,-1){0.23}}
\multiput(108.89,56.57)(0.1,-0.23){2}{\line(0,-1){0.23}}
\multiput(108.66,57.02)(0.11,-0.23){2}{\line(0,-1){0.23}}
\multiput(108.43,57.46)(0.12,-0.22){2}{\line(0,-1){0.22}}
\multiput(108.18,57.9)(0.13,-0.22){2}{\line(0,-1){0.22}}
\multiput(107.91,58.32)(0.13,-0.21){2}{\line(0,-1){0.21}}
\multiput(107.63,58.74)(0.14,-0.21){2}{\line(0,-1){0.21}}
\multiput(107.34,59.15)(0.15,-0.2){2}{\line(0,-1){0.2}}
\multiput(107.03,59.55)(0.1,-0.13){3}{\line(0,-1){0.13}}
\multiput(106.71,59.94)(0.11,-0.13){3}{\line(0,-1){0.13}}
\multiput(106.38,60.32)(0.11,-0.13){3}{\line(0,-1){0.13}}
\multiput(106.04,60.68)(0.11,-0.12){3}{\line(0,-1){0.12}}
\multiput(105.68,61.04)(0.12,-0.12){3}{\line(1,0){0.12}}
\multiput(105.32,61.38)(0.12,-0.11){3}{\line(1,0){0.12}}
\multiput(104.94,61.71)(0.13,-0.11){3}{\line(1,0){0.13}}
\multiput(104.55,62.03)(0.13,-0.11){3}{\line(1,0){0.13}}
\multiput(104.15,62.34)(0.13,-0.1){3}{\line(1,0){0.13}}
\multiput(103.74,62.63)(0.2,-0.15){2}{\line(1,0){0.2}}
\multiput(103.32,62.91)(0.21,-0.14){2}{\line(1,0){0.21}}
\multiput(102.9,63.18)(0.21,-0.13){2}{\line(1,0){0.21}}
\multiput(102.46,63.43)(0.22,-0.13){2}{\line(1,0){0.22}}
\multiput(102.02,63.66)(0.22,-0.12){2}{\line(1,0){0.22}}
\multiput(101.57,63.89)(0.23,-0.11){2}{\line(1,0){0.23}}
\multiput(101.11,64.09)(0.23,-0.1){2}{\line(1,0){0.23}}
\multiput(100.65,64.29)(0.23,-0.1){2}{\line(1,0){0.23}}
\multiput(100.18,64.46)(0.47,-0.18){1}{\line(1,0){0.47}}
\multiput(99.7,64.62)(0.48,-0.16){1}{\line(1,0){0.48}}
\multiput(99.22,64.77)(0.48,-0.15){1}{\line(1,0){0.48}}
\multiput(98.73,64.9)(0.49,-0.13){1}{\line(1,0){0.49}}
\multiput(98.24,65.02)(0.49,-0.11){1}{\line(1,0){0.49}}
\multiput(97.75,65.11)(0.49,-0.1){1}{\line(1,0){0.49}}
\multiput(97.25,65.2)(0.5,-0.08){1}{\line(1,0){0.5}}
\multiput(96.76,65.26)(0.5,-0.07){1}{\line(1,0){0.5}}
\multiput(96.26,65.31)(0.5,-0.05){1}{\line(1,0){0.5}}
\multiput(95.75,65.34)(0.5,-0.03){1}{\line(1,0){0.5}}
\multiput(95.25,65.36)(0.5,-0.02){1}{\line(1,0){0.5}}
\put(94.75,65.36){\line(1,0){0.5}}
\multiput(94.25,65.34)(0.5,0.02){1}{\line(1,0){0.5}}
\multiput(93.74,65.31)(0.5,0.03){1}{\line(1,0){0.5}}
\multiput(93.24,65.26)(0.5,0.05){1}{\line(1,0){0.5}}
\multiput(92.75,65.2)(0.5,0.07){1}{\line(1,0){0.5}}
\multiput(92.25,65.11)(0.5,0.08){1}{\line(1,0){0.5}}
\multiput(91.76,65.02)(0.49,0.1){1}{\line(1,0){0.49}}
\multiput(91.27,64.9)(0.49,0.11){1}{\line(1,0){0.49}}
\multiput(90.78,64.77)(0.49,0.13){1}{\line(1,0){0.49}}
\multiput(90.3,64.62)(0.48,0.15){1}{\line(1,0){0.48}}
\multiput(89.82,64.46)(0.48,0.16){1}{\line(1,0){0.48}}
\multiput(89.35,64.29)(0.47,0.18){1}{\line(1,0){0.47}}
\multiput(88.89,64.09)(0.23,0.1){2}{\line(1,0){0.23}}
\multiput(88.43,63.89)(0.23,0.1){2}{\line(1,0){0.23}}
\multiput(87.98,63.66)(0.23,0.11){2}{\line(1,0){0.23}}
\multiput(87.54,63.43)(0.22,0.12){2}{\line(1,0){0.22}}
\multiput(87.1,63.18)(0.22,0.13){2}{\line(1,0){0.22}}
\multiput(86.68,62.91)(0.21,0.13){2}{\line(1,0){0.21}}
\multiput(86.26,62.63)(0.21,0.14){2}{\line(1,0){0.21}}
\multiput(85.85,62.34)(0.2,0.15){2}{\line(1,0){0.2}}
\multiput(85.45,62.03)(0.13,0.1){3}{\line(1,0){0.13}}
\multiput(85.06,61.71)(0.13,0.11){3}{\line(1,0){0.13}}
\multiput(84.68,61.38)(0.13,0.11){3}{\line(1,0){0.13}}
\multiput(84.32,61.04)(0.12,0.11){3}{\line(1,0){0.12}}
\multiput(83.96,60.68)(0.12,0.12){3}{\line(0,1){0.12}}
\multiput(83.62,60.32)(0.11,0.12){3}{\line(0,1){0.12}}
\multiput(83.29,59.94)(0.11,0.13){3}{\line(0,1){0.13}}
\multiput(82.97,59.55)(0.11,0.13){3}{\line(0,1){0.13}}
\multiput(82.66,59.15)(0.1,0.13){3}{\line(0,1){0.13}}
\multiput(82.37,58.74)(0.15,0.2){2}{\line(0,1){0.2}}
\multiput(82.09,58.32)(0.14,0.21){2}{\line(0,1){0.21}}
\multiput(81.82,57.9)(0.13,0.21){2}{\line(0,1){0.21}}
\multiput(81.57,57.46)(0.13,0.22){2}{\line(0,1){0.22}}
\multiput(81.34,57.02)(0.12,0.22){2}{\line(0,1){0.22}}
\multiput(81.11,56.57)(0.11,0.23){2}{\line(0,1){0.23}}
\multiput(80.91,56.11)(0.1,0.23){2}{\line(0,1){0.23}}
\multiput(80.71,55.65)(0.1,0.23){2}{\line(0,1){0.23}}
\multiput(80.54,55.18)(0.18,0.47){1}{\line(0,1){0.47}}
\multiput(80.38,54.7)(0.16,0.48){1}{\line(0,1){0.48}}
\multiput(80.23,54.22)(0.15,0.48){1}{\line(0,1){0.48}}
\multiput(80.1,53.73)(0.13,0.49){1}{\line(0,1){0.49}}
\multiput(79.98,53.24)(0.11,0.49){1}{\line(0,1){0.49}}
\multiput(79.89,52.75)(0.1,0.49){1}{\line(0,1){0.49}}
\multiput(79.8,52.25)(0.08,0.5){1}{\line(0,1){0.5}}
\multiput(79.74,51.76)(0.07,0.5){1}{\line(0,1){0.5}}
\multiput(79.69,51.26)(0.05,0.5){1}{\line(0,1){0.5}}
\multiput(79.66,50.75)(0.03,0.5){1}{\line(0,1){0.5}}
\multiput(79.64,50.25)(0.02,0.5){1}{\line(0,1){0.5}}
\put(79.64,49.75){\line(0,1){0.5}}
\multiput(79.64,49.75)(0.02,-0.5){1}{\line(0,-1){0.5}}
\multiput(79.66,49.25)(0.03,-0.5){1}{\line(0,-1){0.5}}
\multiput(79.69,48.74)(0.05,-0.5){1}{\line(0,-1){0.5}}
\multiput(79.74,48.24)(0.07,-0.5){1}{\line(0,-1){0.5}}
\multiput(79.8,47.75)(0.08,-0.5){1}{\line(0,-1){0.5}}
\multiput(79.89,47.25)(0.1,-0.49){1}{\line(0,-1){0.49}}
\multiput(79.98,46.76)(0.11,-0.49){1}{\line(0,-1){0.49}}
\multiput(80.1,46.27)(0.13,-0.49){1}{\line(0,-1){0.49}}
\multiput(80.23,45.78)(0.15,-0.48){1}{\line(0,-1){0.48}}
\multiput(80.38,45.3)(0.16,-0.48){1}{\line(0,-1){0.48}}
\multiput(80.54,44.82)(0.18,-0.47){1}{\line(0,-1){0.47}}
\multiput(80.71,44.35)(0.1,-0.23){2}{\line(0,-1){0.23}}
\multiput(80.91,43.89)(0.1,-0.23){2}{\line(0,-1){0.23}}
\multiput(81.11,43.43)(0.11,-0.23){2}{\line(0,-1){0.23}}
\multiput(81.34,42.98)(0.12,-0.22){2}{\line(0,-1){0.22}}
\multiput(81.57,42.54)(0.13,-0.22){2}{\line(0,-1){0.22}}
\multiput(81.82,42.1)(0.13,-0.21){2}{\line(0,-1){0.21}}
\multiput(82.09,41.68)(0.14,-0.21){2}{\line(0,-1){0.21}}
\multiput(82.37,41.26)(0.15,-0.2){2}{\line(0,-1){0.2}}
\multiput(82.66,40.85)(0.1,-0.13){3}{\line(0,-1){0.13}}
\multiput(82.97,40.45)(0.11,-0.13){3}{\line(0,-1){0.13}}
\multiput(83.29,40.06)(0.11,-0.13){3}{\line(0,-1){0.13}}
\multiput(83.62,39.68)(0.11,-0.12){3}{\line(0,-1){0.12}}
\multiput(83.96,39.32)(0.12,-0.12){3}{\line(0,-1){0.12}}
\multiput(84.32,38.96)(0.12,-0.11){3}{\line(1,0){0.12}}
\multiput(84.68,38.62)(0.13,-0.11){3}{\line(1,0){0.13}}
\multiput(85.06,38.29)(0.13,-0.11){3}{\line(1,0){0.13}}
\multiput(85.45,37.97)(0.13,-0.1){3}{\line(1,0){0.13}}
\multiput(85.85,37.66)(0.2,-0.15){2}{\line(1,0){0.2}}
\multiput(86.26,37.37)(0.21,-0.14){2}{\line(1,0){0.21}}
\multiput(86.68,37.09)(0.21,-0.13){2}{\line(1,0){0.21}}
\multiput(87.1,36.82)(0.22,-0.13){2}{\line(1,0){0.22}}
\multiput(87.54,36.57)(0.22,-0.12){2}{\line(1,0){0.22}}
\multiput(87.98,36.34)(0.23,-0.11){2}{\line(1,0){0.23}}
\multiput(88.43,36.11)(0.23,-0.1){2}{\line(1,0){0.23}}
\multiput(88.89,35.91)(0.23,-0.1){2}{\line(1,0){0.23}}
\multiput(89.35,35.71)(0.47,-0.18){1}{\line(1,0){0.47}}
\multiput(89.82,35.54)(0.48,-0.16){1}{\line(1,0){0.48}}
\multiput(90.3,35.38)(0.48,-0.15){1}{\line(1,0){0.48}}
\multiput(90.78,35.23)(0.49,-0.13){1}{\line(1,0){0.49}}
\multiput(91.27,35.1)(0.49,-0.11){1}{\line(1,0){0.49}}
\multiput(91.76,34.98)(0.49,-0.1){1}{\line(1,0){0.49}}
\multiput(92.25,34.89)(0.5,-0.08){1}{\line(1,0){0.5}}
\multiput(92.75,34.8)(0.5,-0.07){1}{\line(1,0){0.5}}
\multiput(93.24,34.74)(0.5,-0.05){1}{\line(1,0){0.5}}
\multiput(93.74,34.69)(0.5,-0.03){1}{\line(1,0){0.5}}
\multiput(94.25,34.66)(0.5,-0.02){1}{\line(1,0){0.5}}
\put(94.75,34.64){\line(1,0){0.5}}
\multiput(95.25,34.64)(0.5,0.02){1}{\line(1,0){0.5}}
\multiput(95.75,34.66)(0.5,0.03){1}{\line(1,0){0.5}}
\multiput(96.26,34.69)(0.5,0.05){1}{\line(1,0){0.5}}
\multiput(96.76,34.74)(0.5,0.07){1}{\line(1,0){0.5}}
\multiput(97.25,34.8)(0.5,0.08){1}{\line(1,0){0.5}}
\multiput(97.75,34.89)(0.49,0.1){1}{\line(1,0){0.49}}
\multiput(98.24,34.98)(0.49,0.11){1}{\line(1,0){0.49}}
\multiput(98.73,35.1)(0.49,0.13){1}{\line(1,0){0.49}}
\multiput(99.22,35.23)(0.48,0.15){1}{\line(1,0){0.48}}
\multiput(99.7,35.38)(0.48,0.16){1}{\line(1,0){0.48}}
\multiput(100.18,35.54)(0.47,0.18){1}{\line(1,0){0.47}}
\multiput(100.65,35.71)(0.23,0.1){2}{\line(1,0){0.23}}
\multiput(101.11,35.91)(0.23,0.1){2}{\line(1,0){0.23}}
\multiput(101.57,36.11)(0.23,0.11){2}{\line(1,0){0.23}}
\multiput(102.02,36.34)(0.22,0.12){2}{\line(1,0){0.22}}
\multiput(102.46,36.57)(0.22,0.13){2}{\line(1,0){0.22}}
\multiput(102.9,36.82)(0.21,0.13){2}{\line(1,0){0.21}}
\multiput(103.32,37.09)(0.21,0.14){2}{\line(1,0){0.21}}
\multiput(103.74,37.37)(0.2,0.15){2}{\line(1,0){0.2}}
\multiput(104.15,37.66)(0.13,0.1){3}{\line(1,0){0.13}}
\multiput(104.55,37.97)(0.13,0.11){3}{\line(1,0){0.13}}
\multiput(104.94,38.29)(0.13,0.11){3}{\line(1,0){0.13}}
\multiput(105.32,38.62)(0.12,0.11){3}{\line(1,0){0.12}}
\multiput(105.68,38.96)(0.12,0.12){3}{\line(1,0){0.12}}
\multiput(106.04,39.32)(0.11,0.12){3}{\line(0,1){0.12}}
\multiput(106.38,39.68)(0.11,0.13){3}{\line(0,1){0.13}}
\multiput(106.71,40.06)(0.11,0.13){3}{\line(0,1){0.13}}
\multiput(107.03,40.45)(0.1,0.13){3}{\line(0,1){0.13}}
\multiput(107.34,40.85)(0.15,0.2){2}{\line(0,1){0.2}}
\multiput(107.63,41.26)(0.14,0.21){2}{\line(0,1){0.21}}
\multiput(107.91,41.68)(0.13,0.21){2}{\line(0,1){0.21}}
\multiput(108.18,42.1)(0.13,0.22){2}{\line(0,1){0.22}}
\multiput(108.43,42.54)(0.12,0.22){2}{\line(0,1){0.22}}
\multiput(108.66,42.98)(0.11,0.23){2}{\line(0,1){0.23}}
\multiput(108.89,43.43)(0.1,0.23){2}{\line(0,1){0.23}}
\multiput(109.09,43.89)(0.1,0.23){2}{\line(0,1){0.23}}
\multiput(109.29,44.35)(0.18,0.47){1}{\line(0,1){0.47}}
\multiput(109.46,44.82)(0.16,0.48){1}{\line(0,1){0.48}}
\multiput(109.62,45.3)(0.15,0.48){1}{\line(0,1){0.48}}
\multiput(109.77,45.78)(0.13,0.49){1}{\line(0,1){0.49}}
\multiput(109.9,46.27)(0.11,0.49){1}{\line(0,1){0.49}}
\multiput(110.02,46.76)(0.1,0.49){1}{\line(0,1){0.49}}
\multiput(110.11,47.25)(0.08,0.5){1}{\line(0,1){0.5}}
\multiput(110.2,47.75)(0.07,0.5){1}{\line(0,1){0.5}}
\multiput(110.26,48.24)(0.05,0.5){1}{\line(0,1){0.5}}
\multiput(110.31,48.74)(0.03,0.5){1}{\line(0,1){0.5}}
\multiput(110.34,49.25)(0.02,0.5){1}{\line(0,1){0.5}}

\linethickness{0.8mm}
\put(5,50){\line(1,0){45}}
\linethickness{0.8mm}
\put(50,50){\line(1,0){30}}
\linethickness{0.3mm}
\put(35,10){\line(0,1){40}}
\linethickness{0.8mm}
\multiput(100,65)(0.16,0.12){125}{\line(1,0){0.16}}
\linethickness{0.8mm}
\put(110,50){\line(1,0){25}}
\linethickness{0.8mm}
\multiput(100,35)(0.12,-0.16){125}{\line(0,-1){0.16}}

\put(95,50){\makebox(0,0)[cc]{$\Gamma$}}

\put(0,55){\makebox(0,0)[cc]{$p_i$}}

\put(65,55){\makebox(0,0)[cc]{$p_i+k$}}

\put(40,25){\makebox(0,0)[cc]{$k$}}

\put(110,82){\makebox(0,0)[cc]{$p_1$}}

\put(130,55){\makebox(0,0)[cc]{$p_{i-1}$}}

\put(108,10){\makebox(0,0)[cc]{$p_N$}}

\put(115,25){\makebox(0,0)[cc]{$\cdot$}}

\put(118,28){\makebox(0,0)[cc]{$\cdot$}}

\put(115,65){\makebox(0,0)[cc]{$\cdot$}}

\put(118,62){\makebox(0,0)[cc]{$\cdot$}}

\linethickness{0.8mm}
\multiput(108,42)(0.2,-0.12){125}{\line(1,0){0.2}}

\put(132,35){\makebox(0,0)[cc]{$p_{i+1}$}}

\end{picture}

}

\def\figsoftonefield{

\def\JPicScale{0.6}
\ifx\JPicScale\undefined\def\JPicScale{1}\fi
\unitlength \JPicScale mm

}

\def\figsoftone{

\def\JPicScale{0.6}
\ifx\JPicScale\undefined\def\JPicScale{1}\fi
\unitlength \JPicScale mm

}

\def\figprmod{

\def\JPicScale{0.6}
\ifx\JPicScale\undefined\def\JPicScale{1}\fi
\unitlength \JPicScale mm
\begin{picture}(200,70.51)(0,0)
\linethickness{0.3mm}
\put(160.5,49.75){\line(0,1){0.5}}
\multiput(160.49,50.75)(0.01,-0.5){1}{\line(0,-1){0.5}}
\multiput(160.47,51.26)(0.02,-0.5){1}{\line(0,-1){0.5}}
\multiput(160.43,51.76)(0.04,-0.5){1}{\line(0,-1){0.5}}
\multiput(160.38,52.26)(0.05,-0.5){1}{\line(0,-1){0.5}}
\multiput(160.32,52.76)(0.06,-0.5){1}{\line(0,-1){0.5}}
\multiput(160.25,53.26)(0.07,-0.5){1}{\line(0,-1){0.5}}
\multiput(160.16,53.75)(0.09,-0.5){1}{\line(0,-1){0.5}}
\multiput(160.06,54.25)(0.1,-0.49){1}{\line(0,-1){0.49}}
\multiput(159.95,54.74)(0.11,-0.49){1}{\line(0,-1){0.49}}
\multiput(159.83,55.23)(0.12,-0.49){1}{\line(0,-1){0.49}}
\multiput(159.69,55.71)(0.13,-0.49){1}{\line(0,-1){0.49}}
\multiput(159.55,56.19)(0.15,-0.48){1}{\line(0,-1){0.48}}
\multiput(159.39,56.67)(0.16,-0.48){1}{\line(0,-1){0.48}}
\multiput(159.22,57.14)(0.17,-0.47){1}{\line(0,-1){0.47}}
\multiput(159.04,57.61)(0.09,-0.23){2}{\line(0,-1){0.23}}
\multiput(158.85,58.08)(0.1,-0.23){2}{\line(0,-1){0.23}}
\multiput(158.64,58.54)(0.1,-0.23){2}{\line(0,-1){0.23}}
\multiput(158.43,58.99)(0.11,-0.23){2}{\line(0,-1){0.23}}
\multiput(158.2,59.44)(0.11,-0.22){2}{\line(0,-1){0.22}}
\multiput(157.96,59.89)(0.12,-0.22){2}{\line(0,-1){0.22}}
\multiput(157.72,60.33)(0.12,-0.22){2}{\line(0,-1){0.22}}
\multiput(157.46,60.76)(0.13,-0.22){2}{\line(0,-1){0.22}}
\multiput(157.19,61.18)(0.13,-0.21){2}{\line(0,-1){0.21}}
\multiput(156.91,61.6)(0.14,-0.21){2}{\line(0,-1){0.21}}
\multiput(156.62,62.01)(0.14,-0.21){2}{\line(0,-1){0.21}}
\multiput(156.32,62.42)(0.15,-0.2){2}{\line(0,-1){0.2}}
\multiput(156.01,62.81)(0.1,-0.13){3}{\line(0,-1){0.13}}
\multiput(155.69,63.2)(0.11,-0.13){3}{\line(0,-1){0.13}}
\multiput(155.36,63.58)(0.11,-0.13){3}{\line(0,-1){0.13}}
\multiput(155.02,63.96)(0.11,-0.12){3}{\line(0,-1){0.12}}
\multiput(154.68,64.32)(0.12,-0.12){3}{\line(0,-1){0.12}}
\multiput(154.32,64.68)(0.12,-0.12){3}{\line(0,-1){0.12}}
\multiput(153.96,65.02)(0.12,-0.12){3}{\line(1,0){0.12}}
\multiput(153.58,65.36)(0.12,-0.11){3}{\line(1,0){0.12}}
\multiput(153.2,65.69)(0.13,-0.11){3}{\line(1,0){0.13}}
\multiput(152.81,66.01)(0.13,-0.11){3}{\line(1,0){0.13}}
\multiput(152.42,66.32)(0.13,-0.1){3}{\line(1,0){0.13}}
\multiput(152.01,66.62)(0.2,-0.15){2}{\line(1,0){0.2}}
\multiput(151.6,66.91)(0.21,-0.14){2}{\line(1,0){0.21}}
\multiput(151.18,67.19)(0.21,-0.14){2}{\line(1,0){0.21}}
\multiput(150.76,67.46)(0.21,-0.13){2}{\line(1,0){0.21}}
\multiput(150.33,67.72)(0.22,-0.13){2}{\line(1,0){0.22}}
\multiput(149.89,67.96)(0.22,-0.12){2}{\line(1,0){0.22}}
\multiput(149.44,68.2)(0.22,-0.12){2}{\line(1,0){0.22}}
\multiput(148.99,68.43)(0.22,-0.11){2}{\line(1,0){0.22}}
\multiput(148.54,68.64)(0.23,-0.11){2}{\line(1,0){0.23}}
\multiput(148.08,68.85)(0.23,-0.1){2}{\line(1,0){0.23}}
\multiput(147.61,69.04)(0.23,-0.1){2}{\line(1,0){0.23}}
\multiput(147.14,69.22)(0.23,-0.09){2}{\line(1,0){0.23}}
\multiput(146.67,69.39)(0.47,-0.17){1}{\line(1,0){0.47}}
\multiput(146.19,69.55)(0.48,-0.16){1}{\line(1,0){0.48}}
\multiput(145.71,69.69)(0.48,-0.15){1}{\line(1,0){0.48}}
\multiput(145.23,69.83)(0.49,-0.13){1}{\line(1,0){0.49}}
\multiput(144.74,69.95)(0.49,-0.12){1}{\line(1,0){0.49}}
\multiput(144.25,70.06)(0.49,-0.11){1}{\line(1,0){0.49}}
\multiput(143.75,70.16)(0.49,-0.1){1}{\line(1,0){0.49}}
\multiput(143.26,70.25)(0.5,-0.09){1}{\line(1,0){0.5}}
\multiput(142.76,70.32)(0.5,-0.07){1}{\line(1,0){0.5}}
\multiput(142.26,70.38)(0.5,-0.06){1}{\line(1,0){0.5}}
\multiput(141.76,70.43)(0.5,-0.05){1}{\line(1,0){0.5}}
\multiput(141.26,70.47)(0.5,-0.04){1}{\line(1,0){0.5}}
\multiput(140.75,70.49)(0.5,-0.02){1}{\line(1,0){0.5}}
\multiput(140.25,70.5)(0.5,-0.01){1}{\line(1,0){0.5}}
\put(139.75,70.5){\line(1,0){0.5}}
\multiput(139.25,70.49)(0.5,0.01){1}{\line(1,0){0.5}}
\multiput(138.74,70.47)(0.5,0.02){1}{\line(1,0){0.5}}
\multiput(138.24,70.43)(0.5,0.04){1}{\line(1,0){0.5}}
\multiput(137.74,70.38)(0.5,0.05){1}{\line(1,0){0.5}}
\multiput(137.24,70.32)(0.5,0.06){1}{\line(1,0){0.5}}
\multiput(136.74,70.25)(0.5,0.07){1}{\line(1,0){0.5}}
\multiput(136.25,70.16)(0.5,0.09){1}{\line(1,0){0.5}}
\multiput(135.75,70.06)(0.49,0.1){1}{\line(1,0){0.49}}
\multiput(135.26,69.95)(0.49,0.11){1}{\line(1,0){0.49}}
\multiput(134.77,69.83)(0.49,0.12){1}{\line(1,0){0.49}}
\multiput(134.29,69.69)(0.49,0.13){1}{\line(1,0){0.49}}
\multiput(133.81,69.55)(0.48,0.15){1}{\line(1,0){0.48}}
\multiput(133.33,69.39)(0.48,0.16){1}{\line(1,0){0.48}}
\multiput(132.86,69.22)(0.47,0.17){1}{\line(1,0){0.47}}
\multiput(132.39,69.04)(0.23,0.09){2}{\line(1,0){0.23}}
\multiput(131.92,68.85)(0.23,0.1){2}{\line(1,0){0.23}}
\multiput(131.46,68.64)(0.23,0.1){2}{\line(1,0){0.23}}
\multiput(131.01,68.43)(0.23,0.11){2}{\line(1,0){0.23}}
\multiput(130.56,68.2)(0.22,0.11){2}{\line(1,0){0.22}}
\multiput(130.11,67.96)(0.22,0.12){2}{\line(1,0){0.22}}
\multiput(129.67,67.72)(0.22,0.12){2}{\line(1,0){0.22}}
\multiput(129.24,67.46)(0.22,0.13){2}{\line(1,0){0.22}}
\multiput(128.82,67.19)(0.21,0.13){2}{\line(1,0){0.21}}
\multiput(128.4,66.91)(0.21,0.14){2}{\line(1,0){0.21}}
\multiput(127.99,66.62)(0.21,0.14){2}{\line(1,0){0.21}}
\multiput(127.58,66.32)(0.2,0.15){2}{\line(1,0){0.2}}
\multiput(127.19,66.01)(0.13,0.1){3}{\line(1,0){0.13}}
\multiput(126.8,65.69)(0.13,0.11){3}{\line(1,0){0.13}}
\multiput(126.42,65.36)(0.13,0.11){3}{\line(1,0){0.13}}
\multiput(126.04,65.02)(0.12,0.11){3}{\line(1,0){0.12}}
\multiput(125.68,64.68)(0.12,0.12){3}{\line(1,0){0.12}}
\multiput(125.32,64.32)(0.12,0.12){3}{\line(1,0){0.12}}
\multiput(124.98,63.96)(0.12,0.12){3}{\line(0,1){0.12}}
\multiput(124.64,63.58)(0.11,0.12){3}{\line(0,1){0.12}}
\multiput(124.31,63.2)(0.11,0.13){3}{\line(0,1){0.13}}
\multiput(123.99,62.81)(0.11,0.13){3}{\line(0,1){0.13}}
\multiput(123.68,62.42)(0.1,0.13){3}{\line(0,1){0.13}}
\multiput(123.38,62.01)(0.15,0.2){2}{\line(0,1){0.2}}
\multiput(123.09,61.6)(0.14,0.21){2}{\line(0,1){0.21}}
\multiput(122.81,61.18)(0.14,0.21){2}{\line(0,1){0.21}}
\multiput(122.54,60.76)(0.13,0.21){2}{\line(0,1){0.21}}
\multiput(122.28,60.33)(0.13,0.22){2}{\line(0,1){0.22}}
\multiput(122.04,59.89)(0.12,0.22){2}{\line(0,1){0.22}}
\multiput(121.8,59.44)(0.12,0.22){2}{\line(0,1){0.22}}
\multiput(121.57,58.99)(0.11,0.22){2}{\line(0,1){0.22}}
\multiput(121.36,58.54)(0.11,0.23){2}{\line(0,1){0.23}}
\multiput(121.15,58.08)(0.1,0.23){2}{\line(0,1){0.23}}
\multiput(120.96,57.61)(0.1,0.23){2}{\line(0,1){0.23}}
\multiput(120.78,57.14)(0.09,0.23){2}{\line(0,1){0.23}}
\multiput(120.61,56.67)(0.17,0.47){1}{\line(0,1){0.47}}
\multiput(120.45,56.19)(0.16,0.48){1}{\line(0,1){0.48}}
\multiput(120.31,55.71)(0.15,0.48){1}{\line(0,1){0.48}}
\multiput(120.17,55.23)(0.13,0.49){1}{\line(0,1){0.49}}
\multiput(120.05,54.74)(0.12,0.49){1}{\line(0,1){0.49}}
\multiput(119.94,54.25)(0.11,0.49){1}{\line(0,1){0.49}}
\multiput(119.84,53.75)(0.1,0.49){1}{\line(0,1){0.49}}
\multiput(119.75,53.26)(0.09,0.5){1}{\line(0,1){0.5}}
\multiput(119.68,52.76)(0.07,0.5){1}{\line(0,1){0.5}}
\multiput(119.62,52.26)(0.06,0.5){1}{\line(0,1){0.5}}
\multiput(119.57,51.76)(0.05,0.5){1}{\line(0,1){0.5}}
\multiput(119.53,51.26)(0.04,0.5){1}{\line(0,1){0.5}}
\multiput(119.51,50.75)(0.02,0.5){1}{\line(0,1){0.5}}
\multiput(119.5,50.25)(0.01,0.5){1}{\line(0,1){0.5}}
\put(119.5,49.75){\line(0,1){0.5}}
\multiput(119.5,49.75)(0.01,-0.5){1}{\line(0,-1){0.5}}
\multiput(119.51,49.25)(0.02,-0.5){1}{\line(0,-1){0.5}}
\multiput(119.53,48.74)(0.04,-0.5){1}{\line(0,-1){0.5}}
\multiput(119.57,48.24)(0.05,-0.5){1}{\line(0,-1){0.5}}
\multiput(119.62,47.74)(0.06,-0.5){1}{\line(0,-1){0.5}}
\multiput(119.68,47.24)(0.07,-0.5){1}{\line(0,-1){0.5}}
\multiput(119.75,46.74)(0.09,-0.5){1}{\line(0,-1){0.5}}
\multiput(119.84,46.25)(0.1,-0.49){1}{\line(0,-1){0.49}}
\multiput(119.94,45.75)(0.11,-0.49){1}{\line(0,-1){0.49}}
\multiput(120.05,45.26)(0.12,-0.49){1}{\line(0,-1){0.49}}
\multiput(120.17,44.77)(0.13,-0.49){1}{\line(0,-1){0.49}}
\multiput(120.31,44.29)(0.15,-0.48){1}{\line(0,-1){0.48}}
\multiput(120.45,43.81)(0.16,-0.48){1}{\line(0,-1){0.48}}
\multiput(120.61,43.33)(0.17,-0.47){1}{\line(0,-1){0.47}}
\multiput(120.78,42.86)(0.09,-0.23){2}{\line(0,-1){0.23}}
\multiput(120.96,42.39)(0.1,-0.23){2}{\line(0,-1){0.23}}
\multiput(121.15,41.92)(0.1,-0.23){2}{\line(0,-1){0.23}}
\multiput(121.36,41.46)(0.11,-0.23){2}{\line(0,-1){0.23}}
\multiput(121.57,41.01)(0.11,-0.22){2}{\line(0,-1){0.22}}
\multiput(121.8,40.56)(0.12,-0.22){2}{\line(0,-1){0.22}}
\multiput(122.04,40.11)(0.12,-0.22){2}{\line(0,-1){0.22}}
\multiput(122.28,39.67)(0.13,-0.22){2}{\line(0,-1){0.22}}
\multiput(122.54,39.24)(0.13,-0.21){2}{\line(0,-1){0.21}}
\multiput(122.81,38.82)(0.14,-0.21){2}{\line(0,-1){0.21}}
\multiput(123.09,38.4)(0.14,-0.21){2}{\line(0,-1){0.21}}
\multiput(123.38,37.99)(0.15,-0.2){2}{\line(0,-1){0.2}}
\multiput(123.68,37.58)(0.1,-0.13){3}{\line(0,-1){0.13}}
\multiput(123.99,37.19)(0.11,-0.13){3}{\line(0,-1){0.13}}
\multiput(124.31,36.8)(0.11,-0.13){3}{\line(0,-1){0.13}}
\multiput(124.64,36.42)(0.11,-0.12){3}{\line(0,-1){0.12}}
\multiput(124.98,36.04)(0.12,-0.12){3}{\line(0,-1){0.12}}
\multiput(125.32,35.68)(0.12,-0.12){3}{\line(1,0){0.12}}
\multiput(125.68,35.32)(0.12,-0.12){3}{\line(1,0){0.12}}
\multiput(126.04,34.98)(0.12,-0.11){3}{\line(1,0){0.12}}
\multiput(126.42,34.64)(0.13,-0.11){3}{\line(1,0){0.13}}
\multiput(126.8,34.31)(0.13,-0.11){3}{\line(1,0){0.13}}
\multiput(127.19,33.99)(0.13,-0.1){3}{\line(1,0){0.13}}
\multiput(127.58,33.68)(0.2,-0.15){2}{\line(1,0){0.2}}
\multiput(127.99,33.38)(0.21,-0.14){2}{\line(1,0){0.21}}
\multiput(128.4,33.09)(0.21,-0.14){2}{\line(1,0){0.21}}
\multiput(128.82,32.81)(0.21,-0.13){2}{\line(1,0){0.21}}
\multiput(129.24,32.54)(0.22,-0.13){2}{\line(1,0){0.22}}
\multiput(129.67,32.28)(0.22,-0.12){2}{\line(1,0){0.22}}
\multiput(130.11,32.04)(0.22,-0.12){2}{\line(1,0){0.22}}
\multiput(130.56,31.8)(0.22,-0.11){2}{\line(1,0){0.22}}
\multiput(131.01,31.57)(0.23,-0.11){2}{\line(1,0){0.23}}
\multiput(131.46,31.36)(0.23,-0.1){2}{\line(1,0){0.23}}
\multiput(131.92,31.15)(0.23,-0.1){2}{\line(1,0){0.23}}
\multiput(132.39,30.96)(0.23,-0.09){2}{\line(1,0){0.23}}
\multiput(132.86,30.78)(0.47,-0.17){1}{\line(1,0){0.47}}
\multiput(133.33,30.61)(0.48,-0.16){1}{\line(1,0){0.48}}
\multiput(133.81,30.45)(0.48,-0.15){1}{\line(1,0){0.48}}
\multiput(134.29,30.31)(0.49,-0.13){1}{\line(1,0){0.49}}
\multiput(134.77,30.17)(0.49,-0.12){1}{\line(1,0){0.49}}
\multiput(135.26,30.05)(0.49,-0.11){1}{\line(1,0){0.49}}
\multiput(135.75,29.94)(0.49,-0.1){1}{\line(1,0){0.49}}
\multiput(136.25,29.84)(0.5,-0.09){1}{\line(1,0){0.5}}
\multiput(136.74,29.75)(0.5,-0.07){1}{\line(1,0){0.5}}
\multiput(137.24,29.68)(0.5,-0.06){1}{\line(1,0){0.5}}
\multiput(137.74,29.62)(0.5,-0.05){1}{\line(1,0){0.5}}
\multiput(138.24,29.57)(0.5,-0.04){1}{\line(1,0){0.5}}
\multiput(138.74,29.53)(0.5,-0.02){1}{\line(1,0){0.5}}
\multiput(139.25,29.51)(0.5,-0.01){1}{\line(1,0){0.5}}
\put(139.75,29.5){\line(1,0){0.5}}
\multiput(140.25,29.5)(0.5,0.01){1}{\line(1,0){0.5}}
\multiput(140.75,29.51)(0.5,0.02){1}{\line(1,0){0.5}}
\multiput(141.26,29.53)(0.5,0.04){1}{\line(1,0){0.5}}
\multiput(141.76,29.57)(0.5,0.05){1}{\line(1,0){0.5}}
\multiput(142.26,29.62)(0.5,0.06){1}{\line(1,0){0.5}}
\multiput(142.76,29.68)(0.5,0.07){1}{\line(1,0){0.5}}
\multiput(143.26,29.75)(0.5,0.09){1}{\line(1,0){0.5}}
\multiput(143.75,29.84)(0.49,0.1){1}{\line(1,0){0.49}}
\multiput(144.25,29.94)(0.49,0.11){1}{\line(1,0){0.49}}
\multiput(144.74,30.05)(0.49,0.12){1}{\line(1,0){0.49}}
\multiput(145.23,30.17)(0.49,0.13){1}{\line(1,0){0.49}}
\multiput(145.71,30.31)(0.48,0.15){1}{\line(1,0){0.48}}
\multiput(146.19,30.45)(0.48,0.16){1}{\line(1,0){0.48}}
\multiput(146.67,30.61)(0.47,0.17){1}{\line(1,0){0.47}}
\multiput(147.14,30.78)(0.23,0.09){2}{\line(1,0){0.23}}
\multiput(147.61,30.96)(0.23,0.1){2}{\line(1,0){0.23}}
\multiput(148.08,31.15)(0.23,0.1){2}{\line(1,0){0.23}}
\multiput(148.54,31.36)(0.23,0.11){2}{\line(1,0){0.23}}
\multiput(148.99,31.57)(0.22,0.11){2}{\line(1,0){0.22}}
\multiput(149.44,31.8)(0.22,0.12){2}{\line(1,0){0.22}}
\multiput(149.89,32.04)(0.22,0.12){2}{\line(1,0){0.22}}
\multiput(150.33,32.28)(0.22,0.13){2}{\line(1,0){0.22}}
\multiput(150.76,32.54)(0.21,0.13){2}{\line(1,0){0.21}}
\multiput(151.18,32.81)(0.21,0.14){2}{\line(1,0){0.21}}
\multiput(151.6,33.09)(0.21,0.14){2}{\line(1,0){0.21}}
\multiput(152.01,33.38)(0.2,0.15){2}{\line(1,0){0.2}}
\multiput(152.42,33.68)(0.13,0.1){3}{\line(1,0){0.13}}
\multiput(152.81,33.99)(0.13,0.11){3}{\line(1,0){0.13}}
\multiput(153.2,34.31)(0.13,0.11){3}{\line(1,0){0.13}}
\multiput(153.58,34.64)(0.12,0.11){3}{\line(1,0){0.12}}
\multiput(153.96,34.98)(0.12,0.12){3}{\line(1,0){0.12}}
\multiput(154.32,35.32)(0.12,0.12){3}{\line(0,1){0.12}}
\multiput(154.68,35.68)(0.12,0.12){3}{\line(0,1){0.12}}
\multiput(155.02,36.04)(0.11,0.12){3}{\line(0,1){0.12}}
\multiput(155.36,36.42)(0.11,0.13){3}{\line(0,1){0.13}}
\multiput(155.69,36.8)(0.11,0.13){3}{\line(0,1){0.13}}
\multiput(156.01,37.19)(0.1,0.13){3}{\line(0,1){0.13}}
\multiput(156.32,37.58)(0.15,0.2){2}{\line(0,1){0.2}}
\multiput(156.62,37.99)(0.14,0.21){2}{\line(0,1){0.21}}
\multiput(156.91,38.4)(0.14,0.21){2}{\line(0,1){0.21}}
\multiput(157.19,38.82)(0.13,0.21){2}{\line(0,1){0.21}}
\multiput(157.46,39.24)(0.13,0.22){2}{\line(0,1){0.22}}
\multiput(157.72,39.67)(0.12,0.22){2}{\line(0,1){0.22}}
\multiput(157.96,40.11)(0.12,0.22){2}{\line(0,1){0.22}}
\multiput(158.2,40.56)(0.11,0.22){2}{\line(0,1){0.22}}
\multiput(158.43,41.01)(0.11,0.23){2}{\line(0,1){0.23}}
\multiput(158.64,41.46)(0.1,0.23){2}{\line(0,1){0.23}}
\multiput(158.85,41.92)(0.1,0.23){2}{\line(0,1){0.23}}
\multiput(159.04,42.39)(0.09,0.23){2}{\line(0,1){0.23}}
\multiput(159.22,42.86)(0.17,0.47){1}{\line(0,1){0.47}}
\multiput(159.39,43.33)(0.16,0.48){1}{\line(0,1){0.48}}
\multiput(159.55,43.81)(0.15,0.48){1}{\line(0,1){0.48}}
\multiput(159.69,44.29)(0.13,0.49){1}{\line(0,1){0.49}}
\multiput(159.83,44.77)(0.12,0.49){1}{\line(0,1){0.49}}
\multiput(159.95,45.26)(0.11,0.49){1}{\line(0,1){0.49}}
\multiput(160.06,45.75)(0.1,0.49){1}{\line(0,1){0.49}}
\multiput(160.16,46.25)(0.09,0.5){1}{\line(0,1){0.5}}
\multiput(160.25,46.74)(0.07,0.5){1}{\line(0,1){0.5}}
\multiput(160.32,47.24)(0.06,0.5){1}{\line(0,1){0.5}}
\multiput(160.38,47.74)(0.05,0.5){1}{\line(0,1){0.5}}
\multiput(160.43,48.24)(0.04,0.5){1}{\line(0,1){0.5}}
\multiput(160.47,48.74)(0.02,0.5){1}{\line(0,1){0.5}}
\multiput(160.49,49.25)(0.01,0.5){1}{\line(0,1){0.5}}

\linethickness{1mm}
\put(85,50){\line(1,0){35}}
\linethickness{1mm}
\put(160,50){\line(1,0){40}}
\linethickness{1mm}
\put(10,50){\line(1,0){50}}
\linethickness{0.3mm}
\put(140,20){\line(0,1){10}}
\put(140,15){\makebox(0,0)[cc]{$\Phi$}}

\put(65,50){\makebox(0,0)[cc]{$ ~ \qquad  + \quad   \lambda$}}

\put(35,45){\makebox(0,0)[cc]{$i\Delta_F$}}

\put(100,45){\makebox(0,0)[cc]{$i\Delta_F$}}

\put(185,45){\makebox(0,0)[cc]{$i\Delta_F$}}

\put(140,50){\makebox(0,0)[cc]{$\Gamma$}}

\end{picture}

}

\def\figwtgam{
\def\JPicScale{0.6}
\ifx\JPicScale\undefined\def\JPicScale{1}\fi
\unitlength \JPicScale mm

}

\def\figsoftthreefieldtree{

\def\JPicScale{0.5}
\ifx\JPicScale\undefined\def\JPicScale{1}\fi
\unitlength \JPicScale mm
\begin{picture}(135,90)(0,0)
\linethickness{0.3mm}
\put(105.03,48.5){\line(0,1){0.5}}
\multiput(105.02,49.5)(0.01,-0.5){1}{\line(0,-1){0.5}}
\multiput(105,50)(0.02,-0.5){1}{\line(0,-1){0.5}}
\multiput(104.97,50.49)(0.03,-0.5){1}{\line(0,-1){0.5}}
\multiput(104.92,50.99)(0.04,-0.5){1}{\line(0,-1){0.5}}
\multiput(104.87,51.49)(0.06,-0.5){1}{\line(0,-1){0.5}}
\multiput(104.8,51.98)(0.07,-0.49){1}{\line(0,-1){0.49}}
\multiput(104.73,52.47)(0.08,-0.49){1}{\line(0,-1){0.49}}
\multiput(104.64,52.96)(0.09,-0.49){1}{\line(0,-1){0.49}}
\multiput(104.54,53.45)(0.1,-0.49){1}{\line(0,-1){0.49}}
\multiput(104.43,53.94)(0.11,-0.49){1}{\line(0,-1){0.49}}
\multiput(104.31,54.42)(0.12,-0.48){1}{\line(0,-1){0.48}}
\multiput(104.18,54.9)(0.13,-0.48){1}{\line(0,-1){0.48}}
\multiput(104.04,55.38)(0.14,-0.48){1}{\line(0,-1){0.48}}
\multiput(103.89,55.86)(0.15,-0.47){1}{\line(0,-1){0.47}}
\multiput(103.72,56.33)(0.16,-0.47){1}{\line(0,-1){0.47}}
\multiput(103.55,56.79)(0.17,-0.47){1}{\line(0,-1){0.47}}
\multiput(103.37,57.26)(0.09,-0.23){2}{\line(0,-1){0.23}}
\multiput(103.17,57.72)(0.1,-0.23){2}{\line(0,-1){0.23}}
\multiput(102.97,58.17)(0.1,-0.23){2}{\line(0,-1){0.23}}
\multiput(102.76,58.62)(0.11,-0.23){2}{\line(0,-1){0.23}}
\multiput(102.53,59.07)(0.11,-0.22){2}{\line(0,-1){0.22}}
\multiput(102.3,59.51)(0.12,-0.22){2}{\line(0,-1){0.22}}
\multiput(102.06,59.95)(0.12,-0.22){2}{\line(0,-1){0.22}}
\multiput(101.8,60.38)(0.13,-0.21){2}{\line(0,-1){0.21}}
\multiput(101.54,60.8)(0.13,-0.21){2}{\line(0,-1){0.21}}
\multiput(101.27,61.22)(0.14,-0.21){2}{\line(0,-1){0.21}}
\multiput(100.99,61.63)(0.14,-0.21){2}{\line(0,-1){0.21}}
\multiput(100.7,62.04)(0.14,-0.2){2}{\line(0,-1){0.2}}
\multiput(100.4,62.44)(0.15,-0.2){2}{\line(0,-1){0.2}}
\multiput(100.1,62.83)(0.1,-0.13){3}{\line(0,-1){0.13}}
\multiput(99.78,63.21)(0.11,-0.13){3}{\line(0,-1){0.13}}
\multiput(99.46,63.59)(0.11,-0.13){3}{\line(0,-1){0.13}}
\multiput(99.12,63.96)(0.11,-0.12){3}{\line(0,-1){0.12}}
\multiput(98.78,64.33)(0.11,-0.12){3}{\line(0,-1){0.12}}
\multiput(98.43,64.68)(0.12,-0.12){3}{\line(0,-1){0.12}}
\multiput(98.08,65.03)(0.12,-0.12){3}{\line(1,0){0.12}}
\multiput(97.71,65.37)(0.12,-0.11){3}{\line(1,0){0.12}}
\multiput(97.34,65.71)(0.12,-0.11){3}{\line(1,0){0.12}}
\multiput(96.96,66.03)(0.13,-0.11){3}{\line(1,0){0.13}}
\multiput(96.58,66.35)(0.13,-0.11){3}{\line(1,0){0.13}}
\multiput(96.19,66.65)(0.13,-0.1){3}{\line(1,0){0.13}}
\multiput(95.79,66.95)(0.2,-0.15){2}{\line(1,0){0.2}}
\multiput(95.38,67.24)(0.2,-0.14){2}{\line(1,0){0.2}}
\multiput(94.97,67.52)(0.21,-0.14){2}{\line(1,0){0.21}}
\multiput(94.55,67.79)(0.21,-0.14){2}{\line(1,0){0.21}}
\multiput(94.13,68.05)(0.21,-0.13){2}{\line(1,0){0.21}}
\multiput(93.7,68.31)(0.21,-0.13){2}{\line(1,0){0.21}}
\multiput(93.26,68.55)(0.22,-0.12){2}{\line(1,0){0.22}}
\multiput(92.82,68.78)(0.22,-0.12){2}{\line(1,0){0.22}}
\multiput(92.37,69.01)(0.22,-0.11){2}{\line(1,0){0.22}}
\multiput(91.92,69.22)(0.23,-0.11){2}{\line(1,0){0.23}}
\multiput(91.47,69.42)(0.23,-0.1){2}{\line(1,0){0.23}}
\multiput(91.01,69.62)(0.23,-0.1){2}{\line(1,0){0.23}}
\multiput(90.54,69.8)(0.23,-0.09){2}{\line(1,0){0.23}}
\multiput(90.08,69.97)(0.47,-0.17){1}{\line(1,0){0.47}}
\multiput(89.61,70.14)(0.47,-0.16){1}{\line(1,0){0.47}}
\multiput(89.13,70.29)(0.47,-0.15){1}{\line(1,0){0.47}}
\multiput(88.65,70.43)(0.48,-0.14){1}{\line(1,0){0.48}}
\multiput(88.17,70.56)(0.48,-0.13){1}{\line(1,0){0.48}}
\multiput(87.69,70.68)(0.48,-0.12){1}{\line(1,0){0.48}}
\multiput(87.2,70.79)(0.49,-0.11){1}{\line(1,0){0.49}}
\multiput(86.71,70.89)(0.49,-0.1){1}{\line(1,0){0.49}}
\multiput(86.22,70.98)(0.49,-0.09){1}{\line(1,0){0.49}}
\multiput(85.73,71.05)(0.49,-0.08){1}{\line(1,0){0.49}}
\multiput(85.24,71.12)(0.49,-0.07){1}{\line(1,0){0.49}}
\multiput(84.74,71.17)(0.5,-0.06){1}{\line(1,0){0.5}}
\multiput(84.24,71.22)(0.5,-0.04){1}{\line(1,0){0.5}}
\multiput(83.75,71.25)(0.5,-0.03){1}{\line(1,0){0.5}}
\multiput(83.25,71.27)(0.5,-0.02){1}{\line(1,0){0.5}}
\multiput(82.75,71.28)(0.5,-0.01){1}{\line(1,0){0.5}}
\put(82.25,71.28){\line(1,0){0.5}}
\multiput(81.75,71.27)(0.5,0.01){1}{\line(1,0){0.5}}
\multiput(81.25,71.25)(0.5,0.02){1}{\line(1,0){0.5}}
\multiput(80.76,71.22)(0.5,0.03){1}{\line(1,0){0.5}}
\multiput(80.26,71.17)(0.5,0.04){1}{\line(1,0){0.5}}
\multiput(79.76,71.12)(0.5,0.06){1}{\line(1,0){0.5}}
\multiput(79.27,71.05)(0.49,0.07){1}{\line(1,0){0.49}}
\multiput(78.78,70.98)(0.49,0.08){1}{\line(1,0){0.49}}
\multiput(78.29,70.89)(0.49,0.09){1}{\line(1,0){0.49}}
\multiput(77.8,70.79)(0.49,0.1){1}{\line(1,0){0.49}}
\multiput(77.31,70.68)(0.49,0.11){1}{\line(1,0){0.49}}
\multiput(76.83,70.56)(0.48,0.12){1}{\line(1,0){0.48}}
\multiput(76.35,70.43)(0.48,0.13){1}{\line(1,0){0.48}}
\multiput(75.87,70.29)(0.48,0.14){1}{\line(1,0){0.48}}
\multiput(75.39,70.14)(0.47,0.15){1}{\line(1,0){0.47}}
\multiput(74.92,69.97)(0.47,0.16){1}{\line(1,0){0.47}}
\multiput(74.46,69.8)(0.47,0.17){1}{\line(1,0){0.47}}
\multiput(73.99,69.62)(0.23,0.09){2}{\line(1,0){0.23}}
\multiput(73.53,69.42)(0.23,0.1){2}{\line(1,0){0.23}}
\multiput(73.08,69.22)(0.23,0.1){2}{\line(1,0){0.23}}
\multiput(72.63,69.01)(0.23,0.11){2}{\line(1,0){0.23}}
\multiput(72.18,68.78)(0.22,0.11){2}{\line(1,0){0.22}}
\multiput(71.74,68.55)(0.22,0.12){2}{\line(1,0){0.22}}
\multiput(71.3,68.31)(0.22,0.12){2}{\line(1,0){0.22}}
\multiput(70.87,68.05)(0.21,0.13){2}{\line(1,0){0.21}}
\multiput(70.45,67.79)(0.21,0.13){2}{\line(1,0){0.21}}
\multiput(70.03,67.52)(0.21,0.14){2}{\line(1,0){0.21}}
\multiput(69.62,67.24)(0.21,0.14){2}{\line(1,0){0.21}}
\multiput(69.21,66.95)(0.2,0.14){2}{\line(1,0){0.2}}
\multiput(68.81,66.65)(0.2,0.15){2}{\line(1,0){0.2}}
\multiput(68.42,66.35)(0.13,0.1){3}{\line(1,0){0.13}}
\multiput(68.04,66.03)(0.13,0.11){3}{\line(1,0){0.13}}
\multiput(67.66,65.71)(0.13,0.11){3}{\line(1,0){0.13}}
\multiput(67.29,65.37)(0.12,0.11){3}{\line(1,0){0.12}}
\multiput(66.92,65.03)(0.12,0.11){3}{\line(1,0){0.12}}
\multiput(66.57,64.68)(0.12,0.12){3}{\line(1,0){0.12}}
\multiput(66.22,64.33)(0.12,0.12){3}{\line(0,1){0.12}}
\multiput(65.88,63.96)(0.11,0.12){3}{\line(0,1){0.12}}
\multiput(65.54,63.59)(0.11,0.12){3}{\line(0,1){0.12}}
\multiput(65.22,63.21)(0.11,0.13){3}{\line(0,1){0.13}}
\multiput(64.9,62.83)(0.11,0.13){3}{\line(0,1){0.13}}
\multiput(64.6,62.44)(0.1,0.13){3}{\line(0,1){0.13}}
\multiput(64.3,62.04)(0.15,0.2){2}{\line(0,1){0.2}}
\multiput(64.01,61.63)(0.14,0.2){2}{\line(0,1){0.2}}
\multiput(63.73,61.22)(0.14,0.21){2}{\line(0,1){0.21}}
\multiput(63.46,60.8)(0.14,0.21){2}{\line(0,1){0.21}}
\multiput(63.2,60.38)(0.13,0.21){2}{\line(0,1){0.21}}
\multiput(62.94,59.95)(0.13,0.21){2}{\line(0,1){0.21}}
\multiput(62.7,59.51)(0.12,0.22){2}{\line(0,1){0.22}}
\multiput(62.47,59.07)(0.12,0.22){2}{\line(0,1){0.22}}
\multiput(62.24,58.62)(0.11,0.22){2}{\line(0,1){0.22}}
\multiput(62.03,58.17)(0.11,0.23){2}{\line(0,1){0.23}}
\multiput(61.83,57.72)(0.1,0.23){2}{\line(0,1){0.23}}
\multiput(61.63,57.26)(0.1,0.23){2}{\line(0,1){0.23}}
\multiput(61.45,56.79)(0.09,0.23){2}{\line(0,1){0.23}}
\multiput(61.28,56.33)(0.17,0.47){1}{\line(0,1){0.47}}
\multiput(61.11,55.86)(0.16,0.47){1}{\line(0,1){0.47}}
\multiput(60.96,55.38)(0.15,0.47){1}{\line(0,1){0.47}}
\multiput(60.82,54.9)(0.14,0.48){1}{\line(0,1){0.48}}
\multiput(60.69,54.42)(0.13,0.48){1}{\line(0,1){0.48}}
\multiput(60.57,53.94)(0.12,0.48){1}{\line(0,1){0.48}}
\multiput(60.46,53.45)(0.11,0.49){1}{\line(0,1){0.49}}
\multiput(60.36,52.96)(0.1,0.49){1}{\line(0,1){0.49}}
\multiput(60.27,52.47)(0.09,0.49){1}{\line(0,1){0.49}}
\multiput(60.2,51.98)(0.08,0.49){1}{\line(0,1){0.49}}
\multiput(60.13,51.49)(0.07,0.49){1}{\line(0,1){0.49}}
\multiput(60.08,50.99)(0.06,0.5){1}{\line(0,1){0.5}}
\multiput(60.03,50.49)(0.04,0.5){1}{\line(0,1){0.5}}
\multiput(60,50)(0.03,0.5){1}{\line(0,1){0.5}}
\multiput(59.98,49.5)(0.02,0.5){1}{\line(0,1){0.5}}
\multiput(59.97,49)(0.01,0.5){1}{\line(0,1){0.5}}
\put(59.97,48.5){\line(0,1){0.5}}
\multiput(59.97,48.5)(0.01,-0.5){1}{\line(0,-1){0.5}}
\multiput(59.98,48)(0.02,-0.5){1}{\line(0,-1){0.5}}
\multiput(60,47.5)(0.03,-0.5){1}{\line(0,-1){0.5}}
\multiput(60.03,47.01)(0.04,-0.5){1}{\line(0,-1){0.5}}
\multiput(60.08,46.51)(0.06,-0.5){1}{\line(0,-1){0.5}}
\multiput(60.13,46.01)(0.07,-0.49){1}{\line(0,-1){0.49}}
\multiput(60.2,45.52)(0.08,-0.49){1}{\line(0,-1){0.49}}
\multiput(60.27,45.03)(0.09,-0.49){1}{\line(0,-1){0.49}}
\multiput(60.36,44.54)(0.1,-0.49){1}{\line(0,-1){0.49}}
\multiput(60.46,44.05)(0.11,-0.49){1}{\line(0,-1){0.49}}
\multiput(60.57,43.56)(0.12,-0.48){1}{\line(0,-1){0.48}}
\multiput(60.69,43.08)(0.13,-0.48){1}{\line(0,-1){0.48}}
\multiput(60.82,42.6)(0.14,-0.48){1}{\line(0,-1){0.48}}
\multiput(60.96,42.12)(0.15,-0.47){1}{\line(0,-1){0.47}}
\multiput(61.11,41.64)(0.16,-0.47){1}{\line(0,-1){0.47}}
\multiput(61.28,41.17)(0.17,-0.47){1}{\line(0,-1){0.47}}
\multiput(61.45,40.71)(0.09,-0.23){2}{\line(0,-1){0.23}}
\multiput(61.63,40.24)(0.1,-0.23){2}{\line(0,-1){0.23}}
\multiput(61.83,39.78)(0.1,-0.23){2}{\line(0,-1){0.23}}
\multiput(62.03,39.33)(0.11,-0.23){2}{\line(0,-1){0.23}}
\multiput(62.24,38.88)(0.11,-0.22){2}{\line(0,-1){0.22}}
\multiput(62.47,38.43)(0.12,-0.22){2}{\line(0,-1){0.22}}
\multiput(62.7,37.99)(0.12,-0.22){2}{\line(0,-1){0.22}}
\multiput(62.94,37.55)(0.13,-0.21){2}{\line(0,-1){0.21}}
\multiput(63.2,37.12)(0.13,-0.21){2}{\line(0,-1){0.21}}
\multiput(63.46,36.7)(0.14,-0.21){2}{\line(0,-1){0.21}}
\multiput(63.73,36.28)(0.14,-0.21){2}{\line(0,-1){0.21}}
\multiput(64.01,35.87)(0.14,-0.2){2}{\line(0,-1){0.2}}
\multiput(64.3,35.46)(0.15,-0.2){2}{\line(0,-1){0.2}}
\multiput(64.6,35.06)(0.1,-0.13){3}{\line(0,-1){0.13}}
\multiput(64.9,34.67)(0.11,-0.13){3}{\line(0,-1){0.13}}
\multiput(65.22,34.29)(0.11,-0.13){3}{\line(0,-1){0.13}}
\multiput(65.54,33.91)(0.11,-0.12){3}{\line(0,-1){0.12}}
\multiput(65.88,33.54)(0.11,-0.12){3}{\line(0,-1){0.12}}
\multiput(66.22,33.17)(0.12,-0.12){3}{\line(0,-1){0.12}}
\multiput(66.57,32.82)(0.12,-0.12){3}{\line(1,0){0.12}}
\multiput(66.92,32.47)(0.12,-0.11){3}{\line(1,0){0.12}}
\multiput(67.29,32.13)(0.12,-0.11){3}{\line(1,0){0.12}}
\multiput(67.66,31.79)(0.13,-0.11){3}{\line(1,0){0.13}}
\multiput(68.04,31.47)(0.13,-0.11){3}{\line(1,0){0.13}}
\multiput(68.42,31.15)(0.13,-0.1){3}{\line(1,0){0.13}}
\multiput(68.81,30.85)(0.2,-0.15){2}{\line(1,0){0.2}}
\multiput(69.21,30.55)(0.2,-0.14){2}{\line(1,0){0.2}}
\multiput(69.62,30.26)(0.21,-0.14){2}{\line(1,0){0.21}}
\multiput(70.03,29.98)(0.21,-0.14){2}{\line(1,0){0.21}}
\multiput(70.45,29.71)(0.21,-0.13){2}{\line(1,0){0.21}}
\multiput(70.87,29.45)(0.21,-0.13){2}{\line(1,0){0.21}}
\multiput(71.3,29.19)(0.22,-0.12){2}{\line(1,0){0.22}}
\multiput(71.74,28.95)(0.22,-0.12){2}{\line(1,0){0.22}}
\multiput(72.18,28.72)(0.22,-0.11){2}{\line(1,0){0.22}}
\multiput(72.63,28.49)(0.23,-0.11){2}{\line(1,0){0.23}}
\multiput(73.08,28.28)(0.23,-0.1){2}{\line(1,0){0.23}}
\multiput(73.53,28.08)(0.23,-0.1){2}{\line(1,0){0.23}}
\multiput(73.99,27.88)(0.23,-0.09){2}{\line(1,0){0.23}}
\multiput(74.46,27.7)(0.47,-0.17){1}{\line(1,0){0.47}}
\multiput(74.92,27.53)(0.47,-0.16){1}{\line(1,0){0.47}}
\multiput(75.39,27.36)(0.47,-0.15){1}{\line(1,0){0.47}}
\multiput(75.87,27.21)(0.48,-0.14){1}{\line(1,0){0.48}}
\multiput(76.35,27.07)(0.48,-0.13){1}{\line(1,0){0.48}}
\multiput(76.83,26.94)(0.48,-0.12){1}{\line(1,0){0.48}}
\multiput(77.31,26.82)(0.49,-0.11){1}{\line(1,0){0.49}}
\multiput(77.8,26.71)(0.49,-0.1){1}{\line(1,0){0.49}}
\multiput(78.29,26.61)(0.49,-0.09){1}{\line(1,0){0.49}}
\multiput(78.78,26.52)(0.49,-0.08){1}{\line(1,0){0.49}}
\multiput(79.27,26.45)(0.49,-0.07){1}{\line(1,0){0.49}}
\multiput(79.76,26.38)(0.5,-0.06){1}{\line(1,0){0.5}}
\multiput(80.26,26.33)(0.5,-0.04){1}{\line(1,0){0.5}}
\multiput(80.76,26.28)(0.5,-0.03){1}{\line(1,0){0.5}}
\multiput(81.25,26.25)(0.5,-0.02){1}{\line(1,0){0.5}}
\multiput(81.75,26.23)(0.5,-0.01){1}{\line(1,0){0.5}}
\put(82.25,26.22){\line(1,0){0.5}}
\multiput(82.75,26.22)(0.5,0.01){1}{\line(1,0){0.5}}
\multiput(83.25,26.23)(0.5,0.02){1}{\line(1,0){0.5}}
\multiput(83.75,26.25)(0.5,0.03){1}{\line(1,0){0.5}}
\multiput(84.24,26.28)(0.5,0.04){1}{\line(1,0){0.5}}
\multiput(84.74,26.33)(0.5,0.06){1}{\line(1,0){0.5}}
\multiput(85.24,26.38)(0.49,0.07){1}{\line(1,0){0.49}}
\multiput(85.73,26.45)(0.49,0.08){1}{\line(1,0){0.49}}
\multiput(86.22,26.52)(0.49,0.09){1}{\line(1,0){0.49}}
\multiput(86.71,26.61)(0.49,0.1){1}{\line(1,0){0.49}}
\multiput(87.2,26.71)(0.49,0.11){1}{\line(1,0){0.49}}
\multiput(87.69,26.82)(0.48,0.12){1}{\line(1,0){0.48}}
\multiput(88.17,26.94)(0.48,0.13){1}{\line(1,0){0.48}}
\multiput(88.65,27.07)(0.48,0.14){1}{\line(1,0){0.48}}
\multiput(89.13,27.21)(0.47,0.15){1}{\line(1,0){0.47}}
\multiput(89.61,27.36)(0.47,0.16){1}{\line(1,0){0.47}}
\multiput(90.08,27.53)(0.47,0.17){1}{\line(1,0){0.47}}
\multiput(90.54,27.7)(0.23,0.09){2}{\line(1,0){0.23}}
\multiput(91.01,27.88)(0.23,0.1){2}{\line(1,0){0.23}}
\multiput(91.47,28.08)(0.23,0.1){2}{\line(1,0){0.23}}
\multiput(91.92,28.28)(0.23,0.11){2}{\line(1,0){0.23}}
\multiput(92.37,28.49)(0.22,0.11){2}{\line(1,0){0.22}}
\multiput(92.82,28.72)(0.22,0.12){2}{\line(1,0){0.22}}
\multiput(93.26,28.95)(0.22,0.12){2}{\line(1,0){0.22}}
\multiput(93.7,29.19)(0.21,0.13){2}{\line(1,0){0.21}}
\multiput(94.13,29.45)(0.21,0.13){2}{\line(1,0){0.21}}
\multiput(94.55,29.71)(0.21,0.14){2}{\line(1,0){0.21}}
\multiput(94.97,29.98)(0.21,0.14){2}{\line(1,0){0.21}}
\multiput(95.38,30.26)(0.2,0.14){2}{\line(1,0){0.2}}
\multiput(95.79,30.55)(0.2,0.15){2}{\line(1,0){0.2}}
\multiput(96.19,30.85)(0.13,0.1){3}{\line(1,0){0.13}}
\multiput(96.58,31.15)(0.13,0.11){3}{\line(1,0){0.13}}
\multiput(96.96,31.47)(0.13,0.11){3}{\line(1,0){0.13}}
\multiput(97.34,31.79)(0.12,0.11){3}{\line(1,0){0.12}}
\multiput(97.71,32.13)(0.12,0.11){3}{\line(1,0){0.12}}
\multiput(98.08,32.47)(0.12,0.12){3}{\line(1,0){0.12}}
\multiput(98.43,32.82)(0.12,0.12){3}{\line(0,1){0.12}}
\multiput(98.78,33.17)(0.11,0.12){3}{\line(0,1){0.12}}
\multiput(99.12,33.54)(0.11,0.12){3}{\line(0,1){0.12}}
\multiput(99.46,33.91)(0.11,0.13){3}{\line(0,1){0.13}}
\multiput(99.78,34.29)(0.11,0.13){3}{\line(0,1){0.13}}
\multiput(100.1,34.67)(0.1,0.13){3}{\line(0,1){0.13}}
\multiput(100.4,35.06)(0.15,0.2){2}{\line(0,1){0.2}}
\multiput(100.7,35.46)(0.14,0.2){2}{\line(0,1){0.2}}
\multiput(100.99,35.87)(0.14,0.21){2}{\line(0,1){0.21}}
\multiput(101.27,36.28)(0.14,0.21){2}{\line(0,1){0.21}}
\multiput(101.54,36.7)(0.13,0.21){2}{\line(0,1){0.21}}
\multiput(101.8,37.12)(0.13,0.21){2}{\line(0,1){0.21}}
\multiput(102.06,37.55)(0.12,0.22){2}{\line(0,1){0.22}}
\multiput(102.3,37.99)(0.12,0.22){2}{\line(0,1){0.22}}
\multiput(102.53,38.43)(0.11,0.22){2}{\line(0,1){0.22}}
\multiput(102.76,38.88)(0.11,0.23){2}{\line(0,1){0.23}}
\multiput(102.97,39.33)(0.1,0.23){2}{\line(0,1){0.23}}
\multiput(103.17,39.78)(0.1,0.23){2}{\line(0,1){0.23}}
\multiput(103.37,40.24)(0.09,0.23){2}{\line(0,1){0.23}}
\multiput(103.55,40.71)(0.17,0.47){1}{\line(0,1){0.47}}
\multiput(103.72,41.17)(0.16,0.47){1}{\line(0,1){0.47}}
\multiput(103.89,41.64)(0.15,0.47){1}{\line(0,1){0.47}}
\multiput(104.04,42.12)(0.14,0.48){1}{\line(0,1){0.48}}
\multiput(104.18,42.6)(0.13,0.48){1}{\line(0,1){0.48}}
\multiput(104.31,43.08)(0.12,0.48){1}{\line(0,1){0.48}}
\multiput(104.43,43.56)(0.11,0.49){1}{\line(0,1){0.49}}
\multiput(104.54,44.05)(0.1,0.49){1}{\line(0,1){0.49}}
\multiput(104.64,44.54)(0.09,0.49){1}{\line(0,1){0.49}}
\multiput(104.73,45.03)(0.08,0.49){1}{\line(0,1){0.49}}
\multiput(104.8,45.52)(0.07,0.49){1}{\line(0,1){0.49}}
\multiput(104.87,46.01)(0.06,0.5){1}{\line(0,1){0.5}}
\multiput(104.92,46.51)(0.04,0.5){1}{\line(0,1){0.5}}
\multiput(104.97,47.01)(0.03,0.5){1}{\line(0,1){0.5}}
\multiput(105,47.5)(0.02,0.5){1}{\line(0,1){0.5}}
\multiput(105.02,48)(0.01,0.5){1}{\line(0,1){0.5}}

\linethickness{1mm}
\put(30,50){\line(1,0){30}}
\linethickness{1mm}
\multiput(60,90)(0.12,-0.16){125}{\line(0,-1){0.16}}
\linethickness{1mm}
\multiput(104,55)(0.36,0.12){83}{\line(1,0){0.36}}
\linethickness{0.3mm}
\multiput(95,30)(0.12,-0.16){125}{\line(0,-1){0.16}}
\put(35,55){\makebox(0,0)[cc]{$p_1$}}

\put(72,85){\makebox(0,0)[cc]{$p_2$}}

\put(95,80){\makebox(0,0)[cc]{$\cdot$}}

\put(110,70){\makebox(0,0)[cc]{$\cdot$}}

\put(120,55){\makebox(0,0)[cc]{$p_N$}}

\put(110,20){\makebox(0,0)[cc]{$k$}}

\put(80,50){\makebox(0,0)[cc]{$\wt\Gamma$}}

\end{picture}

}

\def\figsoftthreefield{

\def\JPicScale{0.6}
\ifx\JPicScale\undefined\def\JPicScale{1}\fi
\unitlength \JPicScale mm
\begin{picture}(135,90)(0,0)
\linethickness{0.3mm}
\put(105.03,48.5){\line(0,1){0.5}}
\multiput(105.02,49.5)(0.01,-0.5){1}{\line(0,-1){0.5}}
\multiput(105,50)(0.02,-0.5){1}{\line(0,-1){0.5}}
\multiput(104.97,50.49)(0.03,-0.5){1}{\line(0,-1){0.5}}
\multiput(104.92,50.99)(0.04,-0.5){1}{\line(0,-1){0.5}}
\multiput(104.87,51.49)(0.06,-0.5){1}{\line(0,-1){0.5}}
\multiput(104.8,51.98)(0.07,-0.49){1}{\line(0,-1){0.49}}
\multiput(104.73,52.47)(0.08,-0.49){1}{\line(0,-1){0.49}}
\multiput(104.64,52.96)(0.09,-0.49){1}{\line(0,-1){0.49}}
\multiput(104.54,53.45)(0.1,-0.49){1}{\line(0,-1){0.49}}
\multiput(104.43,53.94)(0.11,-0.49){1}{\line(0,-1){0.49}}
\multiput(104.31,54.42)(0.12,-0.48){1}{\line(0,-1){0.48}}
\multiput(104.18,54.9)(0.13,-0.48){1}{\line(0,-1){0.48}}
\multiput(104.04,55.38)(0.14,-0.48){1}{\line(0,-1){0.48}}
\multiput(103.89,55.86)(0.15,-0.47){1}{\line(0,-1){0.47}}
\multiput(103.72,56.33)(0.16,-0.47){1}{\line(0,-1){0.47}}
\multiput(103.55,56.79)(0.17,-0.47){1}{\line(0,-1){0.47}}
\multiput(103.37,57.26)(0.09,-0.23){2}{\line(0,-1){0.23}}
\multiput(103.17,57.72)(0.1,-0.23){2}{\line(0,-1){0.23}}
\multiput(102.97,58.17)(0.1,-0.23){2}{\line(0,-1){0.23}}
\multiput(102.76,58.62)(0.11,-0.23){2}{\line(0,-1){0.23}}
\multiput(102.53,59.07)(0.11,-0.22){2}{\line(0,-1){0.22}}
\multiput(102.3,59.51)(0.12,-0.22){2}{\line(0,-1){0.22}}
\multiput(102.06,59.95)(0.12,-0.22){2}{\line(0,-1){0.22}}
\multiput(101.8,60.38)(0.13,-0.21){2}{\line(0,-1){0.21}}
\multiput(101.54,60.8)(0.13,-0.21){2}{\line(0,-1){0.21}}
\multiput(101.27,61.22)(0.14,-0.21){2}{\line(0,-1){0.21}}
\multiput(100.99,61.63)(0.14,-0.21){2}{\line(0,-1){0.21}}
\multiput(100.7,62.04)(0.14,-0.2){2}{\line(0,-1){0.2}}
\multiput(100.4,62.44)(0.15,-0.2){2}{\line(0,-1){0.2}}
\multiput(100.1,62.83)(0.1,-0.13){3}{\line(0,-1){0.13}}
\multiput(99.78,63.21)(0.11,-0.13){3}{\line(0,-1){0.13}}
\multiput(99.46,63.59)(0.11,-0.13){3}{\line(0,-1){0.13}}
\multiput(99.12,63.96)(0.11,-0.12){3}{\line(0,-1){0.12}}
\multiput(98.78,64.33)(0.11,-0.12){3}{\line(0,-1){0.12}}
\multiput(98.43,64.68)(0.12,-0.12){3}{\line(0,-1){0.12}}
\multiput(98.08,65.03)(0.12,-0.12){3}{\line(1,0){0.12}}
\multiput(97.71,65.37)(0.12,-0.11){3}{\line(1,0){0.12}}
\multiput(97.34,65.71)(0.12,-0.11){3}{\line(1,0){0.12}}
\multiput(96.96,66.03)(0.13,-0.11){3}{\line(1,0){0.13}}
\multiput(96.58,66.35)(0.13,-0.11){3}{\line(1,0){0.13}}
\multiput(96.19,66.65)(0.13,-0.1){3}{\line(1,0){0.13}}
\multiput(95.79,66.95)(0.2,-0.15){2}{\line(1,0){0.2}}
\multiput(95.38,67.24)(0.2,-0.14){2}{\line(1,0){0.2}}
\multiput(94.97,67.52)(0.21,-0.14){2}{\line(1,0){0.21}}
\multiput(94.55,67.79)(0.21,-0.14){2}{\line(1,0){0.21}}
\multiput(94.13,68.05)(0.21,-0.13){2}{\line(1,0){0.21}}
\multiput(93.7,68.31)(0.21,-0.13){2}{\line(1,0){0.21}}
\multiput(93.26,68.55)(0.22,-0.12){2}{\line(1,0){0.22}}
\multiput(92.82,68.78)(0.22,-0.12){2}{\line(1,0){0.22}}
\multiput(92.37,69.01)(0.22,-0.11){2}{\line(1,0){0.22}}
\multiput(91.92,69.22)(0.23,-0.11){2}{\line(1,0){0.23}}
\multiput(91.47,69.42)(0.23,-0.1){2}{\line(1,0){0.23}}
\multiput(91.01,69.62)(0.23,-0.1){2}{\line(1,0){0.23}}
\multiput(90.54,69.8)(0.23,-0.09){2}{\line(1,0){0.23}}
\multiput(90.08,69.97)(0.47,-0.17){1}{\line(1,0){0.47}}
\multiput(89.61,70.14)(0.47,-0.16){1}{\line(1,0){0.47}}
\multiput(89.13,70.29)(0.47,-0.15){1}{\line(1,0){0.47}}
\multiput(88.65,70.43)(0.48,-0.14){1}{\line(1,0){0.48}}
\multiput(88.17,70.56)(0.48,-0.13){1}{\line(1,0){0.48}}
\multiput(87.69,70.68)(0.48,-0.12){1}{\line(1,0){0.48}}
\multiput(87.2,70.79)(0.49,-0.11){1}{\line(1,0){0.49}}
\multiput(86.71,70.89)(0.49,-0.1){1}{\line(1,0){0.49}}
\multiput(86.22,70.98)(0.49,-0.09){1}{\line(1,0){0.49}}
\multiput(85.73,71.05)(0.49,-0.08){1}{\line(1,0){0.49}}
\multiput(85.24,71.12)(0.49,-0.07){1}{\line(1,0){0.49}}
\multiput(84.74,71.17)(0.5,-0.06){1}{\line(1,0){0.5}}
\multiput(84.24,71.22)(0.5,-0.04){1}{\line(1,0){0.5}}
\multiput(83.75,71.25)(0.5,-0.03){1}{\line(1,0){0.5}}
\multiput(83.25,71.27)(0.5,-0.02){1}{\line(1,0){0.5}}
\multiput(82.75,71.28)(0.5,-0.01){1}{\line(1,0){0.5}}
\put(82.25,71.28){\line(1,0){0.5}}
\multiput(81.75,71.27)(0.5,0.01){1}{\line(1,0){0.5}}
\multiput(81.25,71.25)(0.5,0.02){1}{\line(1,0){0.5}}
\multiput(80.76,71.22)(0.5,0.03){1}{\line(1,0){0.5}}
\multiput(80.26,71.17)(0.5,0.04){1}{\line(1,0){0.5}}
\multiput(79.76,71.12)(0.5,0.06){1}{\line(1,0){0.5}}
\multiput(79.27,71.05)(0.49,0.07){1}{\line(1,0){0.49}}
\multiput(78.78,70.98)(0.49,0.08){1}{\line(1,0){0.49}}
\multiput(78.29,70.89)(0.49,0.09){1}{\line(1,0){0.49}}
\multiput(77.8,70.79)(0.49,0.1){1}{\line(1,0){0.49}}
\multiput(77.31,70.68)(0.49,0.11){1}{\line(1,0){0.49}}
\multiput(76.83,70.56)(0.48,0.12){1}{\line(1,0){0.48}}
\multiput(76.35,70.43)(0.48,0.13){1}{\line(1,0){0.48}}
\multiput(75.87,70.29)(0.48,0.14){1}{\line(1,0){0.48}}
\multiput(75.39,70.14)(0.47,0.15){1}{\line(1,0){0.47}}
\multiput(74.92,69.97)(0.47,0.16){1}{\line(1,0){0.47}}
\multiput(74.46,69.8)(0.47,0.17){1}{\line(1,0){0.47}}
\multiput(73.99,69.62)(0.23,0.09){2}{\line(1,0){0.23}}
\multiput(73.53,69.42)(0.23,0.1){2}{\line(1,0){0.23}}
\multiput(73.08,69.22)(0.23,0.1){2}{\line(1,0){0.23}}
\multiput(72.63,69.01)(0.23,0.11){2}{\line(1,0){0.23}}
\multiput(72.18,68.78)(0.22,0.11){2}{\line(1,0){0.22}}
\multiput(71.74,68.55)(0.22,0.12){2}{\line(1,0){0.22}}
\multiput(71.3,68.31)(0.22,0.12){2}{\line(1,0){0.22}}
\multiput(70.87,68.05)(0.21,0.13){2}{\line(1,0){0.21}}
\multiput(70.45,67.79)(0.21,0.13){2}{\line(1,0){0.21}}
\multiput(70.03,67.52)(0.21,0.14){2}{\line(1,0){0.21}}
\multiput(69.62,67.24)(0.21,0.14){2}{\line(1,0){0.21}}
\multiput(69.21,66.95)(0.2,0.14){2}{\line(1,0){0.2}}
\multiput(68.81,66.65)(0.2,0.15){2}{\line(1,0){0.2}}
\multiput(68.42,66.35)(0.13,0.1){3}{\line(1,0){0.13}}
\multiput(68.04,66.03)(0.13,0.11){3}{\line(1,0){0.13}}
\multiput(67.66,65.71)(0.13,0.11){3}{\line(1,0){0.13}}
\multiput(67.29,65.37)(0.12,0.11){3}{\line(1,0){0.12}}
\multiput(66.92,65.03)(0.12,0.11){3}{\line(1,0){0.12}}
\multiput(66.57,64.68)(0.12,0.12){3}{\line(1,0){0.12}}
\multiput(66.22,64.33)(0.12,0.12){3}{\line(0,1){0.12}}
\multiput(65.88,63.96)(0.11,0.12){3}{\line(0,1){0.12}}
\multiput(65.54,63.59)(0.11,0.12){3}{\line(0,1){0.12}}
\multiput(65.22,63.21)(0.11,0.13){3}{\line(0,1){0.13}}
\multiput(64.9,62.83)(0.11,0.13){3}{\line(0,1){0.13}}
\multiput(64.6,62.44)(0.1,0.13){3}{\line(0,1){0.13}}
\multiput(64.3,62.04)(0.15,0.2){2}{\line(0,1){0.2}}
\multiput(64.01,61.63)(0.14,0.2){2}{\line(0,1){0.2}}
\multiput(63.73,61.22)(0.14,0.21){2}{\line(0,1){0.21}}
\multiput(63.46,60.8)(0.14,0.21){2}{\line(0,1){0.21}}
\multiput(63.2,60.38)(0.13,0.21){2}{\line(0,1){0.21}}
\multiput(62.94,59.95)(0.13,0.21){2}{\line(0,1){0.21}}
\multiput(62.7,59.51)(0.12,0.22){2}{\line(0,1){0.22}}
\multiput(62.47,59.07)(0.12,0.22){2}{\line(0,1){0.22}}
\multiput(62.24,58.62)(0.11,0.22){2}{\line(0,1){0.22}}
\multiput(62.03,58.17)(0.11,0.23){2}{\line(0,1){0.23}}
\multiput(61.83,57.72)(0.1,0.23){2}{\line(0,1){0.23}}
\multiput(61.63,57.26)(0.1,0.23){2}{\line(0,1){0.23}}
\multiput(61.45,56.79)(0.09,0.23){2}{\line(0,1){0.23}}
\multiput(61.28,56.33)(0.17,0.47){1}{\line(0,1){0.47}}
\multiput(61.11,55.86)(0.16,0.47){1}{\line(0,1){0.47}}
\multiput(60.96,55.38)(0.15,0.47){1}{\line(0,1){0.47}}
\multiput(60.82,54.9)(0.14,0.48){1}{\line(0,1){0.48}}
\multiput(60.69,54.42)(0.13,0.48){1}{\line(0,1){0.48}}
\multiput(60.57,53.94)(0.12,0.48){1}{\line(0,1){0.48}}
\multiput(60.46,53.45)(0.11,0.49){1}{\line(0,1){0.49}}
\multiput(60.36,52.96)(0.1,0.49){1}{\line(0,1){0.49}}
\multiput(60.27,52.47)(0.09,0.49){1}{\line(0,1){0.49}}
\multiput(60.2,51.98)(0.08,0.49){1}{\line(0,1){0.49}}
\multiput(60.13,51.49)(0.07,0.49){1}{\line(0,1){0.49}}
\multiput(60.08,50.99)(0.06,0.5){1}{\line(0,1){0.5}}
\multiput(60.03,50.49)(0.04,0.5){1}{\line(0,1){0.5}}
\multiput(60,50)(0.03,0.5){1}{\line(0,1){0.5}}
\multiput(59.98,49.5)(0.02,0.5){1}{\line(0,1){0.5}}
\multiput(59.97,49)(0.01,0.5){1}{\line(0,1){0.5}}
\put(59.97,48.5){\line(0,1){0.5}}
\multiput(59.97,48.5)(0.01,-0.5){1}{\line(0,-1){0.5}}
\multiput(59.98,48)(0.02,-0.5){1}{\line(0,-1){0.5}}
\multiput(60,47.5)(0.03,-0.5){1}{\line(0,-1){0.5}}
\multiput(60.03,47.01)(0.04,-0.5){1}{\line(0,-1){0.5}}
\multiput(60.08,46.51)(0.06,-0.5){1}{\line(0,-1){0.5}}
\multiput(60.13,46.01)(0.07,-0.49){1}{\line(0,-1){0.49}}
\multiput(60.2,45.52)(0.08,-0.49){1}{\line(0,-1){0.49}}
\multiput(60.27,45.03)(0.09,-0.49){1}{\line(0,-1){0.49}}
\multiput(60.36,44.54)(0.1,-0.49){1}{\line(0,-1){0.49}}
\multiput(60.46,44.05)(0.11,-0.49){1}{\line(0,-1){0.49}}
\multiput(60.57,43.56)(0.12,-0.48){1}{\line(0,-1){0.48}}
\multiput(60.69,43.08)(0.13,-0.48){1}{\line(0,-1){0.48}}
\multiput(60.82,42.6)(0.14,-0.48){1}{\line(0,-1){0.48}}
\multiput(60.96,42.12)(0.15,-0.47){1}{\line(0,-1){0.47}}
\multiput(61.11,41.64)(0.16,-0.47){1}{\line(0,-1){0.47}}
\multiput(61.28,41.17)(0.17,-0.47){1}{\line(0,-1){0.47}}
\multiput(61.45,40.71)(0.09,-0.23){2}{\line(0,-1){0.23}}
\multiput(61.63,40.24)(0.1,-0.23){2}{\line(0,-1){0.23}}
\multiput(61.83,39.78)(0.1,-0.23){2}{\line(0,-1){0.23}}
\multiput(62.03,39.33)(0.11,-0.23){2}{\line(0,-1){0.23}}
\multiput(62.24,38.88)(0.11,-0.22){2}{\line(0,-1){0.22}}
\multiput(62.47,38.43)(0.12,-0.22){2}{\line(0,-1){0.22}}
\multiput(62.7,37.99)(0.12,-0.22){2}{\line(0,-1){0.22}}
\multiput(62.94,37.55)(0.13,-0.21){2}{\line(0,-1){0.21}}
\multiput(63.2,37.12)(0.13,-0.21){2}{\line(0,-1){0.21}}
\multiput(63.46,36.7)(0.14,-0.21){2}{\line(0,-1){0.21}}
\multiput(63.73,36.28)(0.14,-0.21){2}{\line(0,-1){0.21}}
\multiput(64.01,35.87)(0.14,-0.2){2}{\line(0,-1){0.2}}
\multiput(64.3,35.46)(0.15,-0.2){2}{\line(0,-1){0.2}}
\multiput(64.6,35.06)(0.1,-0.13){3}{\line(0,-1){0.13}}
\multiput(64.9,34.67)(0.11,-0.13){3}{\line(0,-1){0.13}}
\multiput(65.22,34.29)(0.11,-0.13){3}{\line(0,-1){0.13}}
\multiput(65.54,33.91)(0.11,-0.12){3}{\line(0,-1){0.12}}
\multiput(65.88,33.54)(0.11,-0.12){3}{\line(0,-1){0.12}}
\multiput(66.22,33.17)(0.12,-0.12){3}{\line(0,-1){0.12}}
\multiput(66.57,32.82)(0.12,-0.12){3}{\line(1,0){0.12}}
\multiput(66.92,32.47)(0.12,-0.11){3}{\line(1,0){0.12}}
\multiput(67.29,32.13)(0.12,-0.11){3}{\line(1,0){0.12}}
\multiput(67.66,31.79)(0.13,-0.11){3}{\line(1,0){0.13}}
\multiput(68.04,31.47)(0.13,-0.11){3}{\line(1,0){0.13}}
\multiput(68.42,31.15)(0.13,-0.1){3}{\line(1,0){0.13}}
\multiput(68.81,30.85)(0.2,-0.15){2}{\line(1,0){0.2}}
\multiput(69.21,30.55)(0.2,-0.14){2}{\line(1,0){0.2}}
\multiput(69.62,30.26)(0.21,-0.14){2}{\line(1,0){0.21}}
\multiput(70.03,29.98)(0.21,-0.14){2}{\line(1,0){0.21}}
\multiput(70.45,29.71)(0.21,-0.13){2}{\line(1,0){0.21}}
\multiput(70.87,29.45)(0.21,-0.13){2}{\line(1,0){0.21}}
\multiput(71.3,29.19)(0.22,-0.12){2}{\line(1,0){0.22}}
\multiput(71.74,28.95)(0.22,-0.12){2}{\line(1,0){0.22}}
\multiput(72.18,28.72)(0.22,-0.11){2}{\line(1,0){0.22}}
\multiput(72.63,28.49)(0.23,-0.11){2}{\line(1,0){0.23}}
\multiput(73.08,28.28)(0.23,-0.1){2}{\line(1,0){0.23}}
\multiput(73.53,28.08)(0.23,-0.1){2}{\line(1,0){0.23}}
\multiput(73.99,27.88)(0.23,-0.09){2}{\line(1,0){0.23}}
\multiput(74.46,27.7)(0.47,-0.17){1}{\line(1,0){0.47}}
\multiput(74.92,27.53)(0.47,-0.16){1}{\line(1,0){0.47}}
\multiput(75.39,27.36)(0.47,-0.15){1}{\line(1,0){0.47}}
\multiput(75.87,27.21)(0.48,-0.14){1}{\line(1,0){0.48}}
\multiput(76.35,27.07)(0.48,-0.13){1}{\line(1,0){0.48}}
\multiput(76.83,26.94)(0.48,-0.12){1}{\line(1,0){0.48}}
\multiput(77.31,26.82)(0.49,-0.11){1}{\line(1,0){0.49}}
\multiput(77.8,26.71)(0.49,-0.1){1}{\line(1,0){0.49}}
\multiput(78.29,26.61)(0.49,-0.09){1}{\line(1,0){0.49}}
\multiput(78.78,26.52)(0.49,-0.08){1}{\line(1,0){0.49}}
\multiput(79.27,26.45)(0.49,-0.07){1}{\line(1,0){0.49}}
\multiput(79.76,26.38)(0.5,-0.06){1}{\line(1,0){0.5}}
\multiput(80.26,26.33)(0.5,-0.04){1}{\line(1,0){0.5}}
\multiput(80.76,26.28)(0.5,-0.03){1}{\line(1,0){0.5}}
\multiput(81.25,26.25)(0.5,-0.02){1}{\line(1,0){0.5}}
\multiput(81.75,26.23)(0.5,-0.01){1}{\line(1,0){0.5}}
\put(82.25,26.22){\line(1,0){0.5}}
\multiput(82.75,26.22)(0.5,0.01){1}{\line(1,0){0.5}}
\multiput(83.25,26.23)(0.5,0.02){1}{\line(1,0){0.5}}
\multiput(83.75,26.25)(0.5,0.03){1}{\line(1,0){0.5}}
\multiput(84.24,26.28)(0.5,0.04){1}{\line(1,0){0.5}}
\multiput(84.74,26.33)(0.5,0.06){1}{\line(1,0){0.5}}
\multiput(85.24,26.38)(0.49,0.07){1}{\line(1,0){0.49}}
\multiput(85.73,26.45)(0.49,0.08){1}{\line(1,0){0.49}}
\multiput(86.22,26.52)(0.49,0.09){1}{\line(1,0){0.49}}
\multiput(86.71,26.61)(0.49,0.1){1}{\line(1,0){0.49}}
\multiput(87.2,26.71)(0.49,0.11){1}{\line(1,0){0.49}}
\multiput(87.69,26.82)(0.48,0.12){1}{\line(1,0){0.48}}
\multiput(88.17,26.94)(0.48,0.13){1}{\line(1,0){0.48}}
\multiput(88.65,27.07)(0.48,0.14){1}{\line(1,0){0.48}}
\multiput(89.13,27.21)(0.47,0.15){1}{\line(1,0){0.47}}
\multiput(89.61,27.36)(0.47,0.16){1}{\line(1,0){0.47}}
\multiput(90.08,27.53)(0.47,0.17){1}{\line(1,0){0.47}}
\multiput(90.54,27.7)(0.23,0.09){2}{\line(1,0){0.23}}
\multiput(91.01,27.88)(0.23,0.1){2}{\line(1,0){0.23}}
\multiput(91.47,28.08)(0.23,0.1){2}{\line(1,0){0.23}}
\multiput(91.92,28.28)(0.23,0.11){2}{\line(1,0){0.23}}
\multiput(92.37,28.49)(0.22,0.11){2}{\line(1,0){0.22}}
\multiput(92.82,28.72)(0.22,0.12){2}{\line(1,0){0.22}}
\multiput(93.26,28.95)(0.22,0.12){2}{\line(1,0){0.22}}
\multiput(93.7,29.19)(0.21,0.13){2}{\line(1,0){0.21}}
\multiput(94.13,29.45)(0.21,0.13){2}{\line(1,0){0.21}}
\multiput(94.55,29.71)(0.21,0.14){2}{\line(1,0){0.21}}
\multiput(94.97,29.98)(0.21,0.14){2}{\line(1,0){0.21}}
\multiput(95.38,30.26)(0.2,0.14){2}{\line(1,0){0.2}}
\multiput(95.79,30.55)(0.2,0.15){2}{\line(1,0){0.2}}
\multiput(96.19,30.85)(0.13,0.1){3}{\line(1,0){0.13}}
\multiput(96.58,31.15)(0.13,0.11){3}{\line(1,0){0.13}}
\multiput(96.96,31.47)(0.13,0.11){3}{\line(1,0){0.13}}
\multiput(97.34,31.79)(0.12,0.11){3}{\line(1,0){0.12}}
\multiput(97.71,32.13)(0.12,0.11){3}{\line(1,0){0.12}}
\multiput(98.08,32.47)(0.12,0.12){3}{\line(1,0){0.12}}
\multiput(98.43,32.82)(0.12,0.12){3}{\line(0,1){0.12}}
\multiput(98.78,33.17)(0.11,0.12){3}{\line(0,1){0.12}}
\multiput(99.12,33.54)(0.11,0.12){3}{\line(0,1){0.12}}
\multiput(99.46,33.91)(0.11,0.13){3}{\line(0,1){0.13}}
\multiput(99.78,34.29)(0.11,0.13){3}{\line(0,1){0.13}}
\multiput(100.1,34.67)(0.1,0.13){3}{\line(0,1){0.13}}
\multiput(100.4,35.06)(0.15,0.2){2}{\line(0,1){0.2}}
\multiput(100.7,35.46)(0.14,0.2){2}{\line(0,1){0.2}}
\multiput(100.99,35.87)(0.14,0.21){2}{\line(0,1){0.21}}
\multiput(101.27,36.28)(0.14,0.21){2}{\line(0,1){0.21}}
\multiput(101.54,36.7)(0.13,0.21){2}{\line(0,1){0.21}}
\multiput(101.8,37.12)(0.13,0.21){2}{\line(0,1){0.21}}
\multiput(102.06,37.55)(0.12,0.22){2}{\line(0,1){0.22}}
\multiput(102.3,37.99)(0.12,0.22){2}{\line(0,1){0.22}}
\multiput(102.53,38.43)(0.11,0.22){2}{\line(0,1){0.22}}
\multiput(102.76,38.88)(0.11,0.23){2}{\line(0,1){0.23}}
\multiput(102.97,39.33)(0.1,0.23){2}{\line(0,1){0.23}}
\multiput(103.17,39.78)(0.1,0.23){2}{\line(0,1){0.23}}
\multiput(103.37,40.24)(0.09,0.23){2}{\line(0,1){0.23}}
\multiput(103.55,40.71)(0.17,0.47){1}{\line(0,1){0.47}}
\multiput(103.72,41.17)(0.16,0.47){1}{\line(0,1){0.47}}
\multiput(103.89,41.64)(0.15,0.47){1}{\line(0,1){0.47}}
\multiput(104.04,42.12)(0.14,0.48){1}{\line(0,1){0.48}}
\multiput(104.18,42.6)(0.13,0.48){1}{\line(0,1){0.48}}
\multiput(104.31,43.08)(0.12,0.48){1}{\line(0,1){0.48}}
\multiput(104.43,43.56)(0.11,0.49){1}{\line(0,1){0.49}}
\multiput(104.54,44.05)(0.1,0.49){1}{\line(0,1){0.49}}
\multiput(104.64,44.54)(0.09,0.49){1}{\line(0,1){0.49}}
\multiput(104.73,45.03)(0.08,0.49){1}{\line(0,1){0.49}}
\multiput(104.8,45.52)(0.07,0.49){1}{\line(0,1){0.49}}
\multiput(104.87,46.01)(0.06,0.5){1}{\line(0,1){0.5}}
\multiput(104.92,46.51)(0.04,0.5){1}{\line(0,1){0.5}}
\multiput(104.97,47.01)(0.03,0.5){1}{\line(0,1){0.5}}
\multiput(105,47.5)(0.02,0.5){1}{\line(0,1){0.5}}
\multiput(105.02,48)(0.01,0.5){1}{\line(0,1){0.5}}

\linethickness{1mm}
\put(30,50){\line(1,0){30}}
\linethickness{1mm}
\multiput(60,90)(0.12,-0.16){125}{\line(0,-1){0.16}}
\linethickness{1mm}
\multiput(104,55)(0.36,0.12){83}{\line(1,0){0.36}}
\linethickness{0.3mm}
\multiput(95,30)(0.12,-0.16){125}{\line(0,-1){0.16}}
\put(35,55){\makebox(0,0)[cc]{$\eps_1,p_1$}}

\put(73,85){\makebox(0,0)[cc]{$\eps_2,p_2$}}

\put(95,80){\makebox(0,0)[cc]{$\cdot$}}

\put(110,70){\makebox(0,0)[cc]{$\cdot$}}

\put(122,55){\makebox(0,0)[cc]{$\eps_N,p_N$}}

\put(110,20){\makebox(0,0)[cc]{$\ve,k$}}

\put(80,50){\makebox(0,0)[cc]{$\wt\Gamma$}}

\end{picture}

}

\begin{document}

\baselineskip 24pt

\begin{center}
{\Large \bf  Subleading Soft Graviton Theorem for  Loop Amplitudes}

\end{center}

\vskip .6cm
\medskip

\vspace*{4.0ex}

\baselineskip=18pt

\centerline{\large \rm Ashoke Sen}

\vspace*{4.0ex}

\centerline{\large \it Harish-Chandra Research Institute}
\centerline{\large \it  Chhatnag Road, Jhusi,
Allahabad 211019, India}

\centerline{and}

\centerline{\large \it Homi Bhabha National Institute}
\centerline{\large \it Training School Complex, Anushakti Nagar,
    Mumbai 400085, India}

\vspace*{1.0ex}
\centerline{\small E-mail:  sen@mri.ernet.in}

\vspace*{5.0ex}

\centerline{\bf Abstract} \bigskip

Superstring field theory gives expressions for heterotic and type II
string loop amplitudes that are free
from ultraviolet and infrared divergences when the number of non-compact space-time
dimensions is five or more.
We prove the subleading
soft graviton theorem  in these theories to all orders in perturbation theory 
for S-matrix elements of arbitrary number of finite energy external
states but only one external soft
graviton. We also prove the leading soft graviton theorem
for arbitrary
number of finite energy external states and arbitrary number of
soft gravitons. Since our analysis is based on general properties of one particle irreducible
effective action, the results are valid in any theory of quantum
gravity that gives finite result for the S-matrix
order by order in perturbation theory without violating
general coordinate invariance.

\bigskip



\vfill \eject

\baselineskip 18pt

\tableofcontents

\sectiono{Introduction} \label{sintro}

In recent years, soft graviton theorem has been studied from various perspectives
-- perturbative quantum field theory\cite{weinberg1,weinberg2,jackiw1,jackiw2,1103.2981,1404.4091,1404.7749,
1405.1015,1405.1410,1405.2346,1405.3413,1405.3533,
1406.6574,1406.6987,1406.7184,
1407.5936,1407.5982,1408.4179,1410.6406,1412.3699,
1503.04816,1504.01364,1507.08882,
1509.07840,1604.00650,1604.03893,
1607.02700,1611.02172,1611.03137,1611.07534,
1702.02350}, perturbative string theory\cite{ademollo,shapiro,1406.4172,1406.5155,1411.6661,
1502.05258,1505.05854,1507.08829,1511.04921,1601.03457,
1604.03355,1610.03481} and
BMS symmetry\cite{1312.2229,1401.7026,1411.5745,1506.05789,1509.01406,
1605.09094,1608.00685,1612.08294,
1701.00496,1612.05886}. Our goal in this paper will be to give a general proof of
the subleading soft graviton theorem in any perturbative quantum field theory that
includes gravity and
gives S-matrix elements free from infrared and ultraviolet divergences. At present
the only known candidates for such theories are heterotic and type II string field 
theories\cite{1703.06410} 
in backgrounds with five or more non-compact flat space-time dimensions. 

Our strategy will be the same one followed in \cite{1702.03934}, with the difference
that instead of the classical action we work with the one particle irreducible (1PI) 
effective action. We begin with the gauge invariant 
1PI effective action and expand it in powers of all fields including the graviton. 
We then gauge fix it using a Lorentz covariant gauge fixing condition. The resulting
action has manifest Lorentz invariance but not manifest general coordinate invariance.
We now introduce the soft graviton field $S_{\mu\nu}$ 
by covariantizing this action with respect
to the soft graviton field.\footnote{As mentioned in \cite{1702.03934}, for superstring
field theory this procedure would follow from background independence of
string field theory that ensures that switching on a soft graviton mode of the string
field is equivalent to deforming the background target space 
metric used for constructing the
world-sheet conformal field theory by  a soft graviton mode. This is known to be
true for bosonic string field theory\cite{9307088,9311009} but has not yet been proven for
superstring field theory\cite{prepare}.}
This requires replacing the background metric by 
$\eta_{\mu\nu}+2\, S_{\mu\nu}$ and the ordinary derivatives by covariant 
derivatives computed with this background metric. To first subleading order in
soft momentum, there are no additional terms coupling 
the soft graviton to the rest
of the fields. Once this replacement is made, we can compute the amplitude involving
the soft graviton from the Feynman diagrams of the resulting quantum field theory.
Our use of 1PI effective action entails that we need to compute only the tree amplitudes.

The other technical difference from the analysis of \cite{1702.03934} 
is that while covariantizing
the action we take all the fields to carry flat tangent space indices instead of
curved space indices. This allows us to deal with fermions in the same way as the
bosons. We now have to use the vielbein $e_\mu^{~a}$ 
instead of the metric to describe the soft
graviton field, but to first order in $S_{\mu\nu}$ -- which is all we shall
need for our analysis
-- this is done simply by taking $e_\mu^{~a}$ to be $\delta_\mu^{~a}
+ S_\mu^{~a}$ where the indices are raised and lowered by the flat background
metric 
$\eta$.  Since $S_{\mu\nu}=S_{\nu\mu}$,
this choice of $e_\mu^{~a}$ amounts to gauge fixing 
the local Lorentz symmetry from the beginning and
allows us to include superstring field theory in our framework where local Lorentz 
symmetry is gauge fixed from the beginning.  

The rest of the paper is organized as follows. In \S\ref{ssub} we prove the
subleading soft graviton theorem for one external soft graviton but arbitrary 
number of finite energy external states. Some of the technical details of this
analysis are given in appendices \ref{sb} and  
\ref{sa}. In \S\ref{smulti} we prove the leading
soft graviton theorem for arbitrary number of soft gravitons and arbitrary number
of finite energy external states. In all cases our results are valid to all orders in the
perturbation theory.

\sectiono{Subleading soft theorem for one external soft graviton} \label{ssub}

\begin{figure}
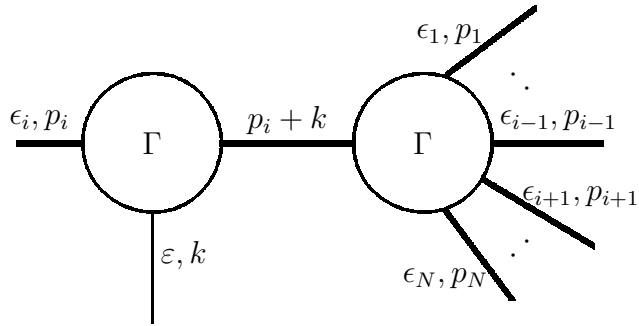


\begin{center}

\figsoftonefield

\end{center}

\caption{Source of the leading contribution to the string loop 
amplitude with one external
soft graviton.  $\eps_i$ and $p_i$ denote the polarization and
momentum of the $i$-th external finite energy particles, while $\ve_{\mu\nu}$
and $k$ denote the polarization and momentum of the external soft graviton.
\label{f1field}
}

\end{figure}

In this section we shall prove the subleading soft graviton theorem for amplitudes
with one external soft graviton, but arbitrary number of finite energy external states.
We begin by describing our notations.

In a Feynman diagram we shall call a line soft if all components of its momentum
are small, nearly on-shell if it carries finite energy but satisfies the on-shell condition
approximately and hard it is is neither soft nor nearly on-shell.
We shall work in backgrounds
where the number of non-compact space-times dimensions is five or more, and
expand the 1PI action in powers of fields
around the extremum describing the vacuum solution
so that there are no tadpoles in the resulting Feynman diagrams. In that
case by standard power counting\cite{sterman} one can show that 
there are no hidden inverse powers of soft momentum coming from the 1PI
vertices with at most one soft external state, 
even in the presence of massless fields, as long as there are no cubic coupling
without derivatives among the massless bosonic fields.
Since we
shall use the vertices computed from the 1PI effective action, we need to draw
only tree
graphs, and for this reason there is a clear labelling of each line as soft,
nearly on-shell or hard.
We use a thin line to denote external soft particle, and a thick line to denote
external or internal particles carrying finite momentum and / or energy. All
internal
lines will denote the full renormalized propagator. 
We also denote by $\Gamma$ the amputated connected
Green's function from which propagators
associated with external legs have been removed -- for three external legs this
coincides with the 1PI vertex. 

The leading contribution to the amplitude, carrying one power of soft momentum $k$
in the denominator, comes from the diagrams of the type shown in Fig.~\ref{f1field}.
We shall use the sign convention that all external momenta enter the diagram so that
incoming (outgoing) particles carry momentum labels with positive (negative) 
energy component.
If $M_i$ denotes the mass of the $i$-th external particle then on-shell condition
gives 
\be 
p_i^2+M_i^2=0, \quad k^2=0\, ,
\ee
together with conditions on polarizations that will be discussed later. Now if we take
the internal particle carrying momentum $p_i+k$ to have the same mass $M_i$, then
the propagator gives a terms proportional to $\{(p_i+k)^2+M_i^2\}^{-1}
= (2p_i\cdot k)^{-1}$. This is responsible for producing the 
inverse power of soft momentum in the amplitude.

\begin{figure}
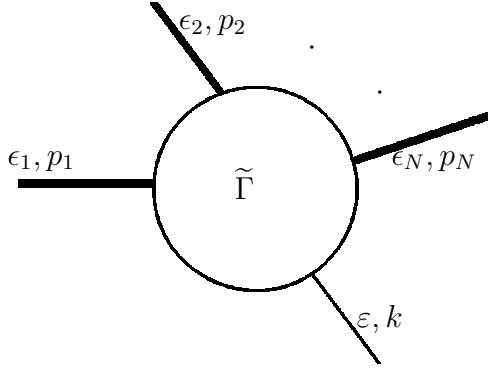


\begin{center}

\figsoftthreefield

\end{center}

\caption{Source of the subleading contribution to the string loop 
amplitude with one external
soft graviton. 
\label{f3field}
}

\end{figure}

The first subleading contribution in powers of soft momentum
comes from the subleading contribution from Fig.~\ref{f1field} as well as the leading 
contribution from Fig.~\ref{f3field}. $\wt\Gamma$ in Fig.~\ref{f3field}  denotes amputated
Green's function from which the contributions of the type shown in Fig.~\ref{f1field} have
been subtracted. As a result $\wt\Gamma$ has no contribution containing inverse
powers of momentum.

For computing the contributions from these diagrams we need to determine the coupling
of the soft graviton to the rest of the fields. This is done by following the procedure
outlined in the introduction. We introduce vielbein $e_\mu^{~a}$ and the inverse
vielbein $E_a^{~\mu}$ in terms of the soft graviton field $S_{\mu\nu}$ 
to first order in $S_{\mu\nu}$ as
\be \label{evrep}
e_\mu^{~a} = \delta_\mu^{~a} + S_\mu^{~a}, \quad E_a^{~\mu} = \delta_a^{~\mu} 
- S_a^{~\mu},
\ee
where all indices are raised and lowered by the flat metric $\eta$. 
Let $\{\Phi_\alpha\}$ denote the collection of all the fields in the theory, 
transforming in some large 
reducible representation of the Lorentz 
group.\footnote{Even though superstring field theory has infinite number of fields,
for any given scattering process we can work with an effective field theory of a
finite number of fields by integrating out fields that are sufficiently heavy so that they
are not produced in the scattering\cite{1703.06410}. 
Therefore we can assume that the
number of fields is finite.} 
Now in the Lorentz invariant gauge fixed 1PI effective action
we replace derivatives of the fields $\Phi_\alpha$ as follows:
\be \label{ex2.3}
\p_{a_1}\ldots \p_{a_n} \Phi_\alpha
\quad \Rightarrow \quad
E_{a_1}^{~\mu_1} \cdots E_{a_n}^{~\mu_n} D_{\mu_1} \ldots D_{\mu_n}
\Phi_\alpha
\ee
where
\be \label{ex2.4}
\OO_1 \, D_\mu \OO_2 \Phi_\alpha \equiv \OO_1 \, \p_\mu \OO_2
\Phi_\alpha + \OO_1 \, {1\over 2} \, \omega_\mu^{ab} 
(J_{ab})_\alpha^{~\gamma}
\OO_2 \Phi_\gamma, \quad \omega_\mu^{ab} \equiv \left(\p^b S_\mu^{~a} -
\p^a S_\mu^{~b}\right)\, .
\ee 
Here the $\OO_i$'s denote any collection of covariant derivative operators, and
$J^{ab}$ are the angular momentum generators, normalized such that  if 
$\Phi$ carries covariant vector indices, then
\be 
(J^{ab})_c^{~d}  = \delta^a_{~c} \eta^{bd} - \delta^b_{~c} \eta^{ad}\, .
\ee
Note that in the expression for $D_\mu$ in \refb{ex2.4} we have not included
the terms involving the 
Christoffel symbol $\Gamma^\rho_{\mu\nu}$, needed for defining 
$D_\mu$ acting on another covariant derivative $D_\nu$ hidden inside 
$\OO_2$. We have provided the
justification of this in appendix \ref{sb}.

First let us evaluate the contribution from Fig.~\ref{f3field}. This analysis will be
more or less identical to the one given in \cite{1702.03934}; so we shall be brief.
Since we are interested
in computing the leading contribution from this graph, we can ignore terms involving
$\omega_\mu^{ab}$ since they involve derivatives of $S_{\mu\nu}$ and therefore have
one or more powers of soft momentum. Therefore for this amplitude 
the effect of coupling the soft graviton can be obtained by replacing the vielbeins
as in \refb{evrep} with $S_{\mu\nu}$ given by the constant polarization tensor
$\ve_{\mu\nu}$. This is equivalent to replacing, in the amplitude without the
soft graviton, the vielbeins as
\be \label{evrep1}
e_\mu^{~a} = \delta_\mu^{~a} + \ve_\mu^{~a}, \quad E_a^{~\mu} = \delta_a^{~\mu} 
- \ve_a^{~\mu}\, .
\ee
Instead of making this replacement inside each vertex and propagator of
$\wt\Gamma$, we can also
make this replacement in the final amplitude written in the constant vielbein
background.
Now since the fields and hence the polarization tensors carry flat tangent space 
indices,
the only place where a vielbein enters in the final expression for the amplitude
is in converting the indices of the external momenta $p_i$ from space-time indices to flat
indices. This can be achieved by using the combination $E_a^{~\mu}p_{i\mu}
= p_{ia}- \ve_a^{~b} p_{ib}$.  Once this is done the indices can be contracted 
with each other by the metric $\eta$
without any reference to the vielbeins.
Therefore the effect of coupling soft graviton in the
amplitude in Fig.~\ref{f3field} is to shift $p_{ia}$ by $- \ve_a^{~b} p_{ib}$. 

In order to express the result in a convenient form, let us
introduce the symbol $\Gamma_{(i)}^\alpha(p_i)$ to be the quantity
such that
\be \label{enot}
\eps_{i,\alpha}  \Gamma_{(i)}^\alpha(p_i) = 
\Gamma(\eps_1, p_1; \ldots ; \eps_N,  p_N)\, ,
\ee
where the right hand side denotes the amputated N-point Green's function
with general off-shell momenta $p_1, \ldots, p_N$ and polarization tensors
$\eps_1, \ldots, \eps_N$ without the external soft photon. 
Therefore the arguments $\eps_1, p_1; \ldots ; \eps_N,  p_N$
other than $\eps_i, p_i$ are hidden in $\Gamma_{(i)}^\alpha(p_i)$.
With this notation, the result of the previous paragraph can be used to express
the amplitude shown in Fig.~\ref{f3field} as
\be \label{exx11}
-\sum_{i=1}^N \ve_{a}^{~b} \, p_{ib} \, \eps_{i,\alpha}\, 
{\p\over \p p_{ia}} \Gamma_{(i)}^\alpha(p_i)\, .
\ee

We now turn to the contribution from Fig.~\ref{f1field}. For this we need to study the
three point coupling between a single soft graviton and two finite energy particles
to the first subleading order in the soft momenta. By our previous argument this
may be obtained by covariantizing the quadratic term in the manifestly Lorentz
invariant, gauge fixed 1PI effective action
without the soft graviton. We begin by writing the general form of the
quadratic part of the 1PI effective action in momentum space: 
\be \label{es2}
S^{(2)}
= {1\over 2} \int {d^D q_1\over (2\pi)^D} \, {d^D q_2\over (2\pi)^D} \,  \Phi_\alpha(q_1) 
\KK^{\alpha\beta}(q_2) \Phi_\beta(q_2)\, (2\pi)^D \delta^{(D)}(q_1+q_2)
\, ,
\ee
where $\Phi_\alpha(q)$ now denotes the Fourier transform of the field $\Phi_\alpha$
introduced earlier and $D$ is the number of non-compact space-time dimensions. 
We shall take $\KK^{\alpha\beta}(q)$ to be symmetric:\footnote{For grassmann 
odd fields there will be an extra minus sign on the right hand
side of \refb{ekab}, but this does not affect the rest of the analysis.}
\be \label{ekab}
\KK^{\alpha\beta}(q) = \KK^{\beta\alpha}(-q)\, .
\ee
In this case the propagator is given by
\be \label{epropdef}
D_F(q)_{\alpha\beta} = i (\KK(q)^{-1})_{\alpha\beta}\, ,
\ee
where $q$ is the momentum flowing from the end carrying the label $\beta$
to the end carrying the label $\alpha$. 
Noting that the derivative operator $\p_\mu$ in position space 
becomes a multiplicative operation by $i \, q_\mu$ in the momentum space, and
using \refb{ex2.3}, \refb{ex2.4}, we see
that   effect of coupling a soft graviton field $S_{\mu\nu}=\ve_{\mu\nu} e^{ik.x}$ with
\be \label{eecond}
\ve_{\mu\nu}=\ve_{\nu\mu}, \quad k^\mu \ve_{\mu\nu}=0=k^\nu \ve_{\mu\nu}, 
\quad \eta^{\mu\nu}\ve_{\mu\nu}=0\, ,
\ee
can be obtained by making the following replacement in 
\refb{es2}:
\ben
&& \delta^{(D)}(q_1+q_2) \,  \KK^{\alpha\beta} (q_2) 
\nonumber \\ \to &&
 \delta^{(D)}(q_1+q_2)\,  \KK^{\alpha\beta}  (q_2)
\nonumber \\ &-&
\delta^{(D)}(q_1+q_2+k) \, \left[
\ve_{\mu \nu} q_2^\nu {\p\over \p q_{2\mu}} \KK^{\alpha\beta} (q_2)
+ {1\over 2} (k_a \, \ve_{b\mu} - k_b \, \ve_{a\mu}) {\p\over \p q_{2\mu}}
\KK^{\alpha\gamma}(q_2) 
\left(J^{ab}\right)_{\gamma}^{~\beta} \right]\, . 
\nonumber \\ \een
This gives the part of the
action describing the coupling of a soft graviton field $S_{\mu\nu}=\ve_{\mu\nu}
e^{ik.x}$ to a pair of other fields to be
\ben \label{e2.14}
S^{(3)} &=& {1\over 2} \int {d^D q_1\over (2\pi)^D} \, {d^D q_2\over (2\pi)^D} \,  (2\pi)^D \delta^{(D)}(q_1+q_2+k)
\nonumber \\ && \times \Phi_\alpha(q_1) 
\left[ - \ve_{\mu \nu} q_2^\nu  {\p\over \p q_{2\mu}}\KK^{\alpha\beta} (q_2)
- {1\over 2} (k_a \, \ve_{b\mu} - k_b \, \ve_{a\mu}) {\p\over \p q_{2\mu}}\KK^{\alpha\gamma}(q_2) 
\left(J^{ab}\right)_{\gamma}^{~\beta} 
\right]\Phi_\beta(q_2) \, . \nonumber \\
\een
Therefore the three point vertex of a soft graviton of momentum $k$, a $\Phi_\alpha$
particle of momentum $p$ and a $\Phi_\beta$ particle of momentum $-p-k$ is given by
\ben \label{e4.8}
\Gamma^{(3)\alpha\beta}(\ve, k; p, -p-k)
&=& {i\over 2} \bigg[- \ve_{\mu \nu} (p+k)^\nu  {\p\over \p p_{\mu}}\KK^{\alpha\beta} (-p-k)
- \ve_{\mu \nu} p^\nu  {\p\over \p p_{\mu}}\KK^{\beta\alpha} (p)
\nonumber \\ &&
+ {1\over 2} (k_a \, \ve_{b\mu} - k_b \, \ve_{a\mu}) {\p\over \p p_{\mu}}\KK^{\alpha\gamma}(-p-k) 
\left(J^{ab}\right)_{\gamma}^{~\beta} \nonumber \\ &&
- {1\over 2} (k_a \, \ve_{b\mu} - k_b \, \ve_{a\mu}) {\p\over \p p_{\mu}}\KK^{\beta\gamma}(p) 
\left(J^{ab}\right)_{\gamma}^{~\alpha} 
\bigg] \, .
\een

The contribution from the amplitude shown in 
Fig.~\ref{f1field} may now be expressed as
\be \label{e4.13pre}
\eps_{i,\alpha}\, \Gamma^{(3)\alpha\beta}(\ve, k; p_i, -p_i-k)\, i 
\{\KK^{-1}(-p_i-k)\}_{\beta\delta} 
\Gamma_{(i)}^{\delta}(p_i+k)\, ,
\ee
where $\Gamma_{(i)}$ has been defined in \refb{enot}.
It has been shown in appendix \ref{sa} that as long as $\eps_{i,\alpha}$ and $p_i$
satisfy the on-shell condition
\be \label{eonshell}
\eps_{i,\alpha} \KK^{\alpha\beta} (-p_i) =0\, ,
\ee
\refb{e4.13pre} can be reduced to
\ben \label{ex.5pre}
&& 
(p_i\cdot k)^{-1}\, \ve_{\mu \nu}\,  p_i^\mu \,  p_i^\nu 
\eps_{i,\alpha} \Gamma_{(i)}^{\alpha}(p_i)  
+ (p_i\cdot k)^{-1} \, \ve_{\mu \nu} \, p_i^\mu\, 
 p_i^\nu \, \eps_{i,\alpha}  
\,k_\rho \, {\p\over \p p_{i\rho}} \Gamma_{(i)}^{\alpha}(p_i) 
\nonumber \\ &+&
(p_i\cdot k)^{-1} \, k_a \, \ve_{b\mu} \, p_i^\mu \, 
\eps_{i,\alpha}\, (J^{ab})_\gamma^{~\alpha} \,
\Gamma_{(i)}^\gamma(p_i)\, .
\een
After summing over $i$ and adding the contribution \refb{exx11} from 
Fig.~\ref{f3field}
we get the subleading soft graviton theorem for one soft 
graviton\cite{1404.4091}:
\ben \label{ex.66}
\Gamma(\ve, k; \eps_1, p_1;\ldots ; \eps_N, p_N)
&=& 
\sum_{i=1}^N \, (p_i\cdot k)^{-1}\, \ve_{\mu \nu}\,  p_i^\mu \,  p_i^\nu 
\eps_{i,\alpha} \Gamma_{(i)}^{\alpha}(p_i)  
\nonumber \\ & + &
 \sum_{i=1}^N \,  \left\{
(p_i\cdot k)^{-1} \, \ve_{\mu \nu} \, p_i^\mu\, 
 p_i^\nu  
\,k_\rho - \ve_{\rho b} \, p_i^b
\right\} \, \eps_{i,\alpha} \, {\p\over \p p_{i\rho}} \Gamma_{(i)}^{\alpha}(p_i) 
\nonumber \\ &+&
 \sum_{i=1}^N \,  (p_i\cdot k)^{-1} \, k_a \,\ve_{b\mu} \,
 p_i^\mu \, \eps_{i,\alpha} (J^{ab})_\gamma^{~\alpha} \, 
\Gamma_{(i)}^\gamma(p_i)\, ,
\een
with $\Gamma_{(i)}^\gamma(p_i)$ defined through \refb{enot}.

\sectiono{Leading soft theorem for multiple soft gravitons} \label{smulti}

\begin{figure}
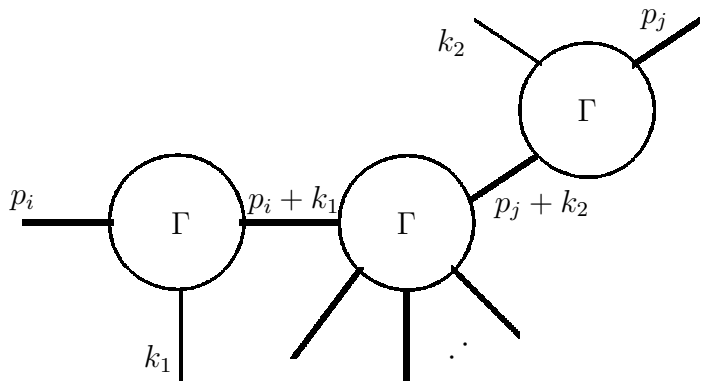


\begin{center}

\figsofttwo

\end{center}

\caption{Two external soft gravitons attached to different external lines carrying finite momenta.
\label{f2tree}
}

\end{figure}

\begin{figure}
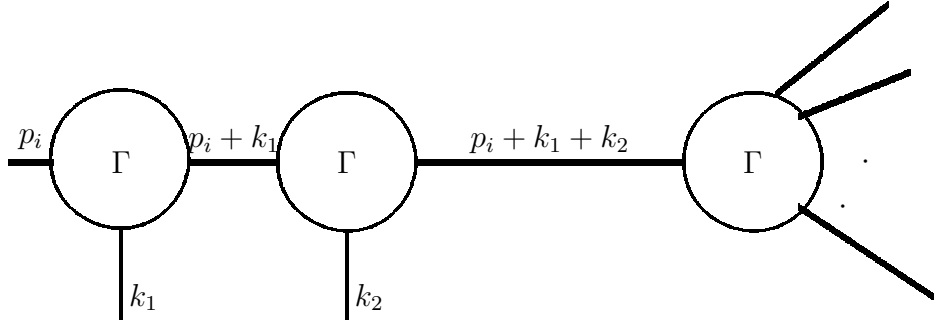


\begin{center}

\figsoftthree

\end{center}

\caption{Two external soft gravitons attached to the same external line 
carrying finite momenta.
\label{f3tree}
}

\end{figure}

We shall now consider amplitudes with multiple soft gravitons.  We shall first
analyze the case where we have two soft external gravitons 
carrying momenta $k_1$
and $k_2$. In this case the leading contribution has
two powers of soft momenta in the denominator, arising from diagrams
where the two soft gravitons attach to different external legs
as in Fig.~\ref{f2tree} or both soft gravitons attach to the same external leg as in
Fig.~\ref{f3tree}.
In either of the diagrams, the product of 
the leading contributions from the three point
vertex and the internal propagator that follows it  is given by
\be
\ve_{\mu\nu} p_i^\mu p_i^\nu \, (p_i\cdot\ell)^{-1}\, ,
\ee
where $p_i$ is the (nearly) on-shell momentum entering the vertex, 
$\ve$ is the polarization of the
soft graviton and $p_i+\ell$ is the momentum carried
by the internal propagator that follows the vertex. The derivation of this is identical to the
derivation of the first term on the right hand side of 
\refb{ex.5pre} and follows easily from the analysis
given in appendix \ref{sa}.

From this point onwards the analysis proceeds as in \cite{1702.03934}.
The contribution from Fig.~\ref{f2tree}
takes the form
\be 
 {1\over p_i\cdot k_1} \varepsilon^{(1)}_{\mu\nu} p_i^\mu p_i^\nu
 \, \, \times \, \, 
  {1\over p_j\cdot k_2} \varepsilon^{(2)}_{\rho\sigma} p_j^\rho p_j^\sigma
 \, \,  \times \, \,     \Gamma(\eps_1, p_1;\ldots ; \eps_N, p_N) \, \,  + \, \,  \hbox{less singular terms}
 \, . \ee
On the other hand the contribution from Fig.~\ref{f3tree} takes the form
\be  \label{eadd1}
 {1\over p_i\cdot k_1} \varepsilon^{(1)}_{\mu\nu} p_i^\mu p_i^\nu
 \, \,  \times \, \, 
  {1\over p_i\cdot (k_1+k_2)} \varepsilon^{(2)}_{\rho\sigma} p_j^\rho p_j^\sigma
  \, \,  \times \, \,     \Gamma(\eps_1, p_1;\ldots ; \eps_N, p_N) \, \,  
  + \, \,  \hbox{less singular terms}
 \, . \ee
There is another contribution where the external soft lines carrying momenta
 $k_1$ and $k_2$ are exchanged in \refb{eadd1}. Adding this to \refb{eadd1} we get
 \be 
  {1\over p_i\cdot k_1} \varepsilon^{(1)}_{\mu\nu} p_i^\mu p_i^\nu
 \, \,  \times \, \, 
  {1\over p_i\cdot k_2} \varepsilon^{(2)}_{\rho\sigma} p_i^\rho p_i^\sigma
  \, \,  \times \, \,       \Gamma(\eps_1, p_1;\ldots ; \eps_N, p_N)  \, \,  
  + \, \,  \hbox{less singular terms}
 \, . \ee
 After summing over all possible insertions of the two soft gravitons on $N$ external lines
 carrying finite momentum, we get\cite{weinberg2}
 \be \label{etwosoftfieldtree}
 \sum_{i=1}^N {1\over p_i\cdot k_1} \varepsilon^{(1)}_{\mu\nu} p_i^\mu p_i^\nu
 \, \,  \times \, \,  \sum_{j=1}^N
  {1\over  p_j\cdot k_2} \varepsilon^{(2)}_{\rho\sigma} p_j^\rho p_j^\sigma
  \, \,  \times \, \,    \Gamma(\eps_1, p_1;\ldots ; \eps_N, p_N) 
   \, \,  + \, \,  \hbox{less singular terms}
 \, . \ee

For $m$ external soft gravitons the leading term will have $m$ powers
of soft momentum in the denominator, coming from diagrams where
each external soft graviton gets attached
to a nearly on-shell line.  After summing over all possible insertions 
we arrive at the
generalization of \refb{etwosoftfieldtree}:
\ben \label{esoftgenfieldtree}
\Gamma(\varepsilon^{(1)},  k_1;
\ldots ; \varepsilon^{(m)}, k_m; \eps_1, p_1;\ldots ; \eps_N, p_N)
&=&\prod_{s=1}^m 
\left[\sum_{i=1}^N {1\over p_i\cdot k_s}  \, \varepsilon^{(s)}_{\mu\nu} 
\, p_i^\mu \, p_i^\nu\right] \, 
\Gamma(\eps_1, p_1;\ldots ; \eps_N, p_N)
\nonumber \\ &&
+ \hbox{less singular terms}\, .
\een

\bigskip

\noindent {\bf Acknowledgement:}
This work was
supported in part by the 
DAE project 12-R\&D-HRI-5.02-0303 and J. C. Bose fellowship of 
the Department of Science and Technology, India.

\bigskip

\appendix

\sectiono{Justification for dropping Christoffel symbols from 
covariant derivatives} \label{sb}

In defining the covariant derivative operator $D_\mu$ appearing in \refb{ex2.3}
using \refb{ex2.4}, we dropped possible terms involving the Christoffel symbol 
$\Gamma^\rho_{\mu\nu}$. These Christoffel symbols arise when $D_\mu$ acts
on another covariant derivative $D_\nu$. In this appendix we shall justify this.

The relevant place where the appearance of these Christoffel symbol terms could
affect our analysis is in \refb{e2.14}. Let us suppose that we have a term in the
action of the form
\be \label{esb0}
\int d^D x \, E_{c_1}^{~\mu_1}\cdots E_{c_r}^{~\mu_r} \, \Phi_{a_1\cdots a_m} D_{\mu_1}\cdots D_{\mu_r} \Psi_{b_1\cdots b_n} 
\ee
with the free indices $a_i, b_i, c_i$ contracted with the flat metric
$\eta$. Here $E_c^{~\mu}$ are the inverse vielbeins and $\Phi$ and
$\Psi$ are appropriate tensor fields carrying tangent space indices.  Upon 
expressing $D_{\mu_i}$ as $\p_{\mu_i}+\cdots$, 
a typical term in $\cdots$ that was ignored in
\refb{ex2.4} is of the form
\be \label{esb1}
-\int d^D x \, E_{c_1}^{~\mu_1}\cdots E_{c_r}^{~\mu_r} \, 
\Phi_{a_1\cdots a_m} 
D_{\mu_1}\cdots D_{\mu_{i-1}} \Gamma_{\mu_i \mu_j}^\rho 
D_{\mu_{i+1}}\cdots 
D_{\mu_{j-1}} D_\rho  \, D_{\mu_{j+1}} \cdots  D_{\mu_r} \Psi_{b_1\cdots b_n} \, .
\ee
In the soft limit $\Gamma^\rho_{\mu_i\mu_j}$ carries a factor of the
soft momentum and a factor of 
soft graviton polarization. Therefore we can replace the rest of the covariant
derivatives by ordinary derivatives,  ignore terms involving derivatives of
$\Gamma$ and replace $E_c^{~\mu}$ by $\delta_c^{~\mu}$. 
With this \refb{esb1} reduces to
\be \label{esb1.9}
-\int d^D x \, \delta_{c_1}^{~\mu_1}\cdots \delta_{c_r}^{~\mu_r} \, 
\Gamma_{\mu_i \mu_j}^\rho 
\Phi_{a_1\cdots a_m} 
\p_\rho 
\p_{\mu_1}\cdots \p_{\mu_{i-1}} \p_{\mu_{i+1}}\cdots 
\p_{\mu_{j-1}} \p_{\mu_{j+1}} \cdots  \p_{\mu_r} \Psi_{b_1\cdots b_n} \, .
\ee

Let us now return to \refb{esb0} and, using integration by parts, express this as
\be\label{esb2}
(-1)^r \int d^D x E_{c_1}^{~\mu_1}\cdots E_{c_r}^{~\mu_r} 
\Psi_{b_1\cdots b_n} D_{\mu_1}\cdots D_{\mu_r} \Phi_{a_1\cdots a_m} \, .
\ee
Note that integration by parts will reverse the order in which the covariant 
derivatives act, but since the commutator of two covariant derivatives is
proportional to the Riemann tensor and carries two powers of soft momentum,
we can ignore the reversal of order. For grassmann odd fields there will be an
additional minus sign in \refb{esb2}, but this will cancel with an additional minus
sign that will appear in going from \refb{esb2a} to \refb{esb3}. The fields will
also carry spinor indices contracted with appropriate Lorentz covariant tensors,
but this does not affect the analysis.
By expanding the expression for
$D_{\mu_i}$ in \refb{esb2}
we shall get the analog of \refb{esb1}, and from this the 
analog of \refb{esb1.9}:
\be\label{esb2a}
-(-1)^r \int d^D x \, \delta_{c_1}^{~\mu_1}\cdots \delta_{c_r}^{~\mu_r} \, 
\Gamma_{\mu_i \mu_j}^\rho  \Psi_{b_1\cdots b_n} 
\p_\rho 
\p_{\mu_1}\cdots \p_{\mu_{i-1}} \p_{\mu_{i+1}}\cdots 
\p_{\mu_{j-1}} \p_{\mu_{j+1}} \cdots  \p_{\mu_r} \Phi_{a_1\cdots a_m} \, .
\ee
We shall now again integrate by parts and ignore derivatives of $\Gamma$
since that will generate two powers of soft momentum. This takes 
\refb{esb2a} to
\be \label{esb3}
-(-1)^r (-1)^{r-1} \int d^D x \, 
\delta_{c_1}^{~\mu_1}\cdots \delta_{c_r}^{~\mu_r} \, 
\Gamma_{\mu_i \mu_j}^\rho 
\Phi_{a_1\cdots a_m} 
\p_\rho 
\p_{\mu_1}\cdots \p_{\mu_{i-1}} \p_{\mu_{i+1}}\cdots 
\p_{\mu_{j-1}} \p_{\mu_{j+1}} \cdots  \p_{\mu_r} \Psi_{b_1\cdots b_n} \, .
\ee
We now see that \refb{esb1.9} and \refb{esb3} cancel each other. This shows that
once we express \refb{esb0} as
\be \label{esb00}
{1\over 2} \int d^D x \, E_{c_1}^{~\mu_1}\cdots E_{c_r}^{~\mu_r} \, \left[
\Phi_{a_1\cdots a_m} D_{\mu_1}\cdots D_{\mu_r} \Psi_{b_1\cdots b_n} 
+ (-1)^r \Psi_{b_1\cdots b_n} D_{\mu_r}\cdots D_{\mu_1} \Phi_{a_1\cdots a_m}
\right]\, ,
\ee
the terms involving Christoffel symbols drop out.

The alert reader may worry that the above derivation assumes that the two
point function computed from the 1PI action has the form of a polynomial in 
derivatives while in practice this is not so. 
We can allay this fear by working in momentum
space. Suppose that in the absence of the soft graviton, the 
quadratic term of the 1PI
effective action involving single powers of $\Phi$ and $\Psi$ takes the form
\be \label{esd1}
\int {d^Dp\over (2\pi)^D} \, \Phi_{a_1\cdots a_m}(-p) f_{c_1\cdots c_r}(p)
\Psi_{b_1\cdots b_n}(p)\, ,
\ee
contracted with $\eta$'s. 
Here $\Phi$ and $\Psi$ are Fourier transforms of the fields that appear in
\refb{esb0} and $f$ is some function of the momentum $p$. Then after
coupling to the soft graviton,  the
unwanted terms given in \refb{esb1.9} have the form
\be \label{esd2}
- {1\over 2} 
\int {d^Dp\over (2\pi)^D} \, \Phi_{a_1\cdots a_m}(-p-k) {\p^2 f_{c_1\cdots c_r}(p)
\over \p p_\mu \p p_\nu }
\Psi_{b_1\cdots b_n}(p) \, (-i p_\rho) \, \Gamma^\rho_{\mu\nu}(k)\, ,
\ee
where $\Gamma$ now denotes the Christoffel symbol computed using soft
graviton in the momentum space. In arriving at \refb{esd2} we have used the
fact that in momentum space $\p_\mu$ is replaced by $i \, p_\mu$.

Now by making a $p\to -p$ change of variables in \refb{esd1} we arrive at a
similar formula with the $\Psi$ and $\Phi$ exchanged 
\be \label{esd11}
\int {d^Dp\over (2\pi)^D} \, \Psi_{b_1\cdots b_n}(-p) f_{c_1\cdots c_r}(-p)
\Phi_{a_1\cdots a_m}(p)\, ,
\ee
Its covariantization will generate the
analog of \refb{esd2}
\be \label{esd3}
-{1\over 2} \int {d^Dp\over (2\pi)^D} \, 
\Psi_{b_1\cdots b_m}(-p-k) {\p^2 f_{c_1\cdots c_r}(-p)
\over \p p_\mu \p p_\nu } \Phi_{a_1\cdots n_n}(p) 
(-i p_\rho) \,  \Gamma^\rho_{\mu\nu}(k)
\, .
\ee
Now making a change of variables $p\to -p-k$ we get
\be\label{esd4}
-{1\over 2} \int {d^Dp\over (2\pi)^D} \, 
\Phi_{a_1\cdots n_n}(-p-k){\p^2 f_{c_1\cdots c_r}(p+k)
\over \p p_\mu \p p_\nu }\Psi_{b_1\cdots b_m}(p) \,  
 (i (p_\rho+k_\rho)) \Gamma^\rho_{\mu\nu}(k) \, .
\ee
Averaging over \refb{esd2} and \refb{esd4}, and using the fact that $\Gamma$
already contains one power of soft momentum, we now easily see that the 
integrand has two powers of $k$. Therefore it vanishes to the first 
subleading order
in the soft momentum $k$.

\sectiono{Derivation of \refb{ex.5pre}} \label{sa}

Our goal in this appendix will be to prove the equality of
\refb{e4.13pre} and  \refb{ex.5pre}. We begin by studying some
properties of the  matrix $\KK^{\alpha\beta}(q)$ appearing in the kinetic term
\refb{es2}, and the propagator $D_F$ defined in \refb{epropdef}.
Let us also define $\Xi(q)$ via
\be  \label{e4.14}
 \Xi(q) =  (q^2 + M^2) \, D_F(q) = i\, (q^2 + M^2) \, \KK(q)^{-1} \, ,
\ee
where $M$ is the mass of the external state that we shall be interested in.
$\Xi(q)$ obviously depends on $M$, but this dependence is not displayed explicitly.
At a generic value of $q$, $\Xi(q)$ has the same rank as that of 
$\KK(q)$ or $D_F(q)$,
i.e.\ the total number of fields. But 
in the limit $q^2+M^2\to 0$, we expect $\Xi(q)$ to approach a finite matrix
of rank that is typically less than 
the total number of 
fields, since only a subset of particles have mass $M$ producing a pole in
the propagator $D_F$ at $q^2+M^2=0$.\footnote{For massless particles the 
propagator may have
double poles in some gauges, {\it e.g.} in a generic covariant gauge the propagator
of a massless gauge field is given by $(\eta^{\mu\nu} - \beta \, k^\mu k^\nu / k^2)/k^2$
for some constant $\beta$. We shall assume that our gauge fixing condition is such 
that we avoid propagators with double poles.} 

Using \refb{e4.14} we get
\be \label{ekxi}
\KK(q) \, \Xi(q) = i \, (q^2 + M^2) \, .
\ee
Differentiation of both sides of \refb{ekxi} with respect to $q_\mu$ gives
\be \label{ediffx}
{\p \KK(q)\over \p q_\mu} \, \Xi(q) +   \KK(q) \, {\p \Xi(q)\over \p q_\mu} 
= 2 \, i\, q^\mu\, .
\ee
Now
suppose $\eps_\alpha$ denotes the polarization of an on-shell state
carrying momentum $q$ and mass $M$. Then we have
\be \label{epol}
\eps_\alpha \KK^{\alpha\beta}(q) = 0\, , \quad
\hbox{at $q^2+M^2=0$}\, .
\ee
Combining this with \refb{ediffx} we get
\be \label{ey1.8n}
\eps_\alpha\, 
\,  \left[ {\p \KK (q)\over \p q_\mu}\,  \Xi(q)\right]^{\alpha}_{~\gamma}  = 2 \, i\, 
\eps_\gamma \, q_\mu 
 \quad
\hbox{at $q^2+M^2=0$}\, .
\ee

Next we shall study the consequence of Lorentz invariance.
First of all, since we use a Lorentz covariant gauge fixing condition, the matrix
$\KK$ and $\Xi$ must be Lorentz covariant:
\be \label{e4.9}
\KK^{\alpha\gamma}(q) (J^{ab})_\gamma^{~\beta} + \KK^{\gamma\beta}(q) (J^{ab})_\gamma^{~\alpha}
= q^a {\p \KK^{\alpha\beta}(q)\over \p q_b} -  q^b {\p \KK^{\alpha\beta}(q)\over \p q_a}
\, ,
\ee
\be \label{e4.9xi}
-\, \Xi_{\alpha\gamma}(q) (J^{ab})_\beta^{~\gamma} - \Xi_{\gamma\beta}(q) (J^{ab})_\alpha^{~\gamma}
= q^a {\p \Xi_{\alpha\beta}(q)\over \p q_b} -  q^b {\p \Xi_{\alpha\beta}(q)\over \p q_a}
\, .
\ee
It is easy to see, using \refb{ey1.8n}, \refb{e4.9} and \refb{e4.9xi} that at 
$q^2+M^2=0$,
\ben \label{ex.2}
\eps_\alpha\, (J^{ab})_\beta^{~\alpha}
\,  \left[ {\p \KK (q)\over \p q_\mu}\,  \Xi(q)\right]^{\beta}_{~\gamma}
&=& \eps_\beta \, \left[\left(q^a {\p^2 \KK(q) \over \p q_b \p q_\mu} - q^b {\p^2\KK 
\over \p q_a\p q_\mu }\right)   
\Xi(q)\right]^{\beta}_{~\gamma} \nonumber \\ 
&&
+ \eps_\beta \, \left[{\p \KK (q)\over \p q_\mu}  
\left(q^a {\p \Xi(q) \over \p q_b } - q^b {\p\Xi(q)
\over \p q_a }\right)  
\right]^{\beta}_{~\gamma}
 \nonumber \\ 
&&
+ 2 \, i\, q^\mu \, \eps_\alpha (J^{ab})_\gamma^{~\alpha}- 2\, i \, \eps_\gamma\, \left(q^a \eta^{\mu b} - q^b \eta^{\mu a}\right)   \, .
\een

Next we turn to the analysis of the three point vertex 
$ \Gamma^{(3)\alpha\beta}(\ve, k; p, -p-k)$ given in \refb{e4.8}.
Using \refb{e4.9} we can simplify the second line of \refb{e4.8} and get
\ben \label{e4.11}
\Gamma^{(3)\alpha\beta}(\ve, k; p, -p-k)
&=& {i\over 2} \Bigg[- \ve_{\mu \nu} (p+k)^\nu  {\p \KK^{\alpha\beta} (-p-k)
\over \p p_{\mu}}
- \ve_{\mu \nu} p^\nu  {\p \KK^{\beta\alpha} (p)\over \p p_{\mu}}
\nonumber \\ &&
- {1\over 2} (k_a \, \ve_{b\mu} - k_b \, \ve_{a\mu}) {\p
\KK^{\gamma\beta}(-p-k) \over \p p_{\mu}}
(J^{ab})_\gamma^{~\alpha}\nonumber \\ &&
+ {1\over 2} (k_a \, \ve_{b\mu} - k_b \, \ve_{a\mu}) {\p\over \p p_{\mu}}
\left\{p^a {\p \KK^{\alpha\beta}(-p-k)\over \p p_b} -  
p^b {\p \KK^{\alpha\beta}(-p-k)\over \p p_a}
\right\}
\nonumber \\ &&
- {1\over 2} (k_a \, \ve_{b\mu} - k_b \, \ve_{a\mu}) {\p\KK^{\beta\gamma}(p) 
\over \p p_{\mu}}\left(J^{ab}\right)_{\gamma}^{~\alpha} 
\Bigg]  \, .
\een
Using \refb{eecond}, \refb{ekab}, expanding $\KK^{\alpha\beta}(-p-k)$ in the
first term in a Taylor series expansion in $k_\rho$,
and keeping terms up to first subleading
order in the soft momentum $k$, we can express \refb{e4.11} as
\ben \label{ethis}
\Gamma^{(3)\alpha\beta}(\ve, k; p, -p-k)
&=& {i\over 2} \bigg[- 2 \,\ve_{\mu \nu} p^\nu  {\p \KK^{\alpha\beta} (-p)\over \p p_{\mu}}
- 2\, \ve_{\mu \nu} p^\nu  k_\sigma  {\p^2 \KK^{\alpha\beta} (-p)\over \p p_\sigma
\p p_{\mu}}
\nonumber \\ &&
+k\cdot p \, \ve_{b\mu}  {\p^2 \KK^{\alpha\beta}(-p)\over\p p_{\mu} \p p_b} 
- (k_a \, \ve_{b\mu} - k_b \, \ve_{a\mu}) {\p \KK^{\gamma\beta}(-p) \over \p p_{\mu}}
\left(J^{ab}\right)_{\gamma}^{~\alpha} 
\bigg] \, . \nonumber \\
\een

We now turn to \refb{e4.13pre}.
Substituting \refb{ethis} into \refb{e4.13pre} we get the
net contribution to \refb{e4.13pre} to first subleading order in the soft
momentum:
\ben \label{e4.13}
&& \eps_{i,\alpha}\, \Gamma^{(3)\alpha\beta}(\ve, k; p_i, -p_i-k)\, i 
\{\KK^{-1}(-p_i-k)\}_{\beta\delta} 
\Gamma_{(i)}^{\delta}(p_i+k)\nonumber \\
&=& {1\over 2} \eps_{i,\alpha} 
\bigg[2 \,\ve_{\mu \nu} p_i^\nu  {\p \KK^{\alpha\beta} (-p_i)\over \p p_{i\mu}}
+  2\, \ve_{\mu \nu} p_i^\nu  k_\sigma  {\p^2 \KK^{\alpha\beta} (-p_i)\over \p p_{i\sigma}
\p p_{i\mu}}
-k\cdot p_i \, \ve_{b\mu}  {\p^2 \KK^{\alpha\beta}(-p_i)\over\p p_{i\mu} \p p_{ib}} 
 \nonumber \\ &&+ (k_a \, \ve_{b\mu} - k_b \, \ve_{a\mu}) 
 {\p \KK^{\gamma\beta}(-p_i) 
\over \p p_{i\mu}}\left(J^{ab}\right)_{\gamma}^{~\alpha} 
\bigg] \{\KK^{-1}(-p_i-k)\}_{\beta\delta} \bigg[1 + k_\rho {\p\over \p p_{i\rho}}\bigg]
 \Gamma_{(i)}^{\delta}(p_i)  \, .\nonumber \\
\een
Replacing the $\KK^{-1}(-p_i-k)$ factor
using \refb{ekxi} after setting $M=M_i$, the mass of the $i$-th external
state, and using $(p_i+k)^2+M_i^2 =2p_i\cdot k$,
we may express \refb{e4.13} as
\ben \label{e4.13a}
&&  -{i\over 2} (2 p_i\cdot k)^{-1} \,  \eps_{i,\alpha} 
\bigg[2 \,\ve_{\mu \nu} p_i^\nu  {\p \KK^{\alpha\beta} (-p_i)\over \p p_{i\mu}}
+  2\, \ve_{\mu \nu} p_i^\nu  k_\sigma  {\p^2 \KK^{\alpha\beta} (-p_i)\over \p p_{i\sigma}
\p p_{i\mu}}
-k\cdot p_i \, \ve_{b\mu}  {\p^2 \KK^{\alpha\beta}(-p_i)\over\p p_{i\mu} \p p_{ib}} 
 \nonumber \\ &&+ (k_a \, \ve_{b\mu} - k_b \, \ve_{a\mu}) {\p \KK^{\gamma\beta}(-p_i) 
\over \p p_{i\mu}}\left(J^{ab}\right)_{\gamma}^{~\alpha} 
\bigg] \Xi_{\beta\delta}(-p_i-k) \bigg[1 + k_\rho {\p\over \p p_{i\rho}}\bigg]
 \Gamma_{(i)}^{\delta}(p_i) \nonumber \\
 &=&  -{i\over 2} (2 p_i\cdot k)^{-1} \,  \eps_{i,\alpha} 
\bigg[2 \,\ve_{\mu \nu} p_i^\nu  {\p\KK^{\alpha\beta} (-p_i)\over \p p_{i\mu}}
+  2\, \ve_{\mu \nu} p_i^\nu  k_\sigma  {\p^2 \KK^{\alpha\beta} (-p_i)\over \p p_{i\sigma}
\p p_{i\mu}}
-k\cdot p_i \, \ve_{b\mu}  {\p^2 \KK^{\alpha\beta}(-p_i)\over\p p_{i\mu} \p p_{ib}} 
 \nonumber \\ &&+ (k_a \, \ve_{b\mu} - k_b \, \ve_{a\mu}) {\p\KK^{\gamma\beta}(-p_i) 
\over \p p_{i\mu}}\left(J^{ab}\right)_{\gamma}^{~\alpha} 
\bigg] \left\{\Xi_{\beta\delta}(-p_i) + k_\sigma {\p\Xi_{\beta\delta}(-p_i)
\over \p p_{i\sigma}}
\right\} 
\bigg[1 + k_\rho {\p\over \p p_{i\rho}}\bigg]
 \Gamma_{(i)}^{\delta}(p_i)  \, . \nonumber \\
\een
If $p_i^2+M_i^2=0$, then the $\eps_{i,\alpha}$ in \refb{e4.13a} is a physical
state of mass $M_i$. Therefore we can now use \refb{ey1.8n} to 
express \refb{e4.13a} as
\ben \label{ex.1}
&& 
(p_i\cdot k)^{-1}\, \ve_{\mu \nu}\,  p_i^\mu \,  p_i^\nu 
\eps_{i,\alpha} \Gamma_{(i)}^{\alpha}(p_i)  - {i\over 2}\, (p_i\cdot k)^{-1}\,
\eps_{i,\alpha}  \, 
\ve_{\mu \nu} p_i^\nu  k_\sigma  {\p^2 \KK^{\alpha\beta} (-p_i)\over \p p_{i\sigma}
\p p_{i\mu}} \,  \Xi_{\beta\delta}(-p_i) \, \Gamma_{(i)}^{\delta}(p_i) 
\nonumber \\ &&
+ {i\over 4} \,  \eps_{i,\alpha}  \, 
\ve_{b\mu}  {\p^2 \KK^{\alpha\beta}(-p_i)\over\p p_{i\mu} \p p_{ib}} 
\, \Xi_{\beta\delta}(-p_i) \, \Gamma_{(i)}^{\delta}(p_i) 
\nonumber \\ &&
-{i}\, \eps_{i,\alpha}  \, (2 p_i\cdot k)^{-1} \, \ve_{\mu \nu} 
 p_i^\nu {\p \KK^{\alpha\beta} (-p_i)\over \p p_{i\mu}}\,  k_\rho {\p
 \Xi_{\beta\delta}(-p_i) \over \p p_{i\rho}}   \, 
\Gamma_{(i)}^{\delta}(p_i) 
\nonumber \\ &&
+ (p_i\cdot k)^{-1} \, \ve_{\mu \nu} \, p_i^\mu\, 
 p_i^\nu \, \eps_{i,\alpha}  
\,k_\rho \, {\p\over \p p_{i\rho}} \Gamma_{(i)}^{\alpha}(p_i) 
\nonumber \\ &&
- {i\over 4} (p_i\cdot k)^{-1} \,  \eps_{i,\alpha} 
(k_a \, \ve_{b\mu} - k_b \, \ve_{a\mu}) {\p \KK^{\gamma\beta}(-p_i) \over \p p_{i\mu}}
\left(J^{ab}\right)_{\gamma}^{~\alpha} \,  \Xi_{\beta\delta}(-p_i)\,  
\Gamma_{(i)}^{\delta}(p_i)  \, .
\een
Using \refb{ex.2} and \refb{eecond}
we can manipulate the last line in \refb{ex.1} and express
\refb{ex.1} as
\ben \label{ex.3}
&& 
(p_i\cdot k)^{-1}\, \ve_{\mu \nu}\,  p_i^\mu \,  p_i^\nu 
\eps_{i,\alpha} \Gamma_{(i)}^{\alpha}(p_i)   - {i\over 2}\, (p_i\cdot k)^{-1}\,
\eps_{i,\alpha}  \, 
\ve_{\mu \nu} p_i^\nu  k_\sigma  {\p^2 \KK^{\alpha\beta} (-p_i)\over \p p_{i\sigma}
\p p_{i\mu}} \,  \Xi_{\beta\delta}(-p_i) \, \Gamma_{(i)}^{\delta}(p_i) 
\nonumber \\ &&
+ {i\over 4} \,  \eps_{i,\alpha}  \, 
\ve_{b\mu}  {\p^2 \KK^{\alpha\beta}(-p_i)\over\p p_{i\mu} \p p_{ib}} 
\, \Xi_{\beta\delta}(-p_i) \, \Gamma_{(i)}^{\delta}(p_i) 
\nonumber \\ &&
-{i}\, \eps_{i,\alpha}  \, (2 p_i\cdot k)^{-1} \, \ve_{\mu \nu} 
 p_i^\nu {\p \KK^{\alpha\beta} (-p_i)\over \p p_{i\mu}}\,  k_\rho {\p
 \Xi_{\beta\delta}(-p_i) \over \p p_{i\rho}}   \, 
\Gamma_{(i)}^{\delta}(p_i) 
\nonumber \\ &&
+ (p_i\cdot k)^{-1} \, \ve_{\mu \nu} \, p_i^\mu\, 
 p_i^\nu \, \eps_{i,\alpha}  
\,k_\rho \, {\p\over \p p_{i\rho}} \Gamma_{(i)}^{\alpha}(p_i) 
\nonumber \\ &&
- {i\over 2} (p_i\cdot k)^{-1} \,  k_a \, \ve_{b\mu} \, \eps_{i,\alpha} 
\left[ \left\{ p_i^a {\p^2 \KK(-p_i)\over \p p_{ib} \p p_{i\mu}}
- p_i^b {\p^2 \KK(-p_i)\over \p p_{ia} \p p_{i\mu}}\right\} \Xi(-p_i)\right]^\alpha_{~\gamma}
 \Gamma_{(i)}^{\gamma}(p_i) 
\nonumber \\ &&
- {i\over 2} (p_i\cdot k)^{-1} \,  k_a \, \ve_{b\mu} \, \eps_{i,\alpha} 
\left[ {\p \KK(-p_i)\over \p p_{i\mu}} \left\{ p_i^a {\p \Xi(-p_i)\over \p p_{ib}}
- p_i^b {\p \Xi(-p_i)\over \p p_{ia}}
\right\}\right]^\alpha_{~\gamma}
 \Gamma_{(i)}^{\gamma}(p_i) 
\nonumber \\ &&
+ (p_i\cdot k)^{-1} \, k_a \, \ve_{b\mu} \, p_i^\mu \, \eps_{i,\alpha} \,
(J^{ab})_\gamma^{~\alpha} \, 
\Gamma_{(i)}^\gamma(p_i) \, .
\een
This can be simplified to
\ben \label{ex.4}
&& 
(p_i\cdot k)^{-1}\, \ve_{\mu \nu}\,  p_i^\mu \,  p_i^\nu 
\eps_{i,\alpha} \Gamma_{(i)}^{\alpha}(p_i)  
- {i\over 4} \,  \eps_{i,\alpha}  \, 
\ve_{b\mu}  {\p^2 \KK^{\alpha\beta}(-p_i)\over\p p_{i\mu} \p p_{ib}} 
\, \Xi_{\beta\delta}(-p_i) \, \Gamma_{(i)}^{\delta}(p_i) 
\nonumber \\ &&
+ (p_i\cdot k)^{-1} \, \ve_{\mu \nu} \, p_i^\mu\, 
 p_i^\nu \, \eps_{i,\alpha}  
\,k_\rho \, {\p\over \p p_{i\rho}} \Gamma_{(i)}^{\alpha}(p_i) 
 - {i\over 2} \, \ve_{b\mu} \, 
 \eps_{i,\alpha} 
\left[ {\p \KK(-p_i)\over \p p_{i\mu}}  \, {\p \Xi(-p_i)\over \p p_{ib}}
\right]^\alpha_{~\gamma}
 \Gamma_{(i)}^{\gamma}(p_i) 
\nonumber \\ &&
+ (p_i\cdot k)^{-1} \, k_a \, \ve_{b\mu} \, p_i^\mu \, 
\eps_{i,\alpha}\, (J^{ab})_\gamma^{~\alpha} \,
\Gamma_{(i)}^\gamma(p_i) \, .
\een
The sum of the second and fourth term of \refb{ex.4} may be written as
\be
 - {i\over 4} \,  \eps_{i,\alpha}  \, 
\ve_{b\mu}\, {\p^2 \over\p p_{i\mu} \p p_{ib}} \left[ \KK^{\alpha\beta}(-p_i)
\, \Xi_{\beta\gamma}(-p_i)\right] \, \Gamma_{(i)}^{\gamma}(p_i) 
 +  {i\over 4} \,  \eps_{i,\alpha}  \, 
\ve_{b\mu}\, \KK^{\alpha\beta}(-p_i) \, {\p^2 \Xi_{\beta\gamma} (-p_i)
\over\p p_{i\mu} \p p_{ib}} 
 \, \Gamma_{(i)}^{\gamma}(p_i) \, .
\ee
Using \refb{ekxi} the first term can be shown to be 
proportional to $\ve_{b\mu}\eta^{b\mu}$ and hence it vanishes
due to \refb{eecond}. 
On the other hand the second term vanishes due to \refb{eonshell}. This
allows us to express \refb{ex.4} as
\ben \label{ex.5}
&& 
(p_i\cdot k)^{-1}\, \ve_{\mu \nu}\,  p_i^\mu \,  p_i^\nu 
\eps_{i,\alpha} \Gamma_{(i)}^{\alpha}(p_i)  
+ (p_i\cdot k)^{-1} \, \ve_{\mu \nu} \, p_i^\mu\, 
 p_i^\nu \, \eps_{i,\alpha}  
\,k_\rho \, {\p\over \p p_{i\rho}} \Gamma_{(i)}^{\alpha}(p_i) 
\nonumber \\ & + &
(p_i\cdot k)^{-1} \, k_a \, \ve_{b\mu} \, p_i^\mu \, 
\eps_{i,\alpha} \, (J^{ab})_\gamma^{~\alpha} \, 
\Gamma_{(i)}^\gamma(p_i) \, .
\een
This proves \refb{ex.5pre}.

\end{document}